\documentclass[twocolumn]{aastex63}
\pdfoutput=1
\usepackage{float, bm, graphicx, amsmath, morefloats, yhmath}
\usepackage[caption=false]{subfig}
\usepackage{pgfplotstable,booktabs} 
\usepackage{threeparttable,threeparttablex,longtable,graphicx}
\PassOptionsToPackage{hyphens}{url}\usepackage{hyperref}
\usepackage{xurl}
\usepackage{soul}
\turnoffeditone
\pgfplotsset{compat=1.18}

\newcommand{\pim}{AT\,2019pim}
\newcommand{\blt}{AT\,2020blt}
\newcommand{\any}{AT\,2021any}
\newcommand{\lfa}{AT\,2021lfa}
\newcommand{\lcr}{AT\,2023lcr}
\newcommand{\afterglowpy}{\texttt{afterglowpy}}
\newcommand{\emcee}{\texttt{emcee}}

\newcommand{\Egiso}{$E_{\gamma,\text{iso}}$}
\newcommand{\Ekiso}{$E_\text{K,iso}$}
\newcommand{\elpd}{$\widehat{\text{elpd}}$}
\newcommand{\thetac}{$\theta_\text{c}$}

\begin{document}

\title{The Nature of Optical Afterglows Without Gamma-ray Bursts: \\ Identification of AT2023lcr and Multiwavelength Modeling}

\author[0009-0001-6911-9144]{Maggie L.~Li}
\correspondingauthor{Maggie L.~Li}
\email{maggieli@caltech.edu}
\affiliation{Department of Astronomy, Cornell University, Ithaca, NY 14853, USA}
\affiliation{Cahill Center for Astrophysics, California Institute of Technology, MC 249-17, 1200 E California Boulevard, Pasadena, CA, 91125, USA}

\author[0000-0002-9017-3567]{Anna Y. Q.~Ho}
\affiliation{Department of Astronomy, Cornell University, Ithaca, NY 14853, USA}

\author[0000-0001-9068-7157]{Geoffrey~Ryan}
\affiliation{Perimeter Institute for Theoretical Physics, Waterloo, Ontario N2L 2Y5, Canada}

\author[0000-0001-8472-1996]{Daniel A.~Perley}
\affiliation{Astrophysics Research Institute, Liverpool John Moores University, IC2, Liverpool Science Park, 146 Brownlow Hill, Liverpool L3 5RF, UK}

\author[0000-0001-5169-4143]{Gavin P.~Lamb}
\affiliation{Astrophysics Research Institute, Liverpool John Moores University, IC2, Liverpool Science Park, 146 Brownlow Hill, Liverpool L3 5RF, UK}

\author[0000-0002-8070-5400]{Nayana A.J.}
\affiliation{National Centre for Radio Astrophysics, Tata Institute of Fundamental Research, Pune University Campus, Ganeshkhind Pune 411007, India}

\author[0000-0002-8977-1498]{Igor~Andreoni}
\affiliation{Joint Space-Science Institute, University of Maryland, College Park, MD 20742, USA}
\affiliation{Department of Astronomy, University of Maryland, College Park, MD 20742, USA}
\affiliation{Astrophysics Science Division, NASA Goddard Space Flight Center, 8800 Greenbelt Rd, Greenbelt, MD 20771, USA}
\affiliation{University of North Carolina at Chapel Hill, 120 E. Cameron Ave., Chapel Hill, NC 27514, USA}

\author[0000-0003-3533-7183]{G. C.~Anupama}
\affiliation{Indian Institute of Astrophysics, II Block Koramangala, Bengaluru 560034, India}

\author[0000-0001-8018-5348]{Eric C.~Bellm}
\affiliation{DIRAC Institute, Department of Astronomy, University of Washington, 3910 15th Avenue NE, Seattle, WA 98195, USA}

\author[0000-0002-9392-9681]{Edo~Berger}
\affiliation{Center for Astrophysics \textbar{} Harvard \& Smithsonian, 60 Garden Street, Cambridge, MA 02138-1516, USA}

\author[0000-0002-7777-216X]{Joshua S.~Bloom}
\affiliation{Department of Astronomy, University of California, Berkeley, CA 94720, USA}
\affiliation{Physics Division, Lawrence Berkeley National Laboratory, 1 Cyclotron Road, MS 50B-4206, Berkeley, CA 94720, USA}

\author[0000-0002-2942-3379]{Eric~Burns}
\affiliation{Department of Physics \& Astronomy, Louisiana State University, Baton Rouge, LA 70803, USA}

\author[0000-0002-4770-5388]{Ilaria~Caiazzo}
\affiliation{Cahill Center for Astrophysics, California Institute of Technology, MC 249-17, 1200 E California Boulevard, Pasadena, CA, 91125, USA}

\author[0000-0002-0844-6563]{Poonam~Chandra}
\affiliation{National Radio Astronomy Observatory, 520 Edgemont Rd, Charlottesville VA 22903, USA}

\author[0000-0002-8262-2924]{Michael W.~Coughlin}
\affiliation{School of Physics and Astronomy, University of Minnesota, Minneapolis, MN 55455, USA}

\author[0000-0002-6871-1752]{Kareem~El-Badry}
\affiliation{Cahill Center for Astrophysics, California Institute of Technology, MC 249-17, 1200 E California Boulevard, Pasadena, CA, 91125, USA}

\author[0000-0002-3168-0139]{Matthew J.~Graham}
\affiliation{Cahill Center for Astrophysics, California Institute of Technology, MC 249-17, 1200 E California Boulevard, Pasadena, CA, 91125, USA}

\author[0000-0002-5619-4938]{Mansi~Kasliwal}
\affiliation{Cahill Center for Astrophysics, California Institute of Technology, MC 249-17, 1200 E California Boulevard, Pasadena, CA, 91125, USA}

\author[0000-0002-3490-146X]{Garrett K.~Keating}
\affiliation{Center for Astrophysics \textbar{} Harvard \& Smithsonian, 60 Garden Street, Cambridge, MA 02138-1516, USA}

\author[0000-0001-5390-8563]{S. R.~Kulkarni}
\affiliation{Cahill Center for Astrophysics, California Institute of Technology, MC 249-17, 1200 E California Boulevard, Pasadena, CA, 91125, USA}

\author[0000-0003-0871-4641]{Harsh~Kumar}
\affiliation{Physics Department, Indian Institute of Technology Bombay, Powai, 400 076, India}
\affiliation{Center for Astrophysics \textbar{} Harvard \& Smithsonian, 60 Garden Street, Cambridge, MA 02138-1516, USA}
\affiliation{The NSF AI Institute for Artificial Intelligence and Fundamental Interactions, USA}

\author[0000-0002-8532-9395]{Frank J.~Masci}
\affiliation{IPAC, California Institute of Technology, 1200 E. California
             Blvd, Pasadena, CA 91125, USA}

\author{Richard A.~Perley}
\affiliation{National Radio Astronomy Observatory, PO Box 0, Socorro, NM 87801, USA}

\author[0000-0003-1227-3738]{Josiah~Purdum}
\affiliation{Caltech Optical Observatories, California Institute of Technology, Pasadena, CA 91125, USA}

\author[0000-0002-1407-7944]{Ramprasad~Rao}
\affiliation{Center for Astrophysics \textbar{} Harvard \& Smithsonian, 60 Garden Street, Cambridge, MA 02138-1516, USA}

\author[0000-0003-4189-9668]{Antonio C.~Rodriguez}
\affiliation{Cahill Center for Astrophysics, California Institute of Technology, MC 249-17, 1200 E California Boulevard, Pasadena, CA, 91125, USA}

\author[0000-0001-7648-4142]{Ben~Rusholme}
\affiliation{IPAC, California Institute of Technology, 1200 E. California
             Blvd, Pasadena, CA 91125, USA}
             
\author[0000-0003-2700-1030]{Nikhil~Sarin}
\affiliation{Nordita, Stockholm University and KTH Royal Institute of Technology, Hannes Alfvens vag 12, SE-106 91 Stockholm, Sweden}
\affiliation{Oskar Klein Centre for Cosmoparticle Physics, Department of Physics, Stockholm University, AlbaNova, Stockholm SE-106 91, Sweden}

\author[0000-0003-1546-6615]{Jesper~Sollerman}
\affiliation{Oskar Klein Centre, Department of Astronomy, Stockholm University, AlbaNova, SE-10691 Stockholm, Sweden}

\author[0000-0002-6428-2700]{Gokul P.~Srinivasaragavan}
\affiliation{Department of Astronomy, University of Maryland, College Park, MD 20742, USA}
\affiliation{Joint Space-Science Institute, University of Maryland, College Park, MD 20742, USA}
\affiliation{Astrophysics Science Division, NASA Goddard Space Flight Center, 8800 Greenbelt Rd, Greenbelt, MD 20771, USA}

\author[0000-0002-7942-8477]{Vishwajeet~Swain}
\affiliation{Department of Physics, Indian Institute of Technology Bombay, Powai, 400 076, India}

\author[0000-0002-0853-3464]{Zachary~Vanderbosch}
\affiliation{Cahill Center for Astrophysics, California Institute of Technology, MC 249-17, 1200 E California Boulevard, Pasadena, CA, 91125, USA}

\begin{abstract}
In the past few years, the improved sensitivity and cadence of wide-field optical surveys have enabled the discovery of several afterglows without associated detected gamma-ray bursts (GRBs). We present the identification, observations, and multi-wavelength modeling of a recent such afterglow (AT\,2023lcr), and model three literature events (AT\,2020blt, AT\,2021any, and AT\,2021lfa) in a consistent fashion. For each event, we consider the following possibilities as to why a GRB was not observed: 1) the jet was off-axis; 2) the jet had a low initial Lorentz factor; and 3) the afterglow was the result of an on-axis classical GRB (on-axis jet with physical parameters typical of the GRB population), but the emission was undetected by gamma-ray satellites. We estimate all physical parameters using \texttt{afterglowpy} and Markov Chain Monte Carlo methods from \texttt{emcee}. We find that AT\,2023lcr, AT\,2020blt, and AT\,2021any are consistent with on-axis classical GRBs, and AT\,2021lfa is consistent with both on-axis low Lorentz factor ($\Gamma_0 \approx 5 - 13$) and off-axis ($\theta_\mathrm{obs}=2\theta_\mathrm{jet}$) high Lorentz factor ($\Gamma_0 \approx 100$) jets.
\end{abstract}

\section{Introduction}
\label{sec:intro}
In the internal-external shocks model for long-duration gamma-ray bursts (LGRBs; \citealt{Piran_2005, Meszaros_2006, Kumar_2015}), the core of a massive star collapses and forms a neutron star or black hole, which launches an ultra-relativistic collimated outflow, or ``jet". The jet's internal collisions produce an initial burst of gamma-rays, called the ``prompt emission", followed by the jet's external collision with the ambient medium, producing an ``afterglow" across the electromagnetic spectrum.

There are several reasons why we should be able to detect afterglows without associated detected GRBs. 
First, the Earth might not be within the jet's opening angle (typically $\theta_c = 5-10^\circ$; \citealt{Ghirlanda_2018}), which is collimated and relativistically beamed (initial $\theta_c = \Gamma_0^{-1}$) \citep{Totani_2002}. In this ``off-axis" GRB scenario, we will miss the prompt emission but still be able to observe an afterglow when the jet decelerates and spreads \citep{Rhoads1997}. Second, a ``dirty fireball" can occur if the outflow is less relativistic than that of a typical GRB ($\Gamma_0 \lesssim 100$) due to a baryon-loaded jet \citep{Dermer_1999, Huang_2002, Rhoads_2003}. In this case, the jet will be below the pair-production threshold for gamma-rays (i.e. the compactness problem; \citealt{Ruderman_1975}), so we would not be able to observe a GRB. Still, we might observe a less energetic prompt emission, such as an X-ray flash \citep{Dermer_1999, Heise_2001, Zhang_2004, Sakamoto_2005, Soderberg_2007}. Third, the source could be an on-axis classical GRB whose prompt emission was undetected by gamma-ray satellites, possibly due to the occultation of the Earth or a weak prompt emission that failed to meet the triggering thresholds of gamma-ray satellites. 

In recent years, high-cadence optical surveys have enabled the discovery of ten likely afterglows without associated detected GRBs, summarized in Table~\ref{tab:orphan-summary}. Prior to the Zwicky Transient Facility (ZTF; \citealt{Graham2019, Bellm2019_ztf, Bellm2019_surveys, Dekany2020, Masci2019}), only one such event, PTF\,11agg \citep{Cenko_2013}, was discovered, found by the Palomar Transient Factory \citep{Law_2009}. Since ZTF's first light, nine other events have been confirmed as afterglows without associated detected GRBs, largely thanks to ZTF's high cadence over a wide field-of-view, enabling the rapid identification of fast transients. Still, 
no convincing dirty fireballs or off-axis LGRB candidates have been discovered.

In this paper, we present the identification, follow-up, and multi-wavelength modeling of one of the most recent such events, {\lcr}. We only consider the afterglow light curve, although an associated Ic-BL supernova was identified at a later time \citep{lcr-supernova-photometry}. As shown in Table~\ref{tab:orphan-summary}, {\lcr} is one of six afterglows discovered in optical-survey data with no detected GRB but with a measured redshift. To put our {\lcr} results into context, we also present multi-wavelength modeling of three afterglows in Table~\ref{tab:orphan-summary}: {\blt}, {\any}, and {\lfa}. For {\pim}, we refer the reader to \citet{Perley_2024}, who used a similar approach to this work; for AT\,2023sva, we refer the reader to \citep{2023sva}. 
All afterglows are modeled using Markov Chain Monte Carlo (MCMC) methods in {\emcee} (version 3.1.4; \citealt{Foreman_Mackey_2013}) and afterglow models from {\afterglowpy} (version 0.7.3; \citealt{Ryan_2020}). 

\begin{deluxetable*}{lcll}
\tablecaption{Summary of afterglows discovered without associated detected GRBs. \label{tab:orphan-summary}} 
\tablewidth{0pt} 
\tablehead{\colhead{Afterglow} & \colhead{Redshift} & \colhead{Ref.} & \colhead{Proposed Models}} 
\tabletypesize{\small} 
\startdata 
PTF\,11agg & - & [1] & on-axis, untriggered GRB [1]; dirty fireball [1]; neutron star merger [2, 3]\\
{\pim} & 1.2596 & [4] & on-axis jet with $\Gamma_0 \approx 30-50$; off-axis GRB with $\Gamma_0 \approx 100$ [4]\\
{\blt} & 2.9 & [5] & on-axis GRB with $\eta_\gamma < 0.3 - 14.5\%$ [6]; on-axis classical GRB [0]\\
{\any} & 2.5131 & [7] & on-axis classical GRB [8; 0]; on-axis moderately dirty fireball [9; 0] \\
{\lfa} & 1.063 & [7, 10] & on-axis jet with $\Gamma_0 \sim 20$ [10, 11, 0]; off-axis GRB with $\Gamma_0 \approx 100$ [0] \\
AT\,2023avj & - & [12, 13] & - \\
AT\,2023azs & - & [14, 15] & - \\
AT\,2023jxk & - & [16, 17] & - \\
AT\,2023lcr & 1.0272 & [0] & on-axis GRB with $\eta_\gamma < 0.95\%$ [0] \\
AT\,2023sva & 2.281 & [18, 19, 20, 21, 22] & slightly off-axis structured jet [22]
\enddata 
\tablerefs{[0] this work, [1] \citet{Cenko_2013}, [2] \citet{Wang_2013}, [3] \citet{Wu_2013}, [4] \citet{Perley_2024}, [5] \citet{Ho_2020}, [6] \citet{Sarin_2022}, [7] \citet{Ho_2022}, [8] \citet{Gupta_2022}, [9] \citet{Xu_2023}, [10] \citet{Lipunov_2022}, [11] \citet{Ye_2024}, [12] \citet{Wang2023_AT2023avj}, [13] \citet{Ho2023_AT2023avj}, [14] \citet{Andreoni2023_AT2023azs}, [15] \citet{Perley2023_AT2023azs}, [16] \citet{Vail2023_AT2023jxk}, [17] \citet{Sfaradi2023_AT2023jxk}, [18] \citet{Vail2023_AT2023sva}, [19] \citet{deUgartePostigo2023_AT2023sva}, [20] \citet{Rhodes2023_AT2023sva}, [21] \citet{Roberts2023_AT2023sva}, [22] \citet{2023sva}.}
\end{deluxetable*}

For each object, we consider the following explanations for why their prompt gamma-ray emission was missed: 1) the jet was off-axis; 2) the jet had a low Lorentz factor ($\Gamma_0 \lesssim 100$); and 3) the afterglow was the result of an on-axis classical GRB but the GRB was undetected by high-energy satellites.
Because of imprecise constraints on the burst time, the possibility of an on-axis classical GRB cannot be ruled out for any of the afterglows on the basis of gamma-ray limits alone \citep{Ho_2020, Ho_2022}.

{\blt} was previously modeled in \citet{Sarin_2022} (hereafter S22); {\any} was previously modeled in \citet{Gupta_2022} (G22) and \citet{Xu_2023} (X23); {\lfa} was previously modeled in \citet{Ye_2024} (Y24). In this work\footnote{Code and data products can be found in the repository \href{https://github.com/liluhua2/afterglowfit-public}{https://github.com/liluhua2/afterglowfit-public}.}, we explore additional jet structures, modeling configurations, and constraints on afterglow behavior, discussing comparisons between the mentioned works and this work in Section~\ref{sec:results}.

This paper is organized as follows: we present observations of {\lcr} in Section~\ref{sec:at2023lcr}. We describe observational features of {\lcr} in Section~\ref{sec:observational-features}. We describe our fitting framework in Section~\ref{sec:method}. In Section~\ref{sec:results}, we present the results of our fitting, discuss preferred models and physical interpretations, and compare our results to past works. Finally, we summarize and discuss implications and future work in Section~\ref{sec:conclusion}.

\section{AT2023lcr Observations}
\label{sec:at2023lcr}

\subsection{Optical Photometry}
{\lcr} was initially reported \citep{Tonry2023,Fulton2023} to the Transient Name Server (TNS) by the ATLAS survey \citep{Tonry2018} as ATLAS23msn. {\lcr} was also detected as part of the ZTF high-cadence partnership survey \citep{Bellm2019_surveys}, with the first detection at\footnote{All times in this paper are in UT.} 06:36:27 on 2023 June 18 (internal name ZTF23aaoohpy) at a position $\alpha= 16$:31:37.416 and $\delta= +$26:21:58.31 (J2000) and a Galactic latitude $b=41.25$ deg \citep{lcr-swain}. {\lcr} was flagged as a transient of interest due to its rapid rise ($>1.2\,$mag\,d$^{-1}$ in $r$ band\footnote{All magnitudes are in AB unless specified otherwise.}), red colors ($g-r=0.29\pm0.08\,$mag, corrected for Milky Way extinction with $E(B-V)=A_V/R_V=0.05$, assuming $R_v = 3.1$; \citealt{Schlegel_1998}), and lack of a bright host-galaxy counterpart. 

The red colors exhibited by AT\,2023lcr were consistent with the synchrotron emission expected from an afterglow-like transient, which has been shown to be a useful discriminant from stellar flares in the Milky Way \citep{Ho_2020}. Liverpool Telescope (LT; \citealt{Steele2004}) IO:O imaging observations were attempted to confirm the synchrotron-like colors and check for rapid fading (as expected for an afterglow) but the telescope was offline due to a power supply problem. Confirmation of the red colors and rapid fading was obtained by ZTF through routine survey operations the following night: the transient faded by approximately one magnitude in both $g$- and $r$-band, and this behavior was flagged \citep{Swain2023} by the ZTFReST pipeline \citep{Andreoni2021}. 

\begin{figure*}[ht]
    \centering
    \subfloat{\includegraphics[width=8.5cm]{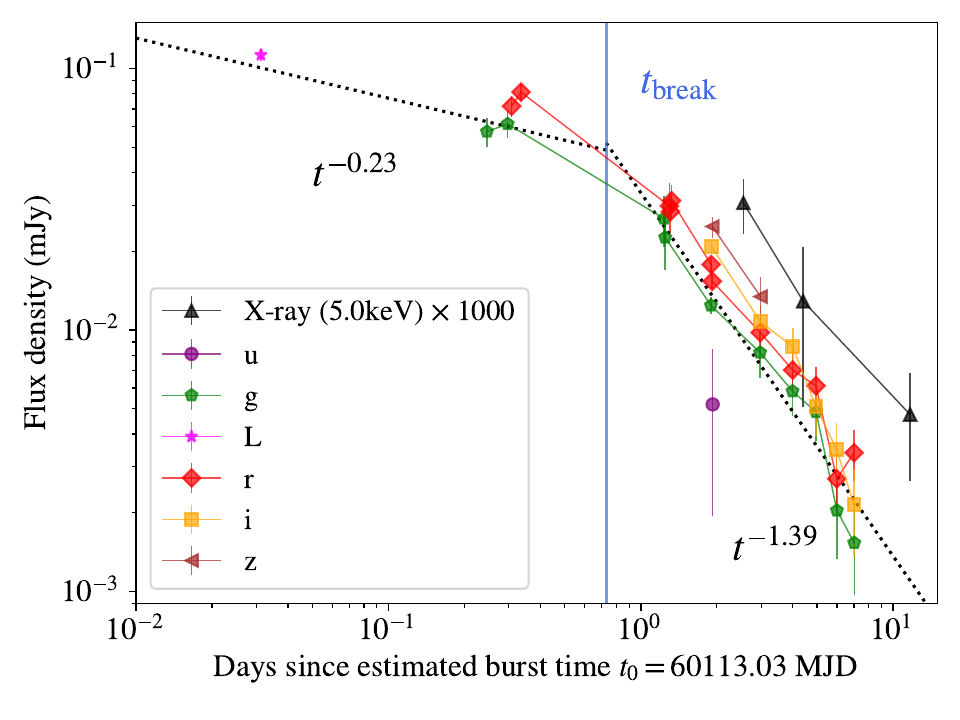}}%
    \qquad
    \subfloat{\includegraphics[width=8.5cm]{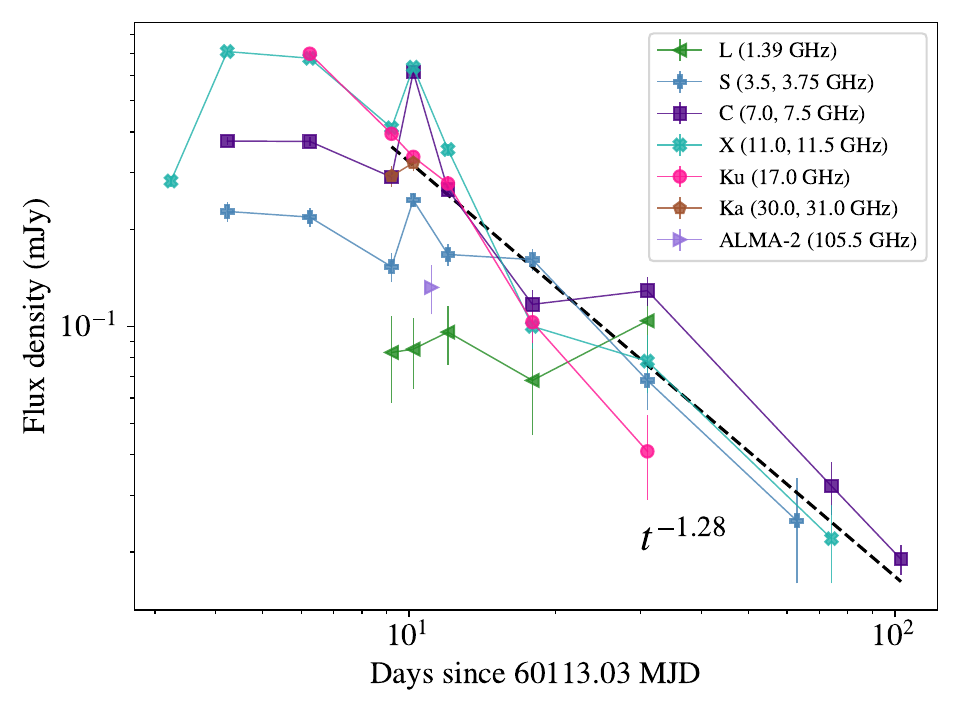}}%
    \caption{\textbf{Left}: The optical light curve of {\lcr} with the best-fit broken power law to $g$, $r$, $i$, and $z$-band observations. The vertical line marks the best-fit break time. \textbf{Right}: The radio light curve of {\lcr} with the best-fit power law to all bands shown. For each band, we select a frequency with observations in the most number of epochs. The early radio emission is likely impacted by interstellar scintillation.}%
    \label{fig:lcr-indices}
\end{figure*}

The Gravitational-wave Optical Transient Observer (GOTO; \citealt{Steeghs2022}) reported \citep{Gompertz2023} an early detection of AT\,2023lcr at 01:27:41 on 2023 June 18 (60113.06089\,MJD), five hours before the first ZTF detection, at $18.77\pm0.06$ in the L-band (400--700\,nm). It was not detected in the previous GOTO epoch at 23:50:30 on 2023 June 17 with a 5-$\sigma$ limiting magnitude of $L>20.3\,$mag, establishing a short window for the onset time of 1 hour and 38 minutes. 

Optical photometric follow-up observations were obtained during the week following the initial detection. Table~\ref{tab:optical-photometry} in Appendix~\ref{tab:optical-photometry} presents the LT, GROWTH-India Telescope (GIT), and Himalayan Chandra Telescope (HCT) photometry. To correct for Milky Way extinction we use $A_V=0.128$\,mag \citep{Schlafly2011}. The full optical light curve of {\lcr} is shown in Figure~\ref{fig:lcr-indices}. Follow-up observations were coordinated using the SkyPortal platform \citep{vanderWalt2019,SkyPortal}.

Later, on 2023 August 12, a James Webb Space Telescope/NIRSPec spectrum was taken \citep{lcr-supernova-spectrum}, which identified a Ic-BL supernova (SN) counterpart for AT\,2023lcr. In this work, we only consider the afterglow. Since the source is at a redshfit $z=1.0272$, we assume that the contribution of the SN is negligible in our observations, supported by the lack of observed flattening in the optical light curve.

\subsection{Optical Spectroscopy}
\begin{figure*}[ht]
    \centering
    \includegraphics[width=\textwidth]{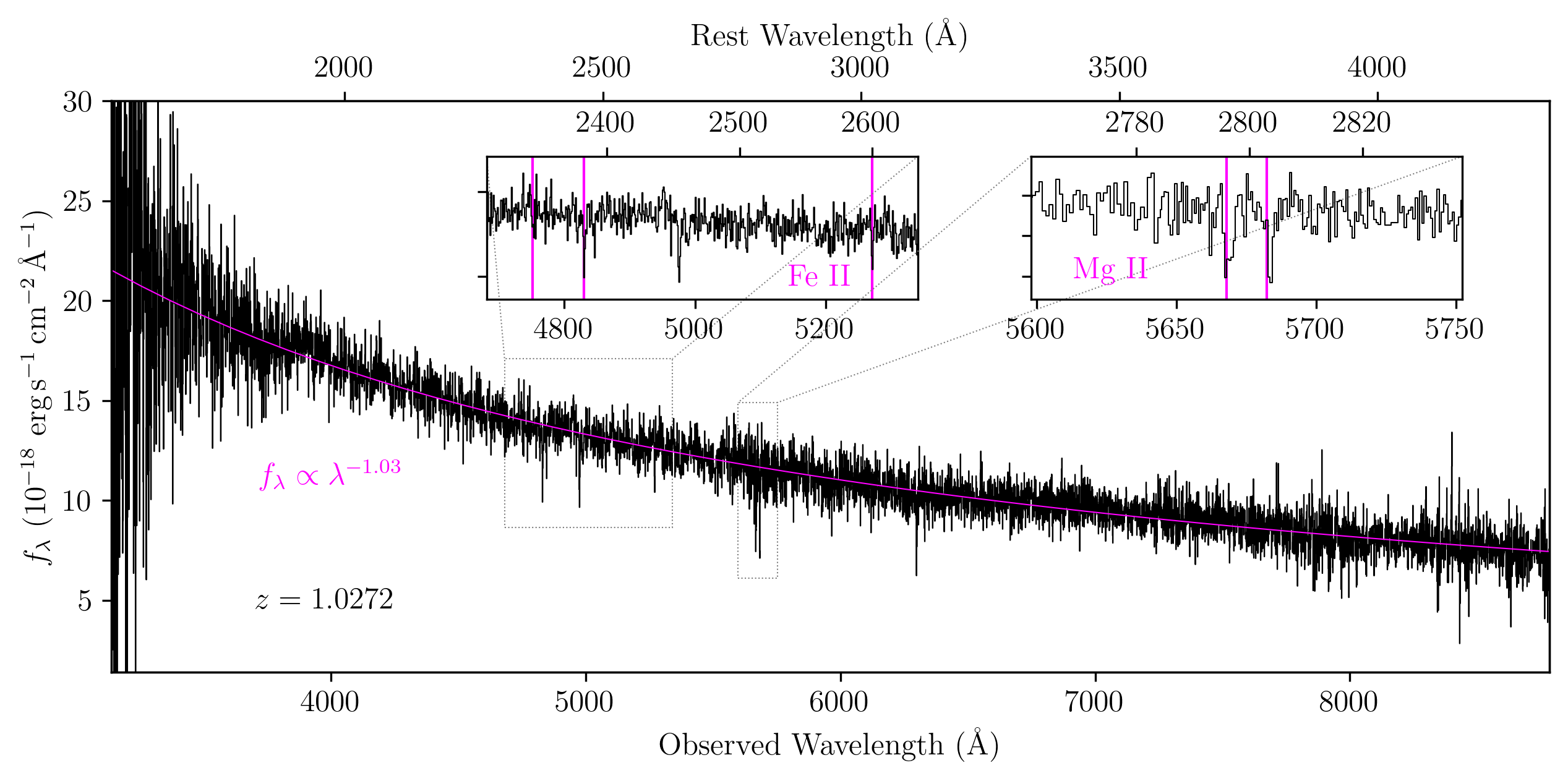}
    \caption{Keck/LRIS spectrum of {\lcr}, with the best-fit power-law index shown as a solid line. The insets show zoom-ins of the regions used to measure the redshift.}
    \label{fig:spec}
\end{figure*}

We triggered observations\footnote{PI K. El-Badry.} using the Low Resolution Imaging Spectrometer (LRIS; \citealt{Oke1995}) on the Keck~I 10-m telescope.
Observations started at 07:22 on 2023 June 20 ($\Delta t = 2.3$\,d from the last GOTO non-detection\footnote{In this work, $\Delta t$ is the observer-frame time in days since the afterglow's last non-detection, unless specified otherwise. For {\lcr}, the last GOTO non-detection was at $60112.99340$\,MJD.}\footnote{In this work, all times are observer-frame unless specified otherwise.}), with exposure times of $2\times 1200$\,s and $3\times 800$\,s in the blue and red arms, respectively. Observations employed the 600/4000 blue grism and 600/7500 red grating, providing continuous wavelength coverage from 3140--8784\,\AA. Data were reduced using LPipe \citep{Perley2019lpipe}. The spectrum (Figure~\ref{fig:spec}) shows a simple continuum, well fit by a power law of $f_\lambda \propto \lambda^{-1}$. 
The signal-to-noise ratio is about 20 per resolution element (although lower blueward of 4000\,\AA). 
We detect clear (but weak) absorption lines at observer-frame wavelengths of 5688 and 5683\,\AA\ which we attribute to redshifted \ion{Mg}{2} $\lambda\lambda$2796, 2803 at $z=1.0272$. Weak absorption from \ion{Fe}{2} $\lambda$2383 and \ion{Fe}{2} $\lambda$2600 at a consistent redshift is also securely detected, and \ion{Fe}{2} $\lambda$2344 is marginally detected. We also detect a possible intervening \ion{Mg}{2} absorber at $z=0.7795$. No other lines are apparent in the spectrum.
Based on this information, we adopt $z=1.0272$ as the redshift of {\lcr}, and the corresponding luminosity distance\footnote{$\Lambda$CDM cosmology of \citet{Planck_2020} is used throughout.} as $D_L=7.0196\,$Gpc. While strictly this redshift is only a lower limit, the absence of any higher-redshift absorption features suggests that a higher-redshift origin is unlikely. The lack of Lyman-$\alpha$ absorption over the spectral range imposes a redshift upper limit of $z<1.6$.

\subsection{X-rays}
\label{sec:obs-xrays}

\begin{table*}
\centering
\caption{0.3-10 keV X-ray observations {\lcr}.            Uncertainties are 68\%.}
\label{tab:xray}
\begin{tabular}{lcccc}
\toprule
  Instrument & Start Date (MJD) &  Exposure Time (ks) & Flux ($10^{-13}$\,erg\,s$^{-1}$\,cm$^{-2}$) & $f_\nu$ (1\,keV; $10^{-2}\,\mu$Jy) \\
\midrule
   Swift/XRT &      60115.58090 &                3.00 &                      $3.70^{+0.88}_{-0.88}$ &           $4.40^{+1.05}_{-1.05}$ \\
   Swift/XRT &      60117.43457 &                3.30 &                      $1.56^{+0.94}_{-0.70}$ &           $1.85^{+1.12}_{-0.83}$ \\
   Swift/XRT &      60124.71261 &                9.10 &                      $0.57^{+0.25}_{-0.21}$ &           $0.68^{+0.30}_{-0.24}$ \\
Chandra/ACIS &      60128.27391 &               18.83 &                      $0.28^{+0.10}_{-0.05}$ &           $0.33^{+0.11}_{-0.06}$ \\
\bottomrule
\end{tabular}
\end{table*}

We triggered observations of {\lcr} with the X-ray Telescope (XRT; \citealt{Burrows2005}) on board the \emph{Neil Gehrels Swift Observatory} through SkyPortal \citep{vanderWalt2019,SkyPortal}.
In total six epochs of observations were obtained under target-of-opportunity programs\footnote{TOO IDs 18987 and 18992, PIs M. Coughlin and D. Malesani.}, from 2023 June 20 to 2023 July 06 ($\Delta t=2$--18\,d). The source was detected in three of those epochs, presented in Table~\ref{tab:xray}.
Fitting the detections in the three epochs simultaneously using the online Swift tool \citep{Evans2007,Evans2009}, with a Galactic hydrogen column density of $N_H=4.12\times10^{20}\,$cm$^{-2}$, we find
a best-fit photon index of $\Gamma=1.8^{+0.8}_{-0.5}$ (90\% confidence interval). To convert from count rate to flux density we take $\Gamma=2$, giving a counts to flux conversion factor (unabsorbed) of $3.93\times10^{-11}$\,erg\,cm$^{-2}$\,ct$^{-1}$.
This photon index corresponds to a spectral index of $\beta_X=1-\Gamma=-1$ where $f_\nu \propto \nu^{\beta_X}$.

{\lcr} was also observed by the Advanced CCD imaging spectrometer (ACIS; \citealt{Garmire2003}) on board the Chandra X-Ray Observatory (Chandra)
\footnote{This paper employs a list of Chandra datasets, obtained by the Chandra X-ray Observatory, contained in Chandra Data Collection~\dataset[doi:10.25574/cdc.364]{https://doi.org/10.25574/cdc.364}.}
under a Director's Discretionary Time proposal\footnote{Proposal Number 24508916, PI A. Martin-Carrillo.}, four days after the final \emph{Swift}/XRT detection. 
We reduced the data using the Chandra Interactive Analysis of Observations (CIAO; \citealt{Fruscione2006}) software package (v4.15). Counts were extracted from {\lcr} using a circle with radius $2''$, and background counts were measured in source-free regions near {\lcr}.
We used \texttt{specextract} to bin the spectrum (with 5 counts per bin). The routine \texttt{sherpa} \citep{sherpa1,sherpa2} was used to fit the spectrum in the range 0.5--6\,keV, with the background subtracted, using a model with photoelectric absorption and a single-component power law (\texttt{xsphabs.abs1 $\times$ powlaw1d.p1}). We set the Galactic hydrogen column density to be the same as for the \emph{Swift} observations. The best-fit power law index was $\Gamma=1.16^{+0.95}_{-0.95}$ (68\% confidence), consistent with the value from \emph{Swift} but with much larger uncertainties, so for consistency we also adopt $\Gamma=2$.
The flux reported in Table~\ref{tab:xray} has been multiplied by a factor of 1.77 to correct from the 0.5--6\,keV band to the 0.3--10\,keV band.
In Table~\ref{tab:xray}, we also present the spectral flux density at 1\,keV assuming a spectral index of $f_\nu \propto \nu^{-1}$.

\subsection{Radio}

\begin{figure}[ht]
    \centering
    \includegraphics[width=\columnwidth]{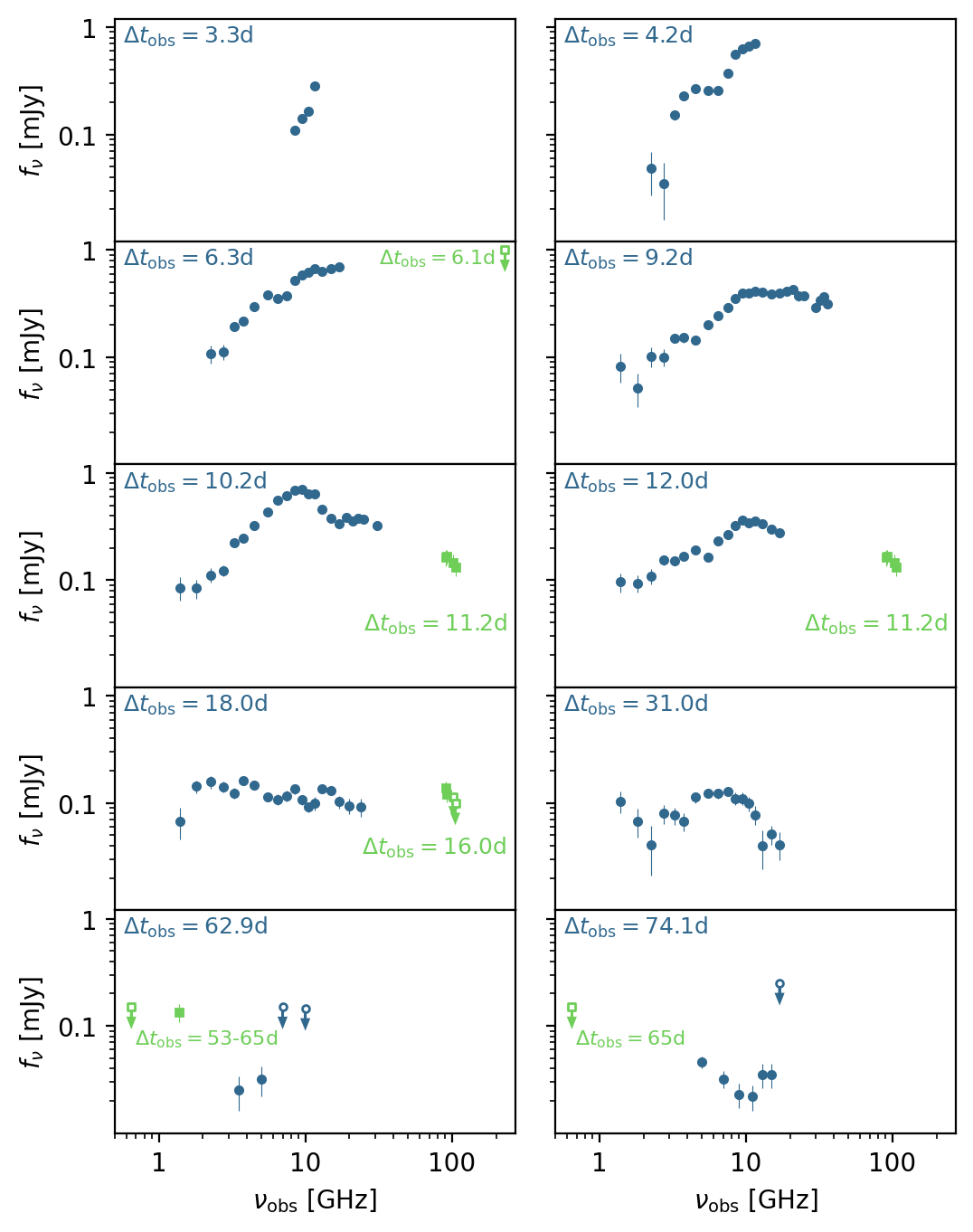}
    \caption{Evolution of the radio spectral energy distribution of {\lcr}. 
    Upper limits (open symbols with arrows) are 5-$\sigma$. 
    VLA (ALMA, GMRT) data are shown as blue circles (green squares).
    Epochs are given in the observer frame. The Ka-band (30\,GHz) observation at $\Delta t_\text{obs} = 10.2$\,d and observations at $\Delta t_\text{obs} = 62.9$\,d were impacted by bad weather.}
    \label{fig:at2023lcr-radio-sed}
\end{figure}

We obtained 11 epochs of observations using the Karl G. Jansky Very Large Array (VLA\footnote{Program IDs 23A-355 and 23A-426, PI D. Perley.}; \citealt{Perley2011}.), spanning 2023 June 21 to 2023 September 29 ($\Delta t = 3$--103\,d) in the L, S, C, X, Ku, and Ka-bands (1--40\,GHz). The primary flux calibrator used was 3C286. Data were calibrated and imaged using standard procedures in the Astronomical Image Processing System (AIPS). Images were typically made in separate windows with a bandwidth of 1\,GHz or 2\,GHz, with adjustments made at lower frequencies due to radio frequency interference excision. Flux-density measurements were performed using \texttt{jmfit}. The second Ka-band epoch was hampered by poor weather conditions, resulting in poor phase stability (with insufficient signal to noise for self-calibration); this measurement should be regarded as a lower limit. Epoch~9 on 2023 August 19 ($\Delta t = 62$\,d) was also hampered by poor phase stability. For Epochs 4--8 ($\Delta t = 9$--31\,d) we obtained C-band observations at the beginning and end of the block in order to search for scintillation.

We obtained observations on epochs 2023 June 29 ($\Delta t = 11$\,d) and 2023 July 04 ($\Delta t = 16$\,d) using the Atacama Large Millimeter/sub-millimeter Array (ALMA) under Director's Discretionary Time\footnote{Program ID 2022.A.00025.T, PI A. Ho.}. Weather conditions in both epochs were excellent.
Data were calibrated and imaged using the automated CASA-based pipeline \citep{CASA2022}. Both observations were in Band 3 (100\,GHz) and yielded a detection with a centroid position of 
$\alpha$ = 16:31:37.419, and $\delta$ = +26:21:58.27 (J2000), consistent with the optical position. The peak flux density of the source was $140\pm14\,\mu$Jy in the first epoch and $94\pm11\,\mu$Jy in the second epoch.

We obtained one epoch of observations with the Submillimeter Array (SMA\footnote{Program ID 2022B-S046, PI E. Berger.}) on 2023 June 24 ($\Delta t = 6$\,d). Observations were conducted between 03:31 and 14:43 UT, using 7 antennas, with an local oscillator frequency of 225.5\,GHz. Weather conditions were favorable (median $\tau_{225\,\mathrm{GHz}} = 0.070$), with good phase stability for all but the first hour of observations. A total of 6.75 hours was spent on source, with 1613+342 and 3C345 used as gain calibrators, and Ceres as the flux calibrator. There was no detection.

We obtained Giant Metrewave Radio Telescope (GMRT) observations from 2023 August 9.63 UT to 2023 August 22.63 UT through a DDT proposal\footnote{Proposal ID DDTC293, PI Nayana AJ.}. The observations were carried out in three frequency bands: band-5 (1000--1460 MHz), band-4 (550--850 MHz), and band-3 (250--500 MHz). The data were collected in standard continuum mode with a time integration of 10 seconds. We used a processing bandwidth of 400 MHz in band-5 and 200 MHz in bands-3 and 4, both split into 2048 channels. 3C286 was used as the flux density and bandpass calibrator while J1609+266 was used as the phase calibrator. Emission was detected at 1.37\,GHz, but not at 0.75 or 0.44\,GHz.

All radio flux density values are provided in Table~\ref{tab:radio} in Appendix~\ref{sec:radio-data}. The radio light curves are shown in Figure~\ref{fig:lcr-indices}, and the evolution of the radio spectral energy distribution is shown in Figure~\ref{fig:at2023lcr-radio-sed}.

\subsection{Search for Gamma-ray Emission}
\label{sec:grb-search}

Throughout the 1 hour and 38 minutes between the last GOTO non-detection and the first GOTO detection, the KONUS instrument on the \emph{Wind} spacecraft \citep{Aptekar1995} was observing the entire sky, with no GRB detection.
For a typical LGRB spectrum, the 90\%-confidence upper limit on the 20--1500\,keV peak flux was reported to be $1.8\times10^{-7}\,$erg\,cm$^{-2}\,$s$^{-1}$ on a 2.944\,s scale \citep{Ridnaia2023}. At the redshift of {\lcr} ($z=1.0272$), this corresponds to an upper limit on the isotropic-equivalent $\gamma$-ray luminosity of $L_{\gamma,\mathrm{iso}}<5.2\times10^{50}\,$erg\,s$^{-1}$.
Assuming a similar scaling over longer time intervals (an observed duration of 40\,s as in \citealt{Perley_2024}) gives a limit on the isotropic-equivalent energy of $E_{\gamma,\mathrm{iso}}<2.1\times10^{52}\,$erg. We perform the same computation at the redshift upper limit of {\lcr} ($z=1.6$), obtaining $L_{\gamma,\mathrm{iso}}< 1.2\times10^{51}\,$erg\,s$^{-1}$ and $E_{\gamma,\mathrm{iso}}<4.9\times10^{52}\,$erg.

The position was visible to the \emph{Fermi} Gamma-ray Burst Monitor (GBM; \citealt{Meegan2009}) for one hour, from 00:23:05 to 01:22:27 on 2023 June 18 with no South Atlantic Anomaly interruptions. A subthreshold search yielded no detections, with a mean upper limit on the peak flux (for the same burst duration) of $8.2\times10^{-8}$\,erg\,cm$^{-2}\,$s$^{-1}$. This corresponds to $L_{\gamma,\mathrm{iso}}<4.8\times10^{50}\,$erg\,s$^{-1}$, or $E_{\gamma,\mathrm{iso}}<9.5\times10^{51}\,$erg.
The position was visible to the \emph{Swift} Burst Alert Telescope (BAT; \citealt{Barthelmy2005}) for 40 minutes, from approximately 00:38 to 01:20.

Given the incomplete coverage of both GBM and Swift BAT, we adopt the more conservative limit from KONUS in what follows.

\section{Observational Features of AT2023lcr}
\label{sec:observational-features}

We find preliminary radio, optical, and X-ray temporal indices for {\lcr}. We also compare {\lcr}'s observational features to those of other $z \approx 1$ afterglows without associated detected GRBs, namely {\pim} at $z=1.2596$ \citep{Perley_2024} and {\lfa} at $z=1.0624$ \citep{Ho_2022}.

\subsection{Optical}

We fit the multi-band optical light curve of AT2023lcr assuming a magnitude offset between each pair of bands that is constant over time, rather than a single overall spectral index.
We treat the GOTO $L$-band point as the average of the $r$- and $g$-band fluxes at that time. We fit the $g$-, $r$-, $i$-, and $z$-band extinction-corrected light curves using the following smoothed broken power law function \citep{Beuermann1999,Zeh2006}:

\begin{equation}
\label{eq:beuermann}
    m(t) = m_c+\frac{2.5}{n}\log_{10}\left[ 
    \frac{(t-t_0)}{t_b-t_0}^{\alpha_1 n} +
    \frac{(t-t_0)}{t_b-t_0}^{\alpha_2 n} \right]
\end{equation} 

\noindent where $m(t)$ is the apparent magnitude as a function of time,
$n$ parameterizes the smoothness of the break ($n=\infty$ is a sharp break),
$\alpha_1$ and $\alpha_2$ are the pre- and post-break temporal indices respectively,
$t_0$ is the time of the explosion in days,
$t_b$ is the time of the break in days,
and $m_c$ is the magnitude at the time of the break assuming $n=\infty$.
We assume no contribution from the host galaxy or the underlying SN. A typical GRB-SN would be both too faint for our observations and redshifted to the NIR. These assumptions are supported by the lack of observed flattening in the optical light curve. We fixed $n=2$ and allowed $t_0$ to vary between the time of the last GOTO non-detection and the first GOTO detection.

From a Markov Chain Monte Carlo (MCMC) fit, we find (see left panel of Figure~\ref{fig:lcr-indices}), a break time $t_b=0.73^{+0.06}_{-0.06}$\,d in the observer frame (68\% confidence). The best-fit broken power-law temporal indices are $\alpha_1=0.23^{+0.04}_{-0.03}$ and $\alpha_2=1.39^{+0.05}_{-0.04}$. We find a best-fit $t_0$ that is $52^{+16}_{-18}$\,min after the GOTO non-detection (and 45\,min before the first GOTO detection), at MJD 60113.02965. 

The optical light curve of {\lcr} is very similar to the early ($\Delta t<10\,$d) optical light curve of {\pim}, although the light curve of {\pim} was better resolved due to TESS coverage. From \citet{Ho_2022}, the optical light curve of {\lfa} had a much steeper temporal decay index than {\lcr}, $t^{-2.5}$ rather than $t^{-1}$, although we caution that the decay index is highly sensitive to the explosion time, which for {\lfa}, is quite uncertain (1.79\,d between the first detection and last non-detection). Using the burst time estimate obtained from results that follow in Section~\ref{sec:lorentz-factor-constraints}, {\lfa} returns a more similar temporal decay index $t^{-1.4}$.

Finally, to measure the spectral index across the optical bands, we use the $ugriz$ photometry from LT in the MJD range 60114.92765--60114.95733 and apply a correction for Milky Way extinction. We find a best-fit spectral index of $\beta_{\mathrm{opt}}=1.20^{+0.16}_{-0.16}$ (68\% confidence) where $f_\nu\propto \nu^{-\beta_\mathrm{opt}}$, consistent with the spectral index measured from the optical spectrum (Figure~\ref{fig:spec}). The spectral index from optical to X-ray bands, as well as within the X-ray band itself (Section~\ref{sec:obs-xrays}), is also close to $f_\nu \propto \nu^{-1}$---this, together with the fairly smooth continuum observed in the spectrum, leads us to conclude that the impact of host-galaxy extinction is negligible.

\subsection{X-rays}
We fit a single power law to the X-ray light curve of {\lcr} at 1\,keV, and find 
a best-fit power law index $\alpha_X=1.47^{+0.17}_{-0.16}$ (68\% confidence) where $f_\nu \propto t^{-\alpha_X}$, consistent with the slope of the optical light curve in the same time range.
Given the similar optical and X-ray spectral indices ($f_\nu \propto \nu^{-1}$), we assume no host galaxy extinction.
There was only one X-ray detection of {\lfa}, at a similar flux to {\lcr}.
The X-ray light curves of {\lcr} and {\pim} have similar temporal decay indices, but the X-ray flux density of {\pim} was an order of magnitude fainter.

\subsection{Radio}

\begin{figure}
    \centering
    \includegraphics[width=\linewidth]{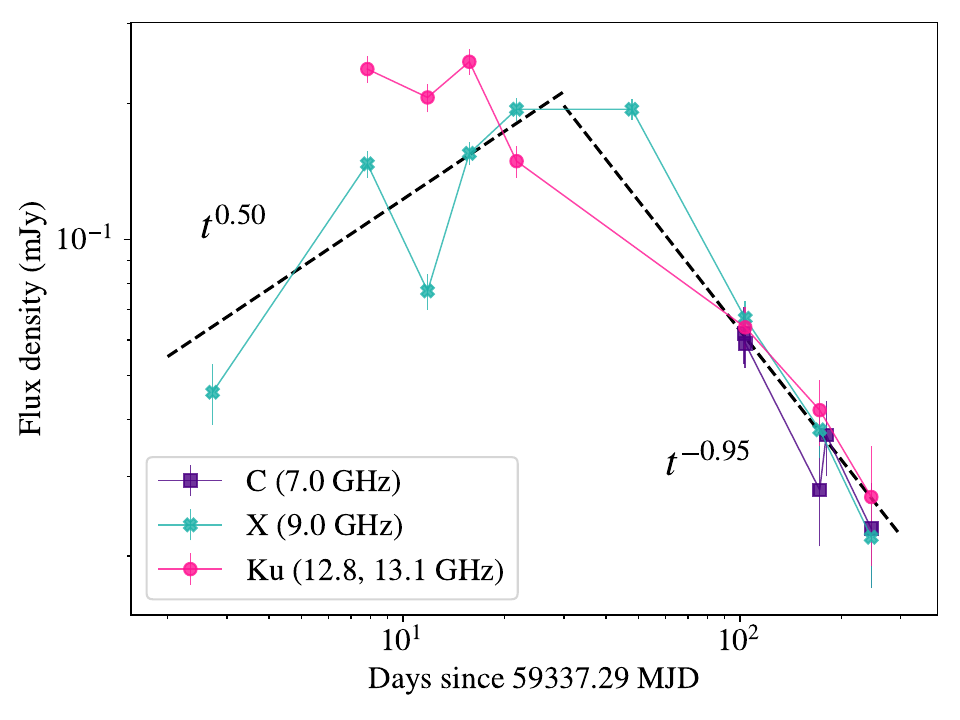}
    \caption{Radio temporal indices for {\lfa}. Indices and the estimated peak are highly affected by interstellar scintillation.}
    \label{fig:lfa-indices}
\end{figure}

The radio light curves of {\pim}, {\lfa}, and {\lcr} reach a similar peak flux density and have roughly a single peak, rising $\sim t^{1}$ then decaying $\sim t^{-1.5}$, although {\lfa}'s radio temporal indices are highly impacted by interstellar scintilation (see Figure~\ref{fig:lfa-indices} for the radio light curve fit of AT\,2021lfa; see Figure 7 in \citet{Perley_2024} for the radio light curve fit of AT\,2019pim). However, the peak for {\lcr} is at $t<10$\,d while the peak for {\pim} and {\lfa} is at tens of days.  Additionally, {\lcr}, {\lfa}, and {\pim} all exhibit significant scintillation at $\nu \lesssim 10$\,GHz. {\lcr} and {\pim} show evidence of scintillation until at least $\Delta t\approx30\,$d, but {\lfa} exhibits scintillation for much longer, until at least $\Delta t\approx100\,$d. The SED evolution is also similar among the three objects, with a hint of self-absorption in the first few days, and a broad peak that passes through $\nu\approx10\,$GHz over the course of the observations. The peak passes through 10\,GHz at around 30\,d for {\pim} and {\lfa}, and at around 10\,d for {\lcr}.

\section{Fitting Framework}
\label{sec:method}

\subsection{Settings}

\begin{table*}[ht]
\normalsize
\centering
\begin{tabular}{cclc}
\hline
Parameter & Unit & Description & Prior (Uniform)\\
\hline
$t_0$ & [MJD] & estimated burst time & - \\

$\theta_\text{v}$ & [rad] & viewing angle & [0, 1.57] \\

$\log_{10}(E_\text{K,iso}/\text{erg})$ & & isotropic equivalent kinetic energy of blast wave along jet axis & [45, 57] \\

$\theta_\text{c}$ & [rad] & half-opening angle of jet core & [0.02, 0.78] \\

$\theta_\text{w}$ & [rad] & wing truncation angle of a structured jet & $[1, 7] \times \theta_\text{c}$ \\

$\log_{10}(n_0/\text{cm}^{-3})$ & & number density of protons in circumburst medium & [-10, 10] \\

$p$ & & power law index of relativistic electron energy distribution & [2, 3] \\

$b$ & & power law index of jet angular energy distribution & [0, 10]\\

$\log_{10}\epsilon_e$ & & fraction of thermal energy in relativistic electrons & [-5, 0]\\

$\log_{10}\epsilon_B$ & & fraction of thermal energy in magnetic field & [-5, 0]\\

$\xi_N$ & & fraction of accelerated electrons & [0, 1] \\

$\log_{10}\Gamma_0$ & & initial Lorentz factor of jet & [0, 3]\\

\hline
\end{tabular}
\caption{Table of priors for {\afterglowpy}. Our choice of priors is uniform and broad to minimize bias. $\theta_\text{w}$ is ignored by the top hat model and $b$ is only used by the power law model. For each afterglow, the prior on $t_0$ spans from its last non-detection to its first detection. We are aware that priors for $\epsilon_e$ and $\epsilon_B$ allow for $\epsilon_e + \epsilon_B > 1$, but none of the fit results are unphysical.}
\label{table:priors}
\end{table*}

We use {\afterglowpy} (version 0.7.3; \citealt{Ryan_2020}) and MCMC methods in {\emcee} (version 3.1.4; \citealt{Foreman_Mackey_2013}) to fit the radio, optical, and X-ray observations of {\lcr} with a set of physical parameters that describe the jet and circumburst medium. To put {\lcr}'s modeling into context, we also perform a consistent analysis on three similarly discovered events, shown in Table~\ref{tab:orphan-summary}: {\blt}, {\any}, and {\lfa}. For {\pim}, see \citet{Perley_2024}, which explored a similarly broad range of scenarios; for AT\,2023sva, see \citep{2023sva}. By contrast, previous works modeling {\blt} \citep{Sarin_2022}, {\any} \citep{Gupta_2022, Xu_2023}, and {\lfa} \citep{Ye_2024} explored more fixed setups. 

All data used in this work for {\blt} and {\any} can be found in \citet{Ho_2020, Ho_2022}. To model {\lfa}, we used all observations from \citet{Ho_2022} and additional observations from \citet{Lipunov_2022}. Optical observations were corrected for Galactic extinction with $E(B-V)=A_V/R_V$, assuming $R_V = 3.1$ \citep{Schlegel_1998}. For all afterglows, we converted X-ray fluxes to a 5\,keV flux density assuming a spectral index of $\beta = -1$, where $f_\nu \propto \nu^\beta$.

Our {\emcee} settings are as follows. To minimize bias, our priors are broad and uniform (see Table~\ref{table:priors}). We use the standard \texttt{EnsembleSampler} from {\emcee}. We perform most runs using 64 walkers and 75,000 iterations, discarding 30,000 iterations as the burn-in. If samples do not appear converged with these settings, we run using 64 walkers and 225,000 iterations, discarding 125,000 as the burn-in. We use a simple Gaussian likelihood for each data point. For AT2020blt, we penalize samples that do not satisfy radio upper limits by finding the log likelihood between the sample-generated radio data point and a zero flux point.

We fit each afterglow to top hat, Gaussian, and power law jet structures found in {\afterglowpy}. The simplest structure is a top hat model, in which energy is constant from the central axis to the edge of the jet:

\begin{equation}
    E(\theta) =
    \left\{
    \begin{array}{cc}
         E_\text{K,iso}, & \theta \leq \theta_\text{c} \\
         0, & \theta > \theta_\text{c}
    \end{array}
    \right\},
\label{eq:top-hat}
\end{equation}

\noindent where {\Ekiso} is the isotropic-equivalent kinetic energy of the outflow along the jet axis and {\thetac} is the half-opening angle of the jet core. The top hat model offers no extended jet structure, unlike Gaussian or power law structures, which particularly affects off-axis or even slightly off-axis observations \citep{Ryan_2020,Cunningham_2020}. For a Gaussian structured jet,

\begin{equation}
    E(\theta) =
    \left\{
    \begin{array}{cc}
         E_\text{K,iso} \exp{(-\frac{\theta^2}{2\theta_\text{c}^2})}, & \theta \leq \theta_\text{w} \\
         0, & \theta > \theta_\text{w}
    \end{array}
    \right\},
\label{eq:gauss}
\end{equation}

\noindent where the jet extends beyond $\theta_\text{c}$ to a ``wing-truncation angle" $\theta_\text{w}$. A power law structured jet has a similar structure:

\begin{equation}
    E(\theta) = E_\text{K,iso} \left(1 + \frac{\theta^2}{b\theta_\text{c}^2}\right)^{-b/2},
\label{eq:power-law}
\end{equation}

\noindent where $b$ is the power law index at which the jet energy decreases. Because there is structure beyond $\theta_\text{c}$ in the Gaussian and power law models, observers are able to view afterglow emission beyond a viewing angle $\theta_\text{v} = \theta_\text{c}$. In this work, we attempt to describe each afterglow with the simplest possible structure (an on-axis top hat jet). We report the result of other structures only if the afterglow is inconsistent with an on-axis top hat jet (by eye, or has a $\chi^2$/DoF significantly worse than other models).

Along with top hat, Gaussian, and power law jet structures, we fit each afterglow to various combinations of multi-wavelength observations, since afterglow emission at different wavelengths can reveal different physics. For example, radio observations can capture ``reverse shock" emission, which traces a shock propagating back through the outgoing jet shell and towards the central engine, typically crossing this region at the deceleration time \citep{Kobayashi_2000,Piran_2005,Laskar_2016}. We fit each model only to optical observations, then only to optical and X-ray observations, then to all radio, optical, and X-ray observations---and report any significant differences in the results. 

\subsection{afterglowpy Limitations}
{\afterglowpy} (version 0.7.3) uses the single-shell approximation \citep{single-shell} to model a blast wave propagating through a homogeneous circumburst medium \citep{Ryan_2020}. {\afterglowpy} is useful for its range of afterglow settings and its implementation of structured jets, but is limited in different ways which could affect our interpretations. For example, if support for Inverse Compton Cooling (ICC) is enabled, {\afterglowpy} overestimates its radiative contribution. Additionally, by default, {\afterglowpy} assumes $\Gamma_0 = \infty$, such that there is no initial coasting phase or deceleration break, which might produce unreliable early-time light curves. On the other hand, for a finite $\Gamma_0$, {\afterglowpy} disables jet spreading, which might produce unreliable late-time light curves. To work around this, we run MCMC multiple times, with each combination: with and without ICC; and with $\Gamma_0 = \infty$ and $\Gamma_0 \neq \infty$. We report any notable differences in the inferred parameters.

In addition to the above, {\afterglowpy} does not support reverse shock physics, which might particularly affect radio observations. 
Also, {\afterglowpy} lacks support for synchrotron self-absorption. In this work, self-absorption affects {\lfa}, in which radio observations $\lesssim$ 21 days from the estimated burst time may be self-absorbed \citep{Ho_2022}. There may also be hints of self-absorption in the first few days of {\lcr}'s SED evolution (see Figure~\ref{fig:at2023lcr-radio-sed} in Section~\ref{sec:at2023lcr}). We include all observations in the fit, but caution that the model may be expected to overpredict the radio luminosity at early times (while lacking the reverse shock may result in underpredicting the radio luminosity, particularly at early times). 

Finally, {\afterglowpy} implements a homogeneous circumburst medium, with no support for a stellar wind medium. Although in principle, a massive star progenitor should have a stellar wind medium, past works have shown that generally, a homogeneous medium fits well to most LGRBs \citep{Schulze_2010, Hjorth_2011}, with some exceptions \citep{Panaitescu_2001}.

\subsection{Goodness of Fit Metrics}
\label{sec:goodness-of-fit}
To quantify the goodness of fit between modeling configurations, we use a reduced $\chi^2$
\begin{equation}
	\chi^2 \equiv \frac{1}{\text{DoF}} \sum \frac{(f_{\text{model}} - f_{\text{obs}})^2}{\sigma^2},
\end{equation}
where $f_{\text{model}}$ is the model generated light curve, $f_{\text{obs}}$ is the observed light curve, $\sigma$ is the uncertainty (systematic and statistical) in observations, and DoF is the difference between the number of observations and the number of MCMC parameters. In the results that follow, we report the minimum $\chi^2$ over 5,000 randomly selected posterior samples. We note that for VLA data, we include a $5\%$ systematic error on the flux densities for the L, S, C, X, and Ku-bands (1-18\,GHz), and a $15\%$ systematic error for the K and Ka-bands (18-40\,GHz).

We also use the Widely Available Information Criterion (WAIC; \citealt{Watanabe_2010, Cunningham_2020}), which can be calculated from MCMC posterior samples. WAIC estimates the ``expected log predictive density" (\elpd), which describes how well a model should fit to new data. We do not normalize our WAIC scores, so a more positive {\elpd} indicates a stronger predictive power. In this paper, we report each {\elpd} calculated over 5,000 randomly selected posterior samples.

\subsection{Radiative Efficiencies}

\begin{table}[ht]
\normalsize
\centering
\begin{tabular}{lll}
\hline
Afterglow & $E_{\gamma,\text{iso}}$ & Reference\\
& ($10^{52}$\,erg) & \\
\hline
{\blt} & $< 1.0$ & \citet{Ho_2020, Ho_2022}\\
& $< 0.1 - 0.6$ & \citet{Sarin_2022}\\
{\any} & $< 14.3$ & \citet{Ho_2022}\\
{\lfa} & $< 0.12$ & \citet{Ho_2022}\\
{\lcr} & $< 2.1$ & \citet{Ridnaia_2023};\\
          &         & this work; for $z=1.0272$\\
       & $< 4.9$ & for $z=1.6$\\
\hline
\end{tabular}
\caption{Radiative energy upper limits for {\blt}, {\any}, {\lfa}, and {\lcr}. \citet{Sarin_2022}'s estimate uses \textit{Fermi} observations, to which the position of {\blt} was not fully visible during the duration between its last non-detection and first detection.}
\label{table:radiative-energies}
\end{table}

To calculate the efficiency of gamma-ray radiation for each event, or ``radiative efficiency", we use

\begin{equation}
    \eta_\gamma = \frac{E_{\gamma,\text{iso}}}{E_\text{K,iso} + E_{\gamma,\text{iso}}},
\label{eq:eff}
\end{equation}

\noindent where {\Ekiso} is the isotropic-equivalent kinetic energy of the jet, found from modeling. {\Egiso} is the observed isotropic-equivalent energy in gamma-rays, or ``radiative energy" from 
flux or fluence upper limits of gamma-ray facilities. We calculate an {\Egiso} upper limit for {\lcr} in Section~\ref{sec:grb-search}. Limits for {\blt}, {\any}, and {\lfa} have been calculated in \citet{Ho_2020}, \citet{Ho_2022}, and \citet{Sarin_2022}. We summarize all {\Egiso} upper limits in Table~\ref{table:radiative-energies}. We note that the upper limit from \citet{Sarin_2022} was based on a subthreshold search of \textit{Fermi} observations, which experienced interruptions between {\blt}'s last non-detection and first detection.

\section{Results and Discussion}
\label{sec:results}

\subsection{Analytical Constraints on the Lorentz Factors}
\label{sec:lorentz-factor-constraints}

We follow \citet{Perley_2024} to obtain analytical estimates on the afterglow bulk Lorentz factors, summarizing results in Table~\ref{table:analytical-lorentz-factors}.
Assuming a uniform medium density, we can obtain a limit on the initial Lorentz factor of the jet from the deceleration time using Equation 16 of \citealt{Meszaros_2006},

\begin{equation}
\Gamma_{2.5} = \left( \frac{10\,\text{s} (1 + z)}{t_\text{dec}} \right)^{3/8} \left(\frac{E_{53}}{n_0}\right)^{1/8},
\label{eq:tdec}
\end{equation}

\noindent where $z$ is the redshift, $n_0$ is the number density of the circumburst medium in cm$^{-3}$, $E_{53} = E_{\text{K,iso}} / 10^{53}$, $\Gamma_{2.5} = \Gamma_0 / 10^{2.5}$, and $t_\text{dec}$ is the deceleration time in seconds\footnote{All times in this work are in the observer frame unless specified otherwise. Rest-frame quantities will be denoted with a subscript, e.g. $t_\text{rise,rest}$ for a rest-frame rise time.}. Typically, $t_\text{dec}$ coincides with the peak of the X-ray afterglow, which occurs on the same timescale as the afterglow rise time. Therefore, we approximate $t_\text{dec} \approx t_\text{rise}$, where $t_\text{rise}$ is the afterglow rise time. Using the difference between the first detection and last non-detection\footnote{The latency is 0.740\,d for {\blt}. 0.015\,d for {\any}, 1.794\,d for {\lfa}, and 0.067\,d for {\lcr}.} as an upper limit on $t_\text{rise}$, we obtain the $\Gamma_0$ lower limits shown in Table~\ref{table:analytical-lorentz-factors}, abbreviating $\kappa = (E_{53}/n_0)^{1/8}$. For {\lfa}, we also obtain an upper limit on $\Gamma_0$, since $t_\text{rise} > 0.13$\,d, which is the time between {\lfa}'s first MASTER detection and ZTF peak detection.

\begin{table}[t]
\small
\centering
\begin{tabular}{llll}
\hline
Afterglow & Rise Time & Spectrum & ISS\\
& ($\Gamma_0 / \kappa$) & ($\Gamma_{\text{avg}}^\dag$) & ($\Gamma_{\text{avg}}^\dag$) \\
\hline
{\blt} & $\gtrsim 19.7$ & - & -\\

{\any} & $\gtrsim 81.2$ & $\gtrsim 45$, $\gtrsim 62$ & $\lesssim 127$, $\lesssim 175$ \\

{\lfa} & $\gtrsim 10.8$, $\lesssim 30.0$ & $\gtrsim 3$, $\gtrsim 27$ & $\lesssim 6.4$, $\lesssim 60$ \\

{\lcr} & $\gtrsim 37.8$& $\gtrsim 14$, $\gtrsim 39$ & $\lesssim 86$, $\lesssim 328$ \\
\hline
\end{tabular}
\caption{Analytical constraints on the Lorentz factors for {\blt}, {\any}, {\lfa}, and {\lcr}. We abbreviate $\kappa = (E_{53} / n_0)^{1/8}$.
$^\dag$Left hand values are average Lorentz factors at the time of first detection. Right hand values are average Lorentz factors at the beginning of deceleration. We assume a typical rest-frame deceleration time $t_\text{dec,rest} \approx t_\text{rise,rest} \approx 200$\,s, but caution that if {\lfa}'s rising phase is due to deceleration, then its right hand values are overestimates.} 
\label{table:analytical-lorentz-factors}
\end{table}

We can also obtain a lower limit on the average Lorentz factor $\Gamma_\text{avg}$ from radio spectra using Equation 5 from \citet{Barniol_Duran_2013}. We assume a full filling factor and use the time of the last non-detection as a lower limit on the explosion time. Since {\blt} lacks multifrequency radio observations, we only perform this calculation on {\any}, {\lfa}, and {\lcr}.

Radio observations of {\any} \citep{Ho_2022} are unlikely to be impacted by synchrotron self-absorption given their observed spectral indices, so we use observations from the its first radio epoch at $\Delta t = 4.91$\,d to estimate $\Gamma_{\text{avg,4.91d}} \gtrsim 3.0$ at $4.91$\,days from its last non-detection. On the other hand, the radio spectra of {\lfa} \citep{Ho_2022} and {\lcr} show possible self-absorption until $\Delta t \approx 20$\,d and $\Delta t \approx 6$\,d, respectively. Using observations from these epochs, we obtain $\Gamma_{\text{avg,22.69d}} \gtrsim 1.1$ for {\lfa} and $\Gamma_\text{avg,6.3d} \gtrsim 2.6$ for {\lcr}. Assuming that their light curve breaks are caused by jet expansion, we can follow \citet{Galama_2003} to extrapolate these estimates to the times of their first detections, obtaining the results in Table~\ref{table:analytical-lorentz-factors}. We extrapolate {\any} and {\lcr} using $\Gamma \propto t^{-1/2}$, which holds for post jet break expansion. {\lfa} exhibits no jet break, so for it we use the pre jet break case $\Gamma \propto t^{-3/8}$.

As discussed in \citet{Perley_2024}, $\Gamma_\text{avg,first}$ is different from $\Gamma_0$, which can still be large if $t_\text{dec}$ is small. To alleviate this uncertainty, we also extrapolate $\Gamma_\text{avg}$ to $t_\text{dec}$ to obtain a closer estimate of $\Gamma_0$, assuming that $\Gamma$ is constant from explosion to the start of deceleration. If we take a typical rest-frame LGRB deceleration time $t_\text{dec,rest} \approx t_\text{rise,rest} \approx 200$\,s \citep{Ghirlanda_2018}, we obtain the larger values for $\Gamma_\text{avg}$ shown along initial estimates in Table~\ref{table:analytical-lorentz-factors}. We caution that if {\lfa}'s rising phase is due to deceleration and not off-axis behavior, then its value is an overestimate.

\begin{table}[t]
\normalsize
\centering
\begin{tabular}{lcccc}
\hline
Afterglow & $T$ & $\nu_0$ & $\theta_{F0}$ & $D$\\
& (days) & (GHz) & ($\mu$arcsec) & ($10^{16}$ cm)\\
\hline

{\any} & 21 & 15 & 2 & $5$\\

{\lfa} & 104 & $<8$ & $<5$ & $<13$\\

{\lcr} & 10 & 9 & 3.5 & $9$\\

\hline
\end{tabular}
\caption{Approximate durations $T$ of strong ISS, critical frequencies ISS $\nu_0$, Fresnel scales $\theta_{F0}$, and physical sizes of the Fresnel scale $D$ of {\any}, {\lfa}, and {\lcr}. Critical ISS frequencies and Fresnel scales are found from \citet{Walker_2001}.}
\label{table:iss-helpers}
\end{table}

Finally, we obtain upper limits on $\Gamma$ from the presence of strong interstellar scintillation (ISS), assuming that is responsible for the observed variability in {\any}, {\lfa}, and {\lcr} ({\blt}'s radio observations are too limited). 
If a source exhibits strong ISS (radio fluctuations greater than a factor $\sim 2$) at a frequency near or less than its critical ISS frequency $\nu_0$ \citep{Walker_2001, Perley_2024}, then the source's size is at most as large as the Fresnel scale $\theta_{F0}$ at its location. Given the source size and estimated explosion time, an upper limit on the Lorentz factor can be calculated.

In Table~\ref{table:iss-helpers}, we list approximate timescales $T$ for strong ISS, critical ISS frequencies $\nu_0$, Fresnel scales $\theta_{F0}$ at $\nu_0$, and physical sizes of the Fresnel scale $D$ given the angular diameter distances of the three afterglows. From these values, we obtain $\Gamma_{\text{avg,21d}} \lesssim 3.4$ for {\any}, $\Gamma_{\text{avg,104d}} \lesssim 1.4$ for {\lfa}, and $\Gamma_{\text{avg,10d}} \lesssim 7.1$ for {\lcr}. Extrapolating to the times of their first detections and deceleration times, we obtain the upper limits found in Table~\ref{table:analytical-lorentz-factors}.

\subsection{AT2023lcr}
\label{subsec:lcr-results}

\begin{figure*}[ht]
    \centering
    \subfloat{\includegraphics[width=8.5cm]{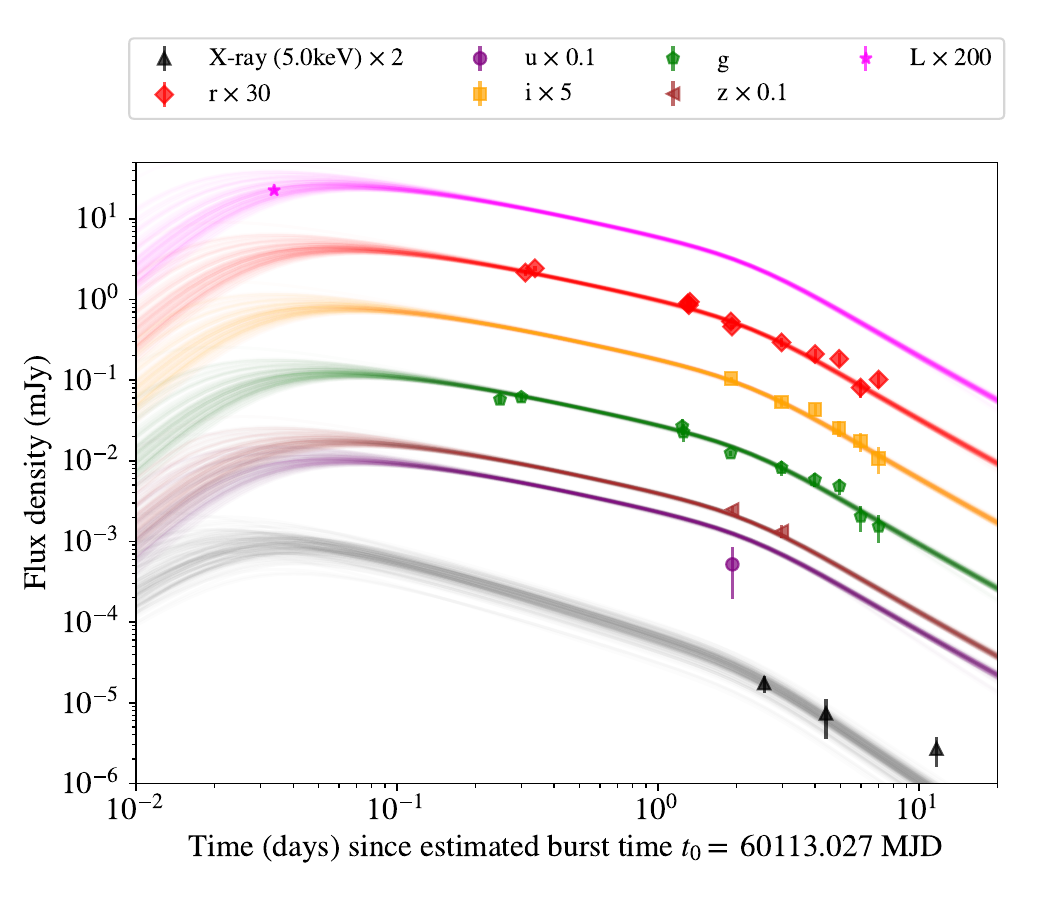}}%
    \qquad
    \subfloat{\includegraphics[width=8.5cm]{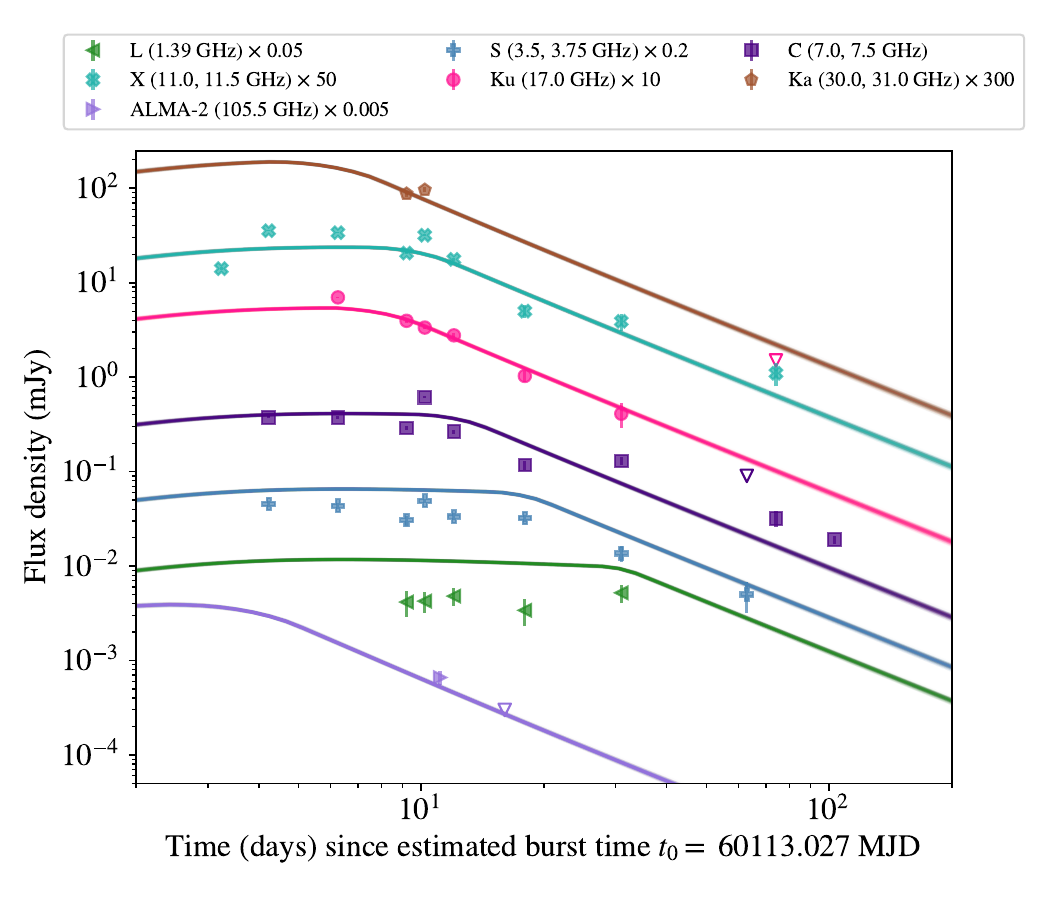}}%
    \caption{On-axis top hat jet with $\Gamma_0 \approx 166$ for {\lcr}, fit to X-ray, optical (left), and radio observations (right). The model is able to reproduce overall trends in all bands, especially the early-time optical $L$-band detection, the optical light curve break, and the X-ray observations. However, the model overestimates the radio L-band (1.39\,GHz) and S-band (3.5, 3.75\,GHz) detections, likely due to {\afterglowpy}’s lack of self-absorption modeling. Plotted are light curves generated from 150 randomly selected posterior samples. Radio upper limits are plotted at 3 $\times$ image RMS.}%
    \label{fig:lcr-results}
\end{figure*}

We present the results of an on-axis top hat jet with a finite $\Gamma_0$ in Table~\ref{tab:lcr-results} and Figure~\ref{fig:lcr-results}, with a corner plot in Figure~\ref{fig:lcr-corner} of Appendix~\ref{app:corner-plots}. The model is able to reproduce key features of all observations, particularly the early-time optical \textit{L}-band detection and the achromatic break $\sim 2$\,days after the estimated burst time. However, the model somewhat overestimates radio observations in the radio L-band (1.39\,GHz) and S-band (3.75\,GHz), which may be due to the susceptibility of low frequency emission to synchrotron self-absorption, which {\afterglowpy} does not model. The model also slightly overestimates the \textit{u}-band detection, which may be due to host-galaxy extinction. Fitting configurations with different {\afterglowpy} settings obtained similar results; see Table~\ref{tab:lcr-selected} in Appendix~\ref{app:selected} for results of selected configurations. 
Optical only and optical-X-ray only fits yielded on-axis low Lorentz factor ($\Gamma_0 \sim 30$) GRBs with typical efficiencies and much higher densities ($n_0 \sim 10^{-1} - 10^{-2}$\,cm$^{-3}$), but had softer light curve breaks at $\Delta t \sim 2$\,d and tended to severely overestimate radio detections.

Our results suggest that {\lcr} is consistent with an on-axis classical GRB, with a highly collimated jet ($\theta_\text{c} \approx 0.02$ rad, or equivalently, $1.15^\circ$) and a low density circumburst environment ($n_0 \sim 10^{-4}$ cm$^{-3}$). Calculations on large catalogues of GRBs find most GRBs to have opening angles $\sim 5^\circ$, with few GRBs populating the $\sim 1^\circ$ opening angle regime \citep{Goldstein_2016, Ghirlanda_2005}. While there are GRBs that have circumburst densities as low as $10^{-5}$ cm$^{-3}$ (see GRB 090423; \citealt{Tanvir_2009, Salvaterra_2009, Laskar_2013b}), most have circumburst densities within $n_0 = 10^{-1} - 10^{2}$ cm$^{-3}$ \citep{Laskar_2013b}. If {\lcr} can be described by a homogeneous medium, then our modeling places {\lcr} in a small opening angle, lower density regime. 
From the median values in Table~\ref{tab:lcr-results}, we obtain a beaming-corrected value of the blast wave energy $E_{K} \approx 2.2 \times 10^{50}$\,erg, which falls within typical LGRB values of $E_{K}$ \citep{Laskar_2013b}. We also note that our posterior $\Gamma_0 \approx 166$ is consistent with analytical estimates from Section~\ref{sec:lorentz-factor-constraints}.

{\lcr} is consistent with an on-axis GRB afterglow, yet
KONUS-\emph{Wind} found no GRB detection while observing the entire night sky during the time between the last GOTO non-detection and first GOTO detection. Therefore, {\lcr} had an isotropic radiative energy below $2.1 \times 10^{52}$\,erg and a possible radiative efficiency $\eta_\gamma < 2.3\%$, which is consistent with typical LGRB efficiencies \citep{Racusin_2011}.

\begin{table}[ht]
\large
\centering
\begin{tabular}{ll}
\hline
Parameter & Result\\
\hline
$t_0$ [MJD] & $60113.03_{-0.02}^{+0.01}$\\
$\theta_\text{v}$ [rad] & $0.00_{-0.00}^{+0.00}$\\
$\log_{10}(E_\text{K,iso}/\text{erg})$ & $53.95_{-0.17}^{+0.22}$\\
$\theta_\text{c}$ [rad] & $0.02_{-0.00}^{+0.00}$\\
$\log_{10}(n_0/\text{cm}^{-3})$ & $-4.37_{-0.36}^{+0.53}$\\
$p$ & $2.13_{-0.01}^{+0.01}$\\
$\log_{10}\epsilon_e$ & $-1.55_{-0.22}^{+0.16}$\\
$\log_{10}\epsilon_B$ & $-0.63_{-0.38}^{+0.32}$\\
$\xi_N$ & $0.63_{-0.24}^{+0.25}$\\
$\log_{10}\Gamma_0$ & $2.22_{-0.09}^{+0.06}$\\
$\eta_\gamma$ ($z = 1.0272$) & $< 1.3 - 3.4\%$ \\
$\eta_\gamma$ ($z = 1.6$) &$< 3.2 - 7.5\%$\\
$\chi^2$/DoF & $14.0$ \\
{\elpd} & $(-1.7 \pm 0.4) \times 10^{2}$ \\
\hline
\end{tabular}
\caption{Final parameters ($68\%$ uncertainty) for the on-axis, top hat, $\Gamma_0 \neq \infty$ configuration for {\lcr}. We calculate $\eta_\gamma$ using the $1\sigma$ distribution of {\Ekiso} and the $E_{\gamma,\text{iso}}$ limits from Table~\ref{table:radiative-energies}. We present the {\elpd} and minimum $\chi^2$/DoF over 5,000 posterior samples; the large $\chi^2$/DoF is due to the overestimation of lower-frequency radio bands (L, S). Ran with 64 walkers and 75,000 iterations; discarded 25,000.}
\label{tab:lcr-results}
\end{table}

\subsection{AT2020blt}
\label{subsec:blt-results}

\begin{figure*}[ht]
    \centering
    \subfloat{\includegraphics[width=8.5cm]{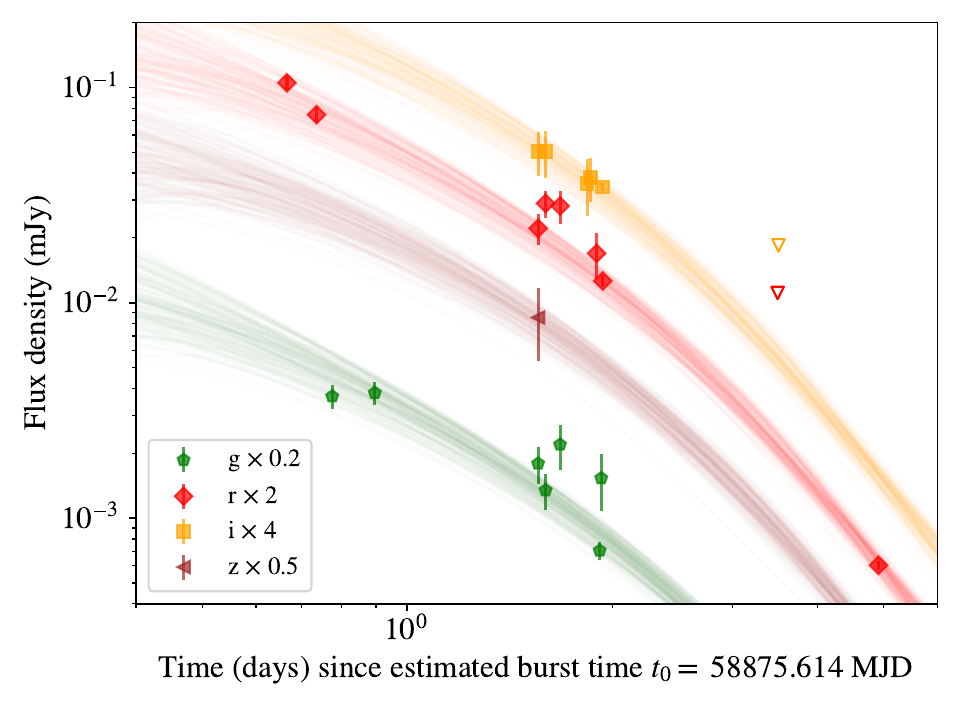}}%
    \qquad
    \subfloat{\includegraphics[width=8.5cm]{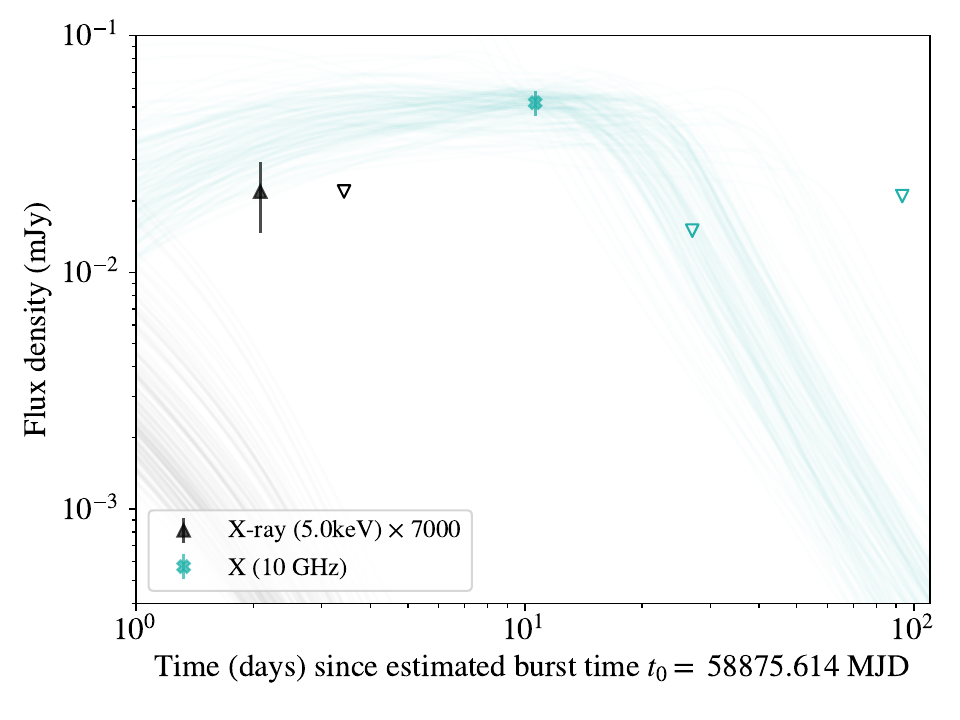}}%
    \caption{On-axis top hat jet with $\Gamma = \infty$ for {\blt}, fit to optical (left), X-ray, and radio observations (right). Plotted are light curves generated from 150 randomly selected posterior samples. The model is consistent with optical and radio observations, but underestimates the X-ray detection by $\sim 1.5$ orders of magnitude, possibly due to unmodeled central engine energy injection. Radio upper limits are plotted at 3 $\times$ image RMS.}%
    \label{fig:blt-results}
\end{figure*}

We present the results of a top hat jet with $\Gamma_0 = \infty$ in Table~\ref{tab:blt-results} and Figure~\ref{fig:blt-results}, with a corner plot in Figure~\ref{fig:blt-corner} of Appendix~\ref{app:corner-plots}. The model is consistent with optical observations, but underpredicts the X-ray detection by 1.5 orders of magnitude. Other {\afterglowpy} configurations produced a similar results, but models with a finite $\Gamma_0$ were inconsistent with the $\sim25$\,d radio non-detection by a factor of 3. We also note that the top hat model we present has the smallest $\chi^2$ and highest predictive power ({\elpd}) of all models attempted for {\blt}. See Table~\ref{tab:blt-selected} in Appendix~\ref{app:selected} for results of selected configurations. We note that modeling {\blt} without radio observations showed no significant improvement in the X-ray discrepancy. Optical only and optical-X-ray only configurations generally yielded on-axis classical GRBs, some with potentially very low efficiencies ($\eta_\gamma \lesssim 0.1 - 1.4\%$; using S22's $E_{\gamma,\text{iso}}$ estimate in Table~\ref{table:radiative-energies}).

\begin{table}[ht]
\normalsize
\centering
\begin{tabular}{lll}
\hline
Parameter & This Work & S22\\
\hline
$t_0$ [MJD] & $58875.61_{-0.05}^{+0.10}$ & $58875.13_{-1.06}^{+0.58}$ \\

$\theta_\text{v}$ [rad] & $0.08_{-0.08}^{+0.08}$ & $0.06_{-0.04}^{+0.05} $\\

$\log_{10}(E_\text{K,iso}/\text{erg})$ & $53.00_{-0.56}^{+0.67}$ & $53.61_{-0.35}^{+0.25}$ \\

$\theta_\text{c}$ [rad] & $0.09_{-0.04}^{+0.06}$ & $0.14_{-0.04}^{+0.04}$ \\

$\theta_\text{w}$ & - & $0.42_{-0.02}^{+0.16}$\\

$\log_{10}(n_0/\text{cm}^{-3})$ & $2.05_{-1.76}^{+1.15}$ & $1.90_{-1.72}^{+1.30}$ \\

$p$ & $2.83_{-0.23}^{+0.13}$ & $2.78_{-0.20}^{+0.14}$ \\

$b$ & - & $5.14_{-2.76}^{+2.89}$\\

$\log_{10}\epsilon_e$ & $-0.63_{-0.62}^{+0.41}$ & $-1.10_{-0.31}^{+0.34}$ \\

$\log_{10}\epsilon_B$ & $-3.56_{-0.86}^{+1.38}$ & $-1.64_{-0.83}^{+0.73}$ \\

$\xi_N$ & $0.46_{-0.31}^{+0.35}$ & $0.67_{-0.83}^{+0.73}$ \\

$\log_{10}\Gamma_0$ & $\infty^\dag$ & $2.70_{-0.43}^{+0.21}$ \\

$\eta_\gamma$ & $<2.1-26.6\%$ & $<0.1-3.2\%$\\

$\eta_\gamma$ (\textit{Fermi})& $<0.2-17.9\%$ & $<0.1-3.2\%$\\

$\chi^2$/DoF & $3.1$ & - \\

{\elpd} & $(1.0 \pm 0.1) \times 10^{2}$ & - \\
\hline
\end{tabular}
\caption{Final parameters ($68\%$ uncertainty) for the top hat, $\Gamma_0 = \infty$ configuration for {\blt}. We calculate $\eta_\gamma$ using the $1\sigma$ distribution of {\Ekiso} and $E_{\gamma,\text{iso}}$ limits from Table~\ref{table:radiative-energies}. We present the {\elpd} and minimum $\chi^2$/DoF over 5,000 posterior samples. We also include the power law fitting results from \citet{Sarin_2022}. Ran with 64 walkers and 225,000 iterations; discarded 125,000.\\
$^\dag$ Not from MCMC.}
\label{tab:blt-results}
\end{table}

From Table~\ref{tab:blt-results}, we obtain $\theta_\text{c} \sim 5^\circ$ and a beaming-corrected $E_\text{K} \sim 4 \times 10^{50}$\,erg, physical parameters typical of GRBs. However, we acknowledge that all parameters have broad uncertainties due to {\blt}'s sparse observations. The modeled circumburst density is somewhat high, with $n_0 \sim 112$ cm$^{-3}$, but we note that LGRBs with densities as high as $n_0 \sim 600$ cm$^{-3}$ (see GRB 050904; \citealt{Cummings_2005, Tagliaferri_2005, Haislip_2006, Laskar_2013b}) have been discovered. We also obtain a radiative efficiency $\eta_\gamma < 2.1-26.6\%$ using the KONUS-\textit{Wind} {\Egiso} limit (see Table~\ref{table:radiative-energies}), which is typical of GRBs, as calculated in \citet{Racusin_2011}. If we use the less conservative \textit{Fermi} {\Egiso} limit, we obtain possibly very low efficiencies $\eta_\gamma < 0.2\%$, lower than $98.5\%$ of LGRB efficiencies reported in \citet{Racusin_2011}. Our high-$\Gamma$ fit is also consistent with the lower limits from Section~\ref{sec:lorentz-factor-constraints}

\begin{figure}[ht]
    \centering
    \subfloat{\includegraphics[width=8.5cm]{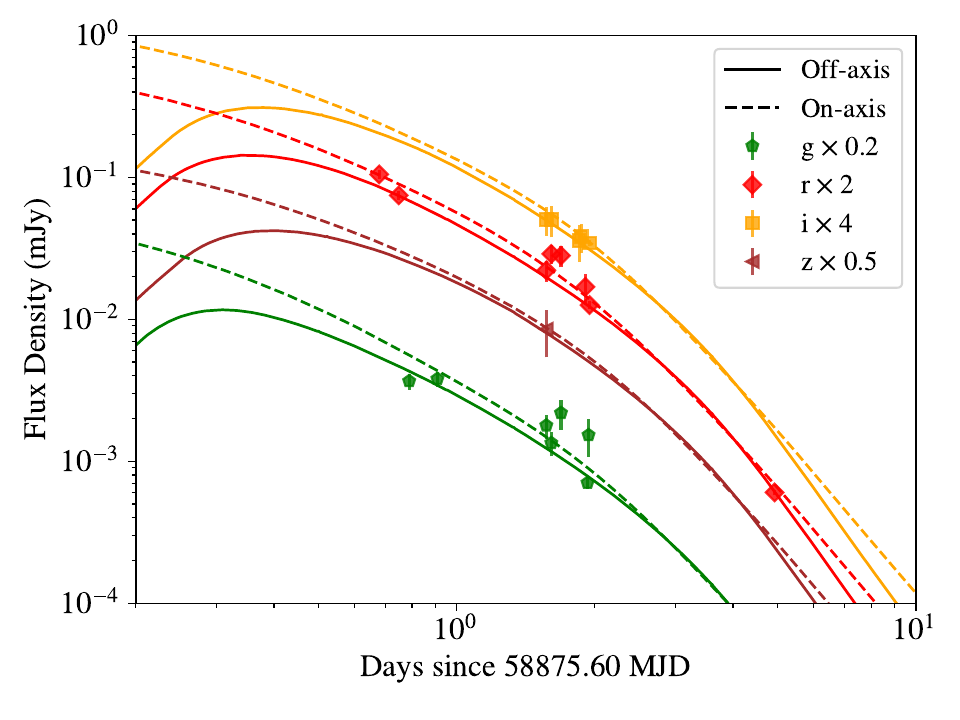}}%
    \qquad
    \subfloat{\includegraphics[width=8.5cm]{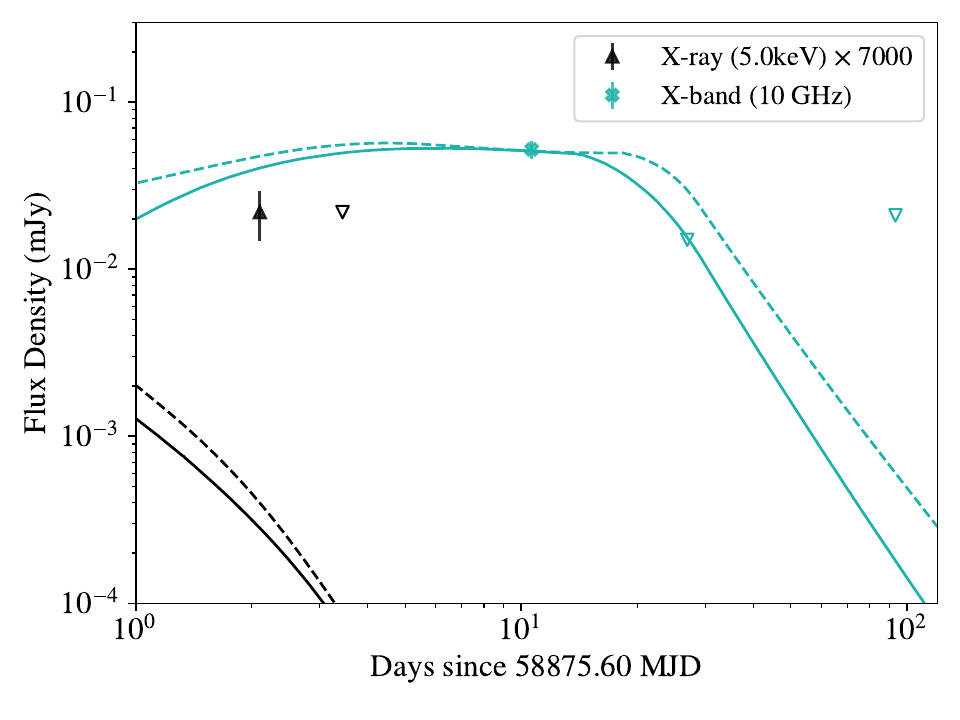}}%
    \caption{The lowest-likelihood off- and on-axis samples for {\blt}, fit to optical (top), X-ray, and radio observations (bottom). As expected, the off-axis model has a later peak; otherwise, the samples produce similar fits. Radio upper limits are plotted at 3 $\times$ image RMS.}%
    \label{fig:blt-off-axis-results}
\end{figure}

We also obtain values of $\theta_\text{v}$ that allow for off-axis solutions. We include a comparison between off- and on-axis fits in Figure \ref{fig:blt-off-axis-results} and Table \ref{tab:blt-off-axis-comparison}. As expected, the off-axis solution places the peak of the light curve at a later time and has a higher blast wave energy than the on-axis fit. Both fits have comparable $\chi^2$ and {\elpd} values, suggesting that {\blt} is consistent with classical on-axis and off-axis GRBs. Ultimately, we lack the early-time data to resolve the viewing angle degeneracy.

To summarize, we find that {\blt} is consistent with off-axis and on-axis classical GRBs. This multimodality may be from sparse observations or from {\emcee} and {\afterglowpy} limitations; in any case, a classical GRB origin cannot be ruled out for {\blt}. We also caution that all fits underestimate the X-ray observation by approximately 1.5 orders of magnitude. This X-ray excess could be from ongoing central engine activity \citep{Zhao_2020} or an insufficient $\chi^2$ penalty, since {\blt} has only a single X-ray observation with a large uncertainty ($f_\text{5\,keV} = 3.14\pm 1.04$\,$\mu$Jy).  

\begin{table}[ht]
\label{tab:blt-off-axis-comparison}
\large
\centering
\begin{tabular}{llll}
\hline
Parameter & Off-axis & On-axis\\
\hline

$\theta_\text{v}$ [rad] & 0.04 & 0.002\\

$\theta_\text{c}$ [rad] & 0.02 & 0.01 \\

$t_0$ [MJD] & 58875.65 & 58875.55 \\

$\log_{10}(E_\text{K,iso}$/\text{erg}) & 54.0 & 52.6\\

$\log_{10} (n_0/\text{cm}^{-3})$ & -1.0 & 0.09\\

$\chi^2$/DoF & 3.1 & 3.3 \\

{\elpd} & $104 \pm 10$ & $103 \pm 9$ \\
\hline
\end{tabular}
\caption{Parameters of lowest-likelihood off- and on-axis solutions. To calculate the {\elpd}, we split the posterior into off-axis and on-axis solution sets and find the {\elpd} over 5,000 samples from each set.}
\end{table}

\subsubsection{Comparison to S22}

S22 use {\afterglowpy} and \texttt{dynesty} \citep{dynesty} to model the optical observations of {\blt} with top hat, power law, and cocooned jet structures, and find that {\blt} is best explained by an on-axis power law jet with physical parameters typical of LGRBs, consistent with this work. Using the upper limit for {\Egiso} from a subthreshold search on \textit{Fermi} (see Table~\ref{table:radiative-energies}), which experienced interruptions between {\blt}'s last non-detection and first detection, S22 explain {\blt} as a low-efficiency burst with $\eta_\gamma < 0.1\%$, smaller than $98.5\%$ of LGRB efficiencies from \citet{Racusin_2011}. This work reports more typical efficiencies $\eta_\gamma<2.1-26.6\%$ using the KONUS-\textit{Wind} upper limit on {\Egiso}, but does obtain estimates as low as $\eta_\gamma < 0.2\%$ with the \textit{Fermi} upper limit, consistent with S22. We note that S22 model only the optical observations of {\blt}; using the \textit{Fermi} estimate, our optical only configurations also produce efficiencies as low as $\eta_\gamma < 0.1\%$. In any case, both S22 and this work indicate that {\blt} is consistent with an on-axis GRB with physical parameters that are fairly typical of the LGRB population.

\subsection{AT2021any}
\label{subsec:any-results}

\begin{figure*}[bt]
    \centering
    \subfloat{\includegraphics[width=8.5cm]{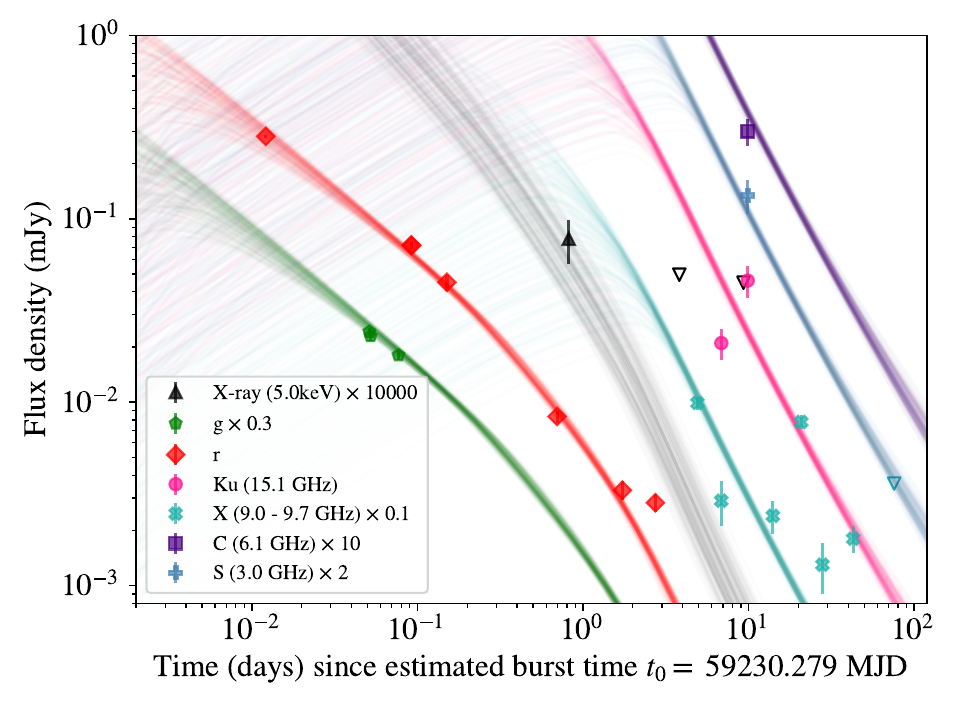}}%
    \qquad
    \subfloat{\includegraphics[width=8.5cm]{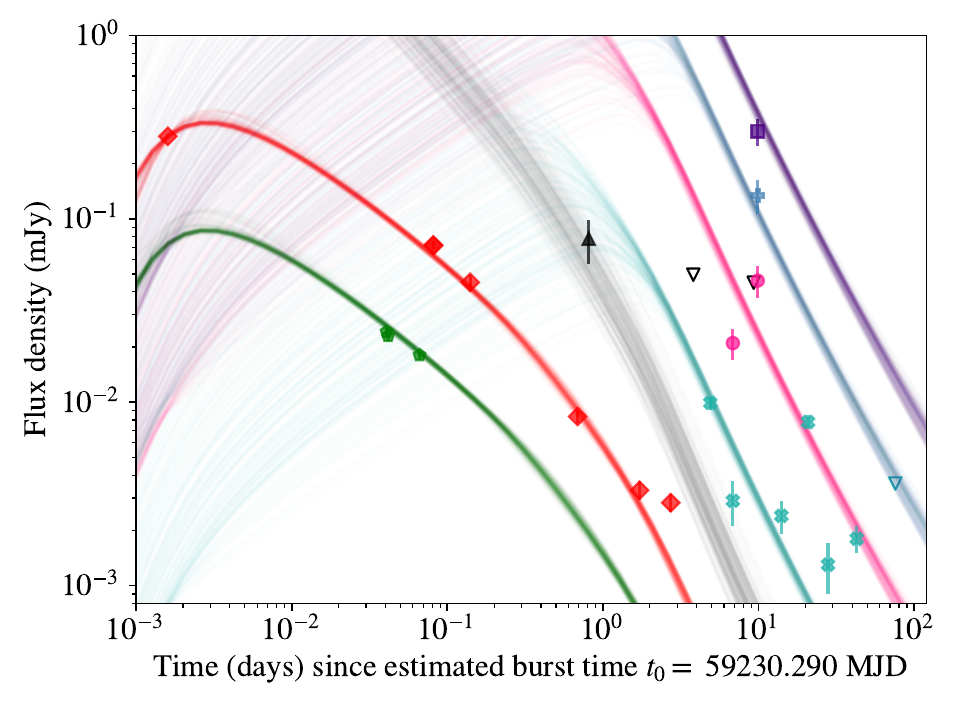}}%
    \caption{\textbf{Left}: On-axis top hat jet with $\Gamma_0 \approx 204$ for {\any} where the burst time was allowed to vary. \textbf{Right}: On-axis top hat jet with $\Gamma_0 \approx 81$ and a fixed $t_0 = 59230.290$ MJD. Both models are consistent with optical and X-ray observations, but struggle with radio X-band detections, possibly due to interstellar scintillation. Plotted are light curves generated from 150 randomly selected posterior samples. Radio upper limits are plotted at 3 $\times$ image RMS.}%
    \label{fig:any-results}
\end{figure*}

\begin{table*}[ht]
\normalsize
\centering
\begin{tabular}{lllll}
\hline
Parameter & Free $t_0$ & Fixed $t_0$ & G22 & X24\\
\hline
$t_0$ [MJD] & $59230.28_{-0.00}^{+0.00}$ & $59230.29$ & $59230.276^\dag$ & $59230.29_{-0.12}^{+0.16}$\\

$\theta_\text{v}$ [rad] & $0.05_{-0.01}^{+0.01}$ & $0.05_{-0.01}^{+0.01}$ & $0.55_{-0.27}^{+0.27}$ & $0.03_{-0.01}^{+0.01}$ \\

$\log_{10}(E_\text{K,iso}/\text{erg})$ & $53.54_{-0.36}^{+0.49}$ & $53.60_{-0.32}^{+0.34}$ & $52.58_{-0.03}^{+0.03}$ & $52.90_{-0.12}^{+0.12}$ \\

$\theta_\text{c}$ [rad] & $0.10_{-0.03}^{+0.03}$ & $0.11_{-0.02}^{+0.03}$ & $0.96_{-0.28}^{+0.17}$ & $0.08_{-0.01}^{+0.01}$ \\

$\log_{10}(n_0/\text{cm}^{-3})$ & $2.41_{-0.68}^{+0.56}$ & $2.57_{-0.43}^{+0.43}$ & $-0.06_{-0.17}^{+0.19}$ & $-0.78_{-0.19}^{+0.19}$ \\

$p$ & $2.01_{-0.01}^{+0.01}$ & $2.01_{-0.00}^{+0.01}$ & $2.30_{-0.05}^{+0.05}$ & $2.39_{-0.02}^{+0.02}$ \\

$\log_{10}\epsilon_e$ & $-0.36_{-0.49}^{+0.26}$ & $-0.33_{-0.37}^{+0.23}$ & $-1^\ddag$ & $-0.94_{-0.05}^{+0.05}$ \\

$\log_{10}\epsilon_B$ & $-4.70_{-0.22}^{+0.41}$ & $-4.78_{-0.16}^{+0.29}$ & $-2.23_{-0.13}^{+0.12}$ & $-2.76_{-0.24}^{+0.24}$ \\

$\xi_N$ & $0.63_{-0.29}^{+0.26}$ & $0.68_{-0.30}^{+0.22}$ & $1^\ddag$ & $1^\ddag$ \\

$\log_{10}\Gamma_0$ & $2.31_{-0.36}^{+0.45}$ & $1.91_{-0.09}^{+0.08}$ & $\infty^\ddag$ & $1.92_{-0.05}^{+0.06}$ \\

$\eta_\gamma$ & $< 11.8 - 48.6\%$ & $< 14.1 - 42.9\%$ & $< 77.8 - 80.1\%^\dag$ & $< 57.7 - 70.3\%^\dag$\\

$\chi^2$/DoF & $17.0$ & $13.3$ & -\\

{\elpd} & $(-1.3 \pm 4.3) \times 10^{2}$ & $10.3 \pm 31.3$ & - & -\\
\hline
\end{tabular}
\caption{Final parameters ($68\%$ uncertainty) for the on-axis, top hat, $\Gamma_0 \neq \infty$, configurations for {\any}. We calculate $\eta_\gamma$ using the $1\sigma$ distribution of {\Ekiso} and the $E_{\gamma,\text{iso}}$ limits from Table~\ref{table:radiative-energies}. We present the {\elpd} and minimum $\chi^2$/DoF over 5,000 posterior samples. The large values of $\chi^2$/DoF are due to poor fitting in the radio X band (9.0 - 9.7\,GHz). The fixed $t_0$ model has a more predictive {\elpd} because it has one less free parameter. We also include top hat configurations from \citet{Gupta_2022} and \citet{Xu_2023}. Ran with 64 walkers and 75,000 iterations; discarded 25,000.\\
$^\dag$ G22 and X24 report smaller efficiencies assuming a typical GRB energy fluence threshold $\leq 10^{-6}$\,erg\,cm$^{-1}$; we use a fluence threshold from KONUS-\textit{Wind}, which is potentially conservative.\\
$^\ddag$ Fitting settings.}
\label{tab:any-results}
\end{table*}

We present the results of an on-axis top hat jet with a finite $\Gamma_0$ in Table~\ref{tab:any-results} and Figure~\ref{fig:any-results}, with a corner plot in Figure~\ref{fig:any-corner} of Appendix~\ref{app:corner-plots}. Since the first detection and last non-detection of {\any} are only 22 minutes apart, we also fit {\any} to an on-axis top hat jet with a fixed burst time $t_0 = 59230.290$\,MJD, which we note is an arbitrary choice that lies between the last non-detection and first detection. We present the results of this configuration in Table~\ref{tab:any-results} and Figure~\ref{fig:any-results}, with a corner plot in Figure~\ref{fig:any-t0-corner} of Appendix~\ref{app:corner-plots}. The free $t_0$ model places the peak of the light curve before the first ZTF detection, while the fixed $t_0$ model places it after. This is expected given the earlier burst time found for the free $t_0$ model in Table~\ref{tab:any-results}. There are no other notable differences and both configurations are able to reproduce all observations. Other {\afterglowpy} configurations are also consistent with observations, except those with ICC enabled, which fail to account for the X-ray observation. See Table~\ref{tab:any-selected} in Appendix~\ref{app:selected} for results of selected configurations. Optical only and optical-X-ray only fits allowed for potentially off-axis and low-efficiency solutions ($\eta_\gamma \lesssim 0.2 - 0.7\%$), all with typical Lorentz factors. However, the potentially off-axis solutions are highly ambiguous given the lack of early-time observations for {\any}.

As shown in Table~\ref{tab:any-results}, both models also find low-$\Gamma_0$ and high-$\Gamma_0$ solutions. The free $t_0$ model finds $\Gamma_0 \approx 204_{-115}^{+371}$, while the fixed $t_0$ model finds $\Gamma_0 \approx 81_{-15}^{+17}$, indicating that {\any} is possibly consistent with both a moderate and ultra-relativistic jet. Ultimately, we lack the higher-cadence early optical data to resolve this degeneracy.

\begin{figure*}[ht]
    \centering
    \subfloat{\includegraphics[width=8.5cm]{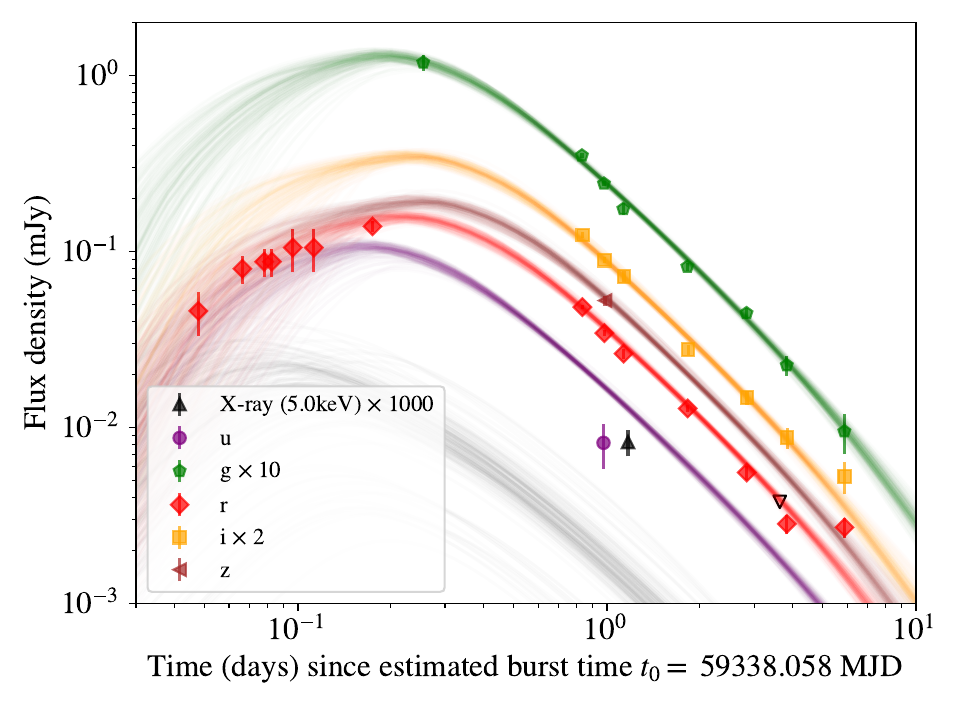}}%
    \qquad
    \subfloat{\includegraphics[width=8.5cm]{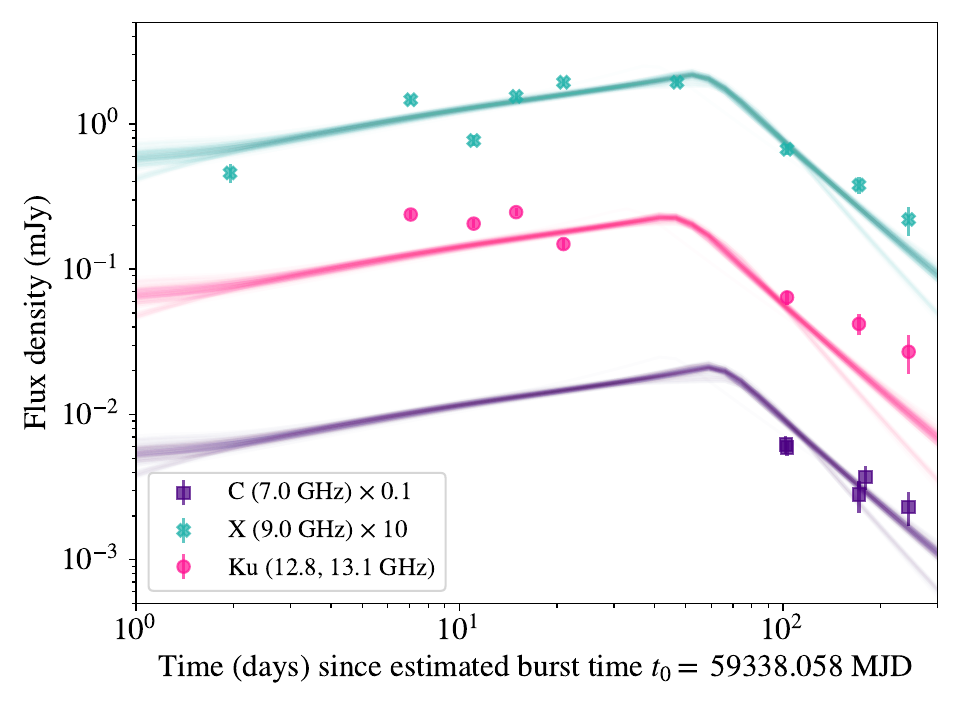}}%
    \qquad
    \subfloat{\includegraphics[width=8.5cm]{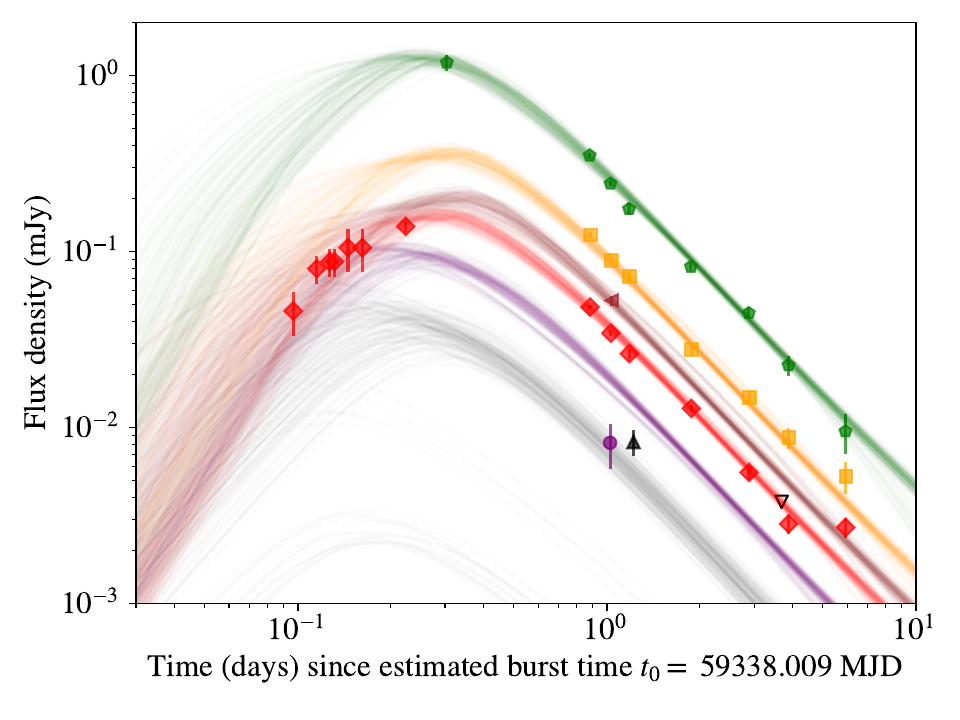}}%
    \qquad
    \subfloat{\includegraphics[width=8.5cm]{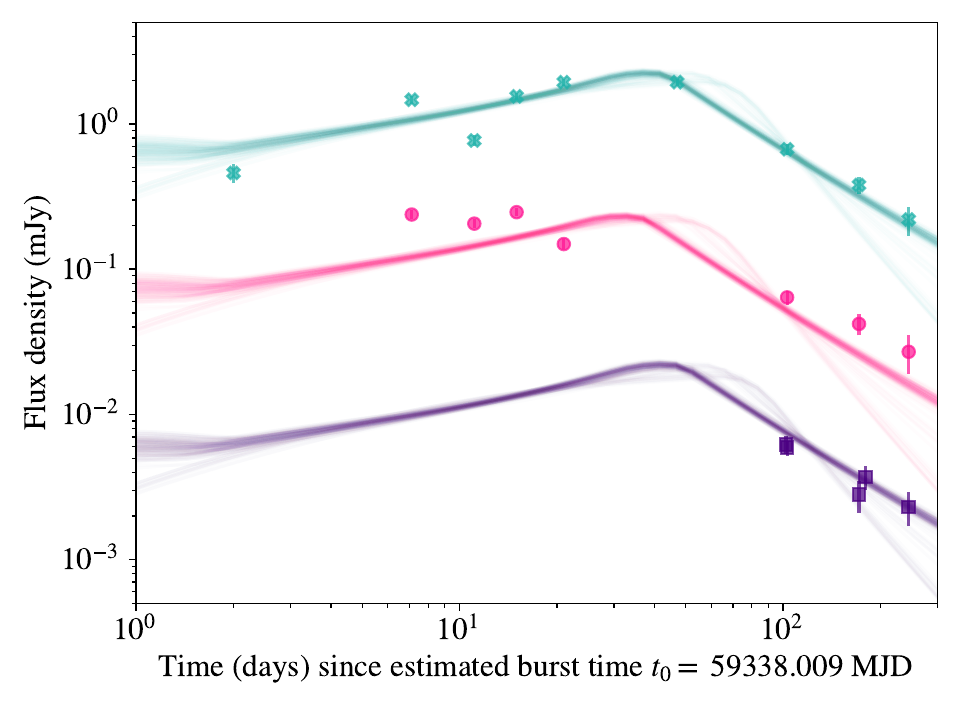}}%
    \caption{\textbf{Top}: On-axis top hat jet with $\Gamma_0 \approx 11.5$ for {\lfa}, fit to X-ray, optical (left), and radio observations (right). The model is consistent with optical and radio observations, but underestimates the X-ray detection by an order of magnitude. \textbf{Bottom}: On-axis Gaussian jet with $\Gamma_0 \approx 6.0$ for {\lfa}, which is consistent with all observations. Plotted are light curves generated from 150 randomly selected posterior samples.}%
    \label{fig:lfa-results}
\end{figure*}

\begin{table*}[ht]
\normalsize
\centering
\begin{tabular}{llll}
\hline
Parameter & Top hat & Gaussian & Y24 \\
\hline
$t_0$ [MJD] & $59338.06_{-0.02}^{+0.01}$ & $59338.01_{-0.02}^{+0.03}$ & $59337.92_{-0.04}^{+0.08}$\\

$\theta_\text{v}$ [rad] & $0.11_{-0.01}^{+0.01}$ & $0.06_{-0.05}^{+0.04}$ & $0.53_{-0.19}^{+0.18}$\\

$\log_{10}(E_\text{K,iso}/\text{erg})$ & $51.76_{-0.23}^{+0.46}$ & $52.91_{-0.52}^{+0.45}$ & $54.77_{-0.39}^{+0.43}$\\

$\theta_\text{c}$ [rad] & $0.23_{-0.02}^{+0.02}$ & $0.08_{-0.03}^{+0.05}$ & $0.66_{-0.24}^{+0.21}$\\

$\theta_\text{w}$ & - & $0.33_{-0.14}^{+0.25}$ & - \\

$\log_{10}(n_0/\text{cm}^{-3})$ & $1.11_{-0.39}^{+0.42}$ & $3.62_{-0.88}^{+0.63}$ & $1.04_{-0.84}^{+0.70}$\\

$p$ & $2.53_{-0.05}^{+0.06}$ & $2.17_{-0.04}^{+0.07}$ & $3.09_{-0.03}^{+0.03}$\\

$\log_{10}\epsilon_e$ & $-0.32_{-0.45}^{+0.23}$ & $-0.28_{-0.54}^{+0.21}$ & $-1.18_{-0.33}^{+0.32}$\\

$\log_{10}\epsilon_B$ & $-1.62_{-0.39}^{+0.30}$ & $-3.52_{-0.61}^{+0.78}$ & $-4.47_{-0.38}^{+0.70}$\\

$\xi_N$ & $0.48_{-0.27}^{+0.31}$ & $0.11_{-0.06}^{+0.06}$ & $0.70_{-0.27}^{+0.22}$\\

$\log_{10}\Gamma_0$ & $1.06_{-0.05}^{+0.06}$ & $0.78_{-0.06}^{+0.12}$ & $\approx 1.3^{\dag}$ \\

$\eta_\gamma$ & $<6.7-26.2\%$ & $<0.5-4.7\%$ & $< 0.01 - 0.05\%$ \\

$\chi^2$/DoF & $6.0$ & $4.7$ & - \\

{\elpd} & $77.0 \pm 39.4$ & $36.7 \pm 43.6$ & - \\
\hline
\end{tabular}
\caption{Final parameters ($68\%$ uncertainty) for the on-axis, finite $\Gamma_0$ jets for {\lfa}. We also include results of the top hat configuration from \citet{Ye_2024}. We calculate $\eta_\gamma$ with respect to the $E_{\gamma,\text{iso}}$ limits from Table~\ref{table:radiative-energies}. We present the {\elpd} and minimum $\chi^2$/DoF over 5,000 posterior samples. Ran with 64 walkers and 225,000 iterations; discarded 125,000.\\
$^\dag$ Not from MCMC.}
\label{tab:lfa-results}
\end{table*}

Both models find physical parameters typical of the LGRB population, including an opening angle $\theta_\text{c} \approx 6^\circ$ and beaming-corrected $E_K \approx 2 \times 10^{51}$\,erg. The densities found are somewhat high at $n_0 \sim 300$ cm$^{-3}$, but as discussed in Section~\ref{subsec:blt-results}, this is not physically implausible. The radiative efficiencies found are also typical of LGRBs \citep{Racusin_2011}. Given the uncertainty on $\Gamma_0$ in both the {\afterglowpy} fits and the analytical constraints in Section~\ref{sec:lorentz-factor-constraints}, we conclude that an on-axis classical GRB origin cannot be ruled out for {\any}.

\subsubsection{Comparison to G22 and X23}
\label{g22-x23}
G22 use {\afterglowpy} and {\emcee} to model the optical and X-ray observations of {\any}, while X23 use GRB evolution models from \citet{Huang_2002_grbmodel1}, \citet{Huang_2006_grbmodel2}, \citet{Geng_2013}, and \citet{Xu_2022}. Both G22 and X23 set $\xi_N = 1$, and G22 uses a $\Gamma_0 = \infty$ {\afterglowpy} configuration. The results from these works are shown in Table~\ref{tab:any-results}.  G22 finds that {\any} is consistent with an on-axis classical GRB, while X23 explains {\any} as an on-axis GRB with a moderate Lorentz factor $\Gamma_0 = 68$.
Differences in physical parameters are likely due to 
differences in fitting configurations. Notably, both G22 and X23 set $\xi_N = 1$. Past works have shown that different values of $\xi_N$ can significantly change other physical parameters \citep{Cunningham_2020}, so this is expected; the discrepancies between our results and G22/X24 are consistent with the expected effects of decreasing $\xi_N$ as discussed in \citet{Cunningham_2020}. However, the physical conclusion is robust: an on-axis classical GRB cannot be ruled out for {\any}.

\subsection{AT2021lfa}
We present the results of an on-axis top hat jet with a finite $\Gamma_0$ in Table~\ref{tab:lfa-results} and Figure~\ref{fig:lfa-results}, with a corner plot in Figure~\ref{fig:lfa-corner} of Appendix~\ref{app:corner-plots}. The model is consistent with optical observations, but underestimates the X-ray detection by $\sim 1$ order of magnitude, and overestimates Ku-band (13\,GHz) detections at $\Delta t \gtrsim 110$\,d by around a factor of 3. We also include in Table~\ref{tab:lfa-results}, Figure~\ref{fig:lfa-results}, and Figure~\ref{fig:lfa-corner-gauss} of Appendix~\ref{app:corner-plots} the results of a finite $\Gamma_0$ Gaussian jet, which is able to reproduce all observations, but slightly underestimates late-time Ku-band detections.

From Table~\ref{tab:lfa-results}, we obtain a beaming-corrected $E_K \sim 10^{50}$\,erg and opening angles $\theta_\text{c} \approx 4.6^\circ$ and $13.2^\circ$ respectively for the Gaussian and top hat models, typical of the LGRB population \citep{Ghirlanda_2018}. The Gaussian model also prefers a much denser environment 
($\approx 4000$\,cm$^{-3}$). Both models have very low Lorentz factors, with $\Gamma_0 \approx 11$ for the top hat model and $\Gamma_0 \approx 6$ for the Gaussian model. The Gaussian model also allows for slightly off-axis ($\theta_\text{v} \sim \theta_\text{c}$) solutions. The Gaussian model also obtains a possibly low efficiency $\eta_\gamma < 0.5\%$, which is smaller than $98.5\%$ of LGRB efficiencies in \citet{Racusin_2011} but consistent with the $\lesssim 1\%$ efficiencies of internal shocks models \citep{Kumar_1999}. 

All other {\afterglowpy} configurations (see Table~\ref{tab:lfa-selected} in Appendix~\ref{app:selected} for results of selected configurations) produce similar results but still have a strong preference for a very low Lorentz factor jet (typically, $\Gamma_0 \approx 5 - 20$), consistent with the analytical $\Gamma$ limits calculated in Section~\ref{sec:lorentz-factor-constraints} and the $\Gamma_0 = 20 \pm 10$ estimate found from \citet{Lipunov_2022}. We note that $\Gamma_0 = \infty$ configurations struggled to reproduce the MASTER observations.

We also fit to a range of data subsets. Optical only and optical-X-ray only configurations obtained on- and off-axis solutions, still with low Lorentz factors ($\Gamma_0 \approx 5 - 20$). We also ran fits that excluded the rising phase, which obtained both classical GRB solutions and on-axis, low Lorentz factor solutions. However, we note that the on-axis classical GRB fits underestimated the Ku-band (13.0\,GHz) detections at $\Delta t \gtrsim 110$\,d by around a factor of 3.

From our fitting, {\lfa} is consistent with on-axis and possibly off-axis low Lorentz factor jets. We explore an off-axis solution in more detail in Section~\ref{sec:off-axis-lfa}. In any case, the immediate results indicate a strong preference for a jet with $\Gamma_0 = 5 - 13$, which is remarkably small. The overwhelming majority of classical GRBs report having $\Gamma_0 \gtrsim 100$, with previous calculations on large catalogues of classical GRBs indicating a median $\Gamma_0 = 320$ for a homogeneous circumburst medium \citep{Ghirlanda_2018}.

\begin{table}[t]
    \centering
    \large
    \begin{tabular}{lcc}
    \hline
    & $\alpha$ & $\beta$ \\
    \hline

    ISM, slow & 0.97 & 0.65 \\
    ISM, fast & 1.22 & 1.15 \\
    Wind, slow & 1.47 & 0.65 \\
    Wind, fast & 1.22 & 1.15 \\
    Jet, slow & 2.30 & 0.65 \\

    \hline
    \end{tabular}
    \caption{Approximate values of the optical temporal index $\alpha$ and the optical SED index $\beta$ for various afterglow models using a typical $p = 2.3$. Values were estimated using standard closure relations from \cite{Zhang_2004b}. We assume $\nu_m < \nu < \nu_c$ for slow cooling cases and $\nu > \nu_m$ for fast cooling cases. }
    \label{tab:closure_relations_lfa}
\end{table}

Some stellar wind LGRBs with successful prompt emissions have had Lorentz factors as small as $\Gamma_0 \approx 20$ \citep{Ghirlanda_2018}. Using standard closure relations from Table 1 of \citet{Zhang_2004b}, we can determine if {\lfa} is consistent with having a wind medium. For {\lfa}, the optical SED index is $\beta = 0.32 \pm 0.46$ \citep{Ho_2022}, but the optical temporal index $\alpha$ is uncertain, especially given the large latency (1.79\,d) between the last GOTO non-detection (MJD = 59336.311) and first GOTO detection (MJD = 59338.105). Fitting a single power law to the optical light curve, we find $\alpha \in (1.2, 3.8)$. Considering a typical electron energy power law index $p \sim 2.3$ \citep{Zhang_2004b}, we obtain values for $\alpha$ and $\beta$ in Table~\ref{tab:closure_relations_lfa}, finding that a stellar wind origin for {\lfa} cannot be ruled out. 


{\lfa} could also be the result of a dirty fireball. 
We estimate the baryon loading of {\lfa} with $E_\text{K,iso} = M \Gamma_0 c^2$ \citep{Ghirlanda_2018}. From Table~\ref{tab:lfa-results}, we find $M \approx 2.8 \times 10^{-4}$\,M$_\odot$ for the top hat model and $M \approx 7.6 \times 10^{-3}$\,M$_\odot$ for the Gaussian model, larger than typical LGRB baryon loading values (typically $10^{-6}$\,M$_\odot$; \citealt{Ghirlanda_2018}) and somewhat larger than the expected baryon loading content required to efficiently produce gamma-rays ($\lesssim 10^{-4}$\,M$_\odot$; \citealt{Piran_2005}), indicating that {\lfa} could be a dirty fireball with strong baryon loading \citep{Rhoads_2003, Huang_2002}, and thus low gamma-ray photon production.

\subsubsection{Comparison to Y24}
\label{y24}
Y24 use {\afterglowpy} and {\emcee} to model all multi-wavelength observations of {\lfa}. Similarly to S22, Y24 fits the burst time $t_0$ independently from other parameters, obtaining $t_{0,\text{Y24}} = 59337.92_{-0.04}^{+0.08}$\,MJD, consistent within $1\sigma$ of our estimated burst times. Otherwise, Y24’s physical parameters differ significantly from ours. Notably, Y24 obtain an opening angle $\theta_\text{c} \approx 38^\circ$ and a beaming-corrected blast wave energy $E_K \approx 1.3 \times 10^{54}$\,erg, which is 4 orders of magnitude greater than our estimate. These values are larger than those of the vast majority of the LGRB population \citep{Ghirlanda_2005, Laskar_2013b, Goldstein_2016}. 
The discrepancies between Y24 and this work are likely due to a difference in fitting configurations; Y24 use an {\afterglowpy} configuration in which $\Gamma_0 = \infty$ and different priors. Despite configuring $\Gamma_0 = \infty$, Y24 constrains $\Gamma \approx 18$ from Equation \ref{eq:tdec} using values from their MCMC fit, which is somewhat larger than our results, but still a remarkably low Lorentz factor.

\subsubsection{Off-axis Interpretation}
\label{sec:off-axis-lfa}

\begin{table}[ht]
\normalsize
\centering
\begin{tabular}{lll}
\hline
Parameter & Initial & Prior (Uniform) \\
\hline
$t_0$ [MJD] & 59338.09 & [59338.05, 59338.10$^\dag$]\\

$\theta_\text{v}$ [rad] & 0.16 & $[1, 6] \times \theta_\text{c}$\\

$\log_{10}(E_\text{K,iso}/\text{erg})$ & 53.14 & [45, 57]\\

$\theta_\text{c}$ [rad] & 0.09 & [0.02, 0.78]\\

$\theta_\text{w}$ [rad] & 0.15 & $[1, 7] \times \theta_\text{c}$\\

$\log_{10}(n_0/\text{cm}^{-3})$ & -3.73 & [-10, 10]\\

$p$ & 2.79 & [2, 3]\\

$b$ & 5 & [0, 10]\\

$\log_{10}\epsilon_e$ & -1.59 & [-5, 0]\\

$\log_{10}\epsilon_B$ & -1.79 & [-5, 0]\\

$\xi_N$ & 0.10 & [0, 1]\\

$\log_{10}\Gamma_0$ & 2.30 & [2, 5]\\

\hline
\end{tabular}
\caption{Values around which the walkers were initialized and priors for the forced off-axis fit. We are aware that priors for $\epsilon_e$ and $\epsilon_B$ allow for $\epsilon_e + \epsilon_B > 1$, but none of the fit results are unphysical. $^\dag$In this table, we truncate the time of the first detection (MJD = $59338.1054282$), but use all decimal places in our MCMC analysis.}
\label{tab:off-axis-prior}
\end{table}

{\lfa} has a rest-frame rise time $\gtrsim 5600$\,s, slower than all upper limits and observed LGRB rise times from \citet{Ghirlanda_2018} and \citet{Hasco_t_2014}. Given the relation between $L_{\gamma,\text{iso}}$ and rise time, the rise time of AT\,2021lfa would imply that any associated LGRB has $L_{\gamma,\text{iso}} \lesssim 10^{47}$\,erg\,s$^{-1}$ \citep{Ghirlanda_2018} which is consistent with the limit $L_{\gamma,\text{iso}} < 2.6 \times 10^{51}$\,erg\,s$^{-1}$ from \citet{Ho_2022}.


Out of all discovered afterglows without associated detected GRBs, {\pim} is the only other event with a confirmed comparably long rest-frame rise time between $1800 - 7200$\,s. If the rise times of {\pim} and {\lfa} are due to deceleration viewed on-axis, then it is likely that they are the result of low Lorentz factor jets, as explored in this work and \citet{Perley_2024}. On the other hand, their long rise times may be due to being viewed off-axis, in which case a high-$\Gamma_0$ jet is possible. Indeed, this degeneracy is present in \citet{Perley_2024}, where an on-axis low-$\Gamma_0$ jet and a slightly off-axis high-$\Gamma_0$ jet are both found as viable solutions for {\pim}. We note that if {\lfa}'s radio fluctuations are due to interstellar scintillation, then {\lfa}'s scintillation timescale would be $\sim 102$\,d, favoring a low-$\Gamma_0$ interpretation (see Section~\ref{table:analytical-lorentz-factors}). However, fluctuations could be explained by other effects, such as circumstellar density variations between the early and late time emission.

\begin{figure*}[ht]
    \centering
    \subfloat{\includegraphics[width=8.5cm]{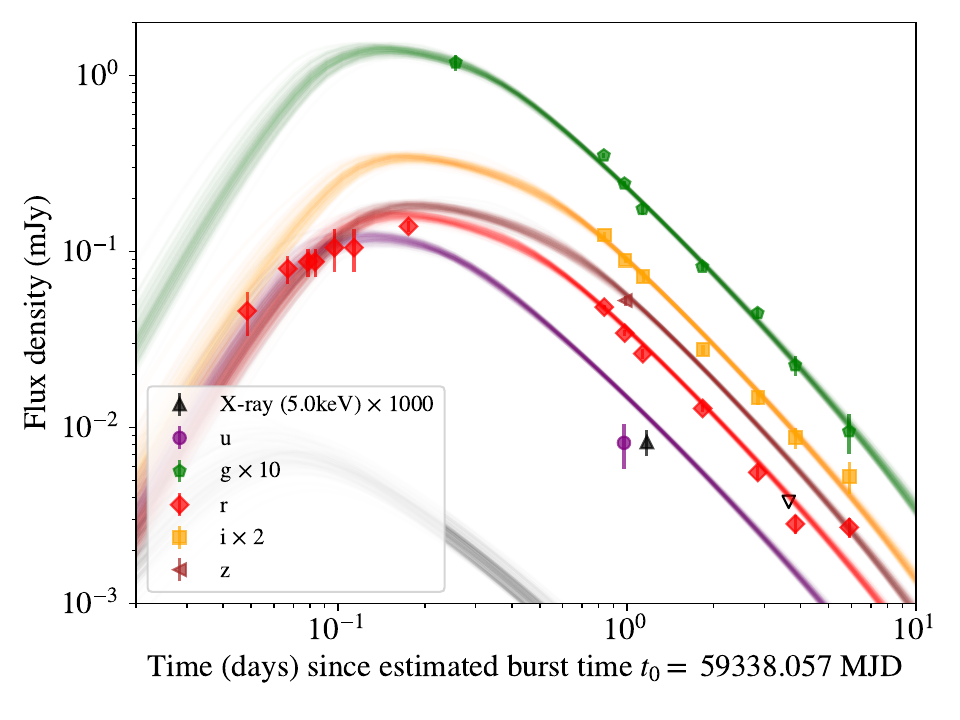}}%
    \qquad
    \subfloat{\includegraphics[width=8.5cm]{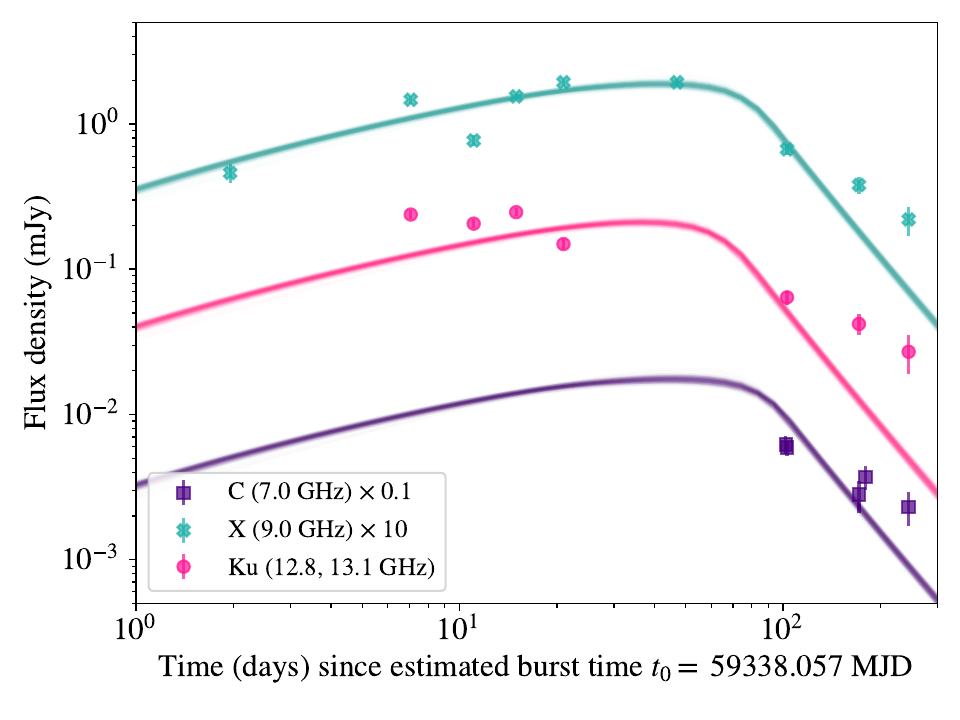}}%
    \qquad
    \subfloat{\includegraphics[width=8.5cm]{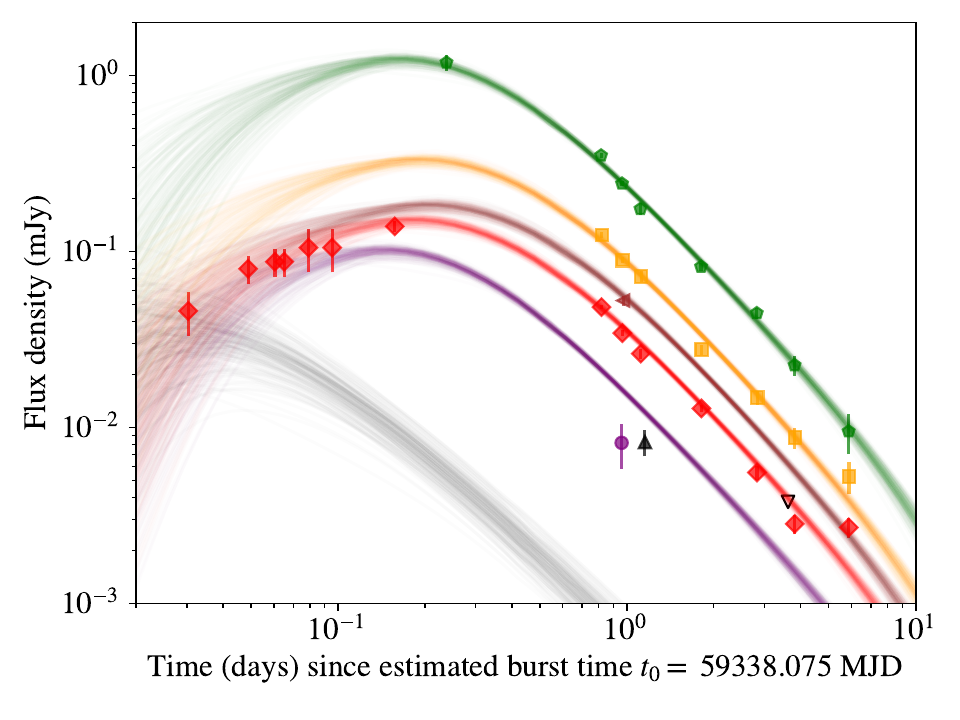}}%
    \qquad
    \subfloat{\includegraphics[width=8.5cm]{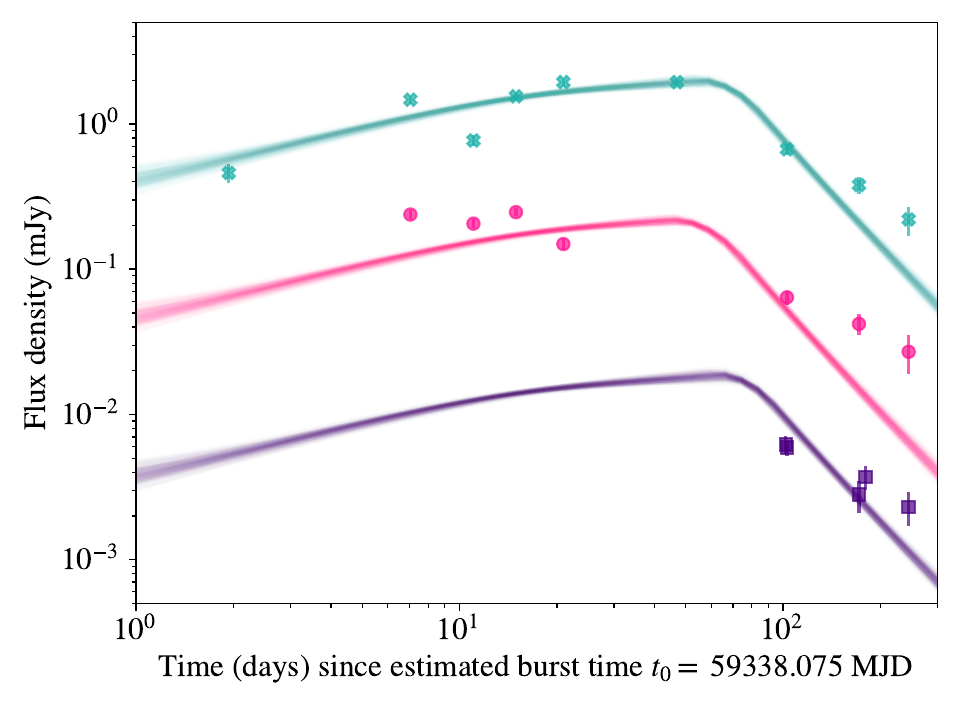}}%
    \caption{\textbf{Top}: Off-axis Gaussian jet with $\Gamma_0 \approx 110$ for {\lfa}, fit to X-ray, optical (left), and radio observations (right). The model struggles with the finer features of the rising phase detections. \textbf{Bottom}: Off-axis power law jet with $\Gamma_0 \approx 155$ for {\lfa}, which is consistent with all optical observations. Both the Gaussian and power law models overestimate the X-ray detection and late-time radio observations. Plotted are light curves generated from 150 randomly selected posterior samples.}%
    \label{fig:lfa-off-axis-results}
\end{figure*}

Motivated by {\lfa}’s slow rise time, \citet{Perley_2024}, and off-axis solutions present in the posterior of the former analysis, we explore an off-axis fit for {\lfa}. First, we find a plausible off-axis high-$\Gamma_0$ solution by manually varying {\afterglowpy} jet parameters, around which we set our priors, shown in Table~\ref{tab:off-axis-prior}. We run Gaussian and power law configurations, since only structured jets will be able to capture very off-axis ($\theta_\text{v} \gtrsim 2 \times \theta_\text{c}$) emission. We fit with a finite $\Gamma_0$, no ICC, and use all radio, optical, and X-ray observations. We also ran fits with $\Gamma_0 = \infty$, but they were unsuccessful, typically overestimating optical and radio light curves, especially the rising phase MASTER detections. We present the results of our fitting in Table~\ref{tab:lfa-off-axis-results} and Figure~\ref{fig:lfa-off-axis-results}, with corner plots in Figures~\ref{fig:lfa-off-axis-corner-gauss} and \ref{fig:lfa-off-axis-corner-power} in Appendix~\ref{app:corner-plots}. 

\begin{table}[ht]
\normalsize
\centering
\begin{tabular}{lll}
\hline
Parameter & Gaussian & Power law\\
\hline
$t_0$ [MJD] & $59338.06_{-0.00}^{+0.01}$ & $59338.08_{-0.01}^{+0.01}$ \\

$\theta_\text{v}$ [rad] & $0.06_{-0.01}^{+0.01}$ & $0.13_{-0.01}^{+0.01}$\\

$\log_{10}(E_\text{K,iso}/\text{erg})$ & $53.14_{-0.27}^{+0.85}$ & $52.41_{-0.23}^{+0.39}$\\

$\theta_\text{c}$ [rad] & $0.03_{-0.00}^{+0.00}$ & $0.10_{-0.01}^{+0.01}$\\

$\theta_\text{w}$ [rad] & $0.15_{-0.06}^{+0.07}$ & $0.11_{-0.01}^{+0.01}$\\

$\log_{10}(n_0/\text{cm}^{-3})$ & $-2.59_{-0.39}^{+0.77}$ & $-0.16_{-0.43}^{+0.45}$\\

$p$ & $2.96_{-0.04}^{+0.02}$ & $2.79_{-0.05}^{+0.05}$\\

$b$ & - & $4.50_{-3.38}^{+3.74}$\\

$\log_{10}\epsilon_e$ & $-1.37_{-0.84}^{+0.27}$ & $-0.75_{-0.39}^{+0.22}$\\

$\log_{10}\epsilon_B$ & $-0.53_{-0.83}^{+0.31}$ & $-1.29_{-0.39}^{+0.27}$\\

$\xi_N$ & $0.44_{-0.38}^{+0.37}$ & $0.53_{-0.31}^{+0.31}$\\

$\log_{10}\Gamma_0$ & $2.04_{-0.03}^{+0.05}$ & $2.19_{-0.14}^{+0.26}$\\

$\eta_\gamma$ & $<0.1-1.6\%$ & $<1.9-7.3\%$\\

$\chi^2$/DoF & $5.5$ & $5.5$\\

{\elpd} & $(1.1 \pm 0.3) \times 10^{2}$ & $(1.2 \pm 0.3) \times 10^{2}$\\

\hline
\end{tabular}
\caption{Final parameters ($68\%$ uncertainty) for the off-axis, finite $\Gamma_0$ jets for {\lfa}. We calculate $\eta_\gamma$ using the $1\sigma$ distribution of {\Ekiso} and the $E_{\gamma,\text{iso}}$ limits from Table~\ref{table:radiative-energies}. We present the {\elpd} and minimum $\chi^2$/DoF over 5,000 posterior samples. Ran with 64 walkers and 225,000 iterations; discarded 125,000.}
\label{tab:lfa-off-axis-results}
\end{table}

Both models are consistent with optical observations, but the Gaussian model struggles with the finer features of the rising phase $r$-band detections. The Gaussian fit also includes an optical light curve break at $\sim 0.5$ days, which is not present in the power law fit or the previous on-axis solutions. The models are also consistent with the radio emission at $\Delta t \lesssim 100$\,d, similar to the on-axis fits, but underestimate late-time observations at $\gtrsim 110$\,d by varying orders of magnitude, at most $\sim 0.5$. By contrast, the on-axis Gaussian solution (see Figure~\ref{fig:lcr-results}) is generally consistent with late-time radio detections. 
This discrepancy may favor a wind environment, which would result in shallower light curves.
Additionally, the Gaussian and power law jets underestimate the X-ray detection by $\sim 1.5$ and $\sim 1$ orders of magnitude, respectively. In comparison, the on-axis Gaussian solution is able to reproduce to X-ray detection, although the on-axis top hat model struggles by $\sim 1$ order of magnitude. As previously mentioned, observed X-ray excesses could be due to an ongoing central engine activity \citep{Zhao_2020} or, since {\lfa} only has a single X-ray detection, an insufficient $\chi^2$ penalty.

From Table~\ref{tab:lfa-off-axis-results}, we obtain $\theta_\text{c} \approx 1.7^\circ$ and $\theta_\text{v} \approx 3.4^\circ$ for the Gaussian solution, which is very off-axis ($\theta_\text{v} \approx 2 \times \theta_\text{c}$). For the power law fit, we obtain $\theta_\text{c} \approx 5.7^\circ$ and $\theta_\text{v} \approx 7.5^\circ$, which is less off-axis ($\theta_\text{v} \approx 1.3 \times \theta_\text{c}$). The Gaussian model obtains a beaming-corrected $E_\text{K} \sim 6 \times 10^{49}$\,erg while the power law model obtains a somewhat greater $E_\text{K} \sim 10^{50}$\,erg, both within typical ranges of LGRB kinetic energies \citep{Yi_2017, Ghirlanda_2018}. Both models obtain typical Lorentz factors, the Gaussian jet with $\Gamma_0 \approx 110$ and power law jet with $\Gamma_0 \approx 155$. Lastly, the Gaussian model obtains a possibly low efficiency $\eta_\gamma < 0.1\%$, smaller than $98.5\%$ of bursts in \citet{Racusin_2011}.

Generally, the off-axis solutions obtain comparable beaming-corrected kinetic energies ($\sim 10^{50}$\,erg), smaller opening angles, and smaller densities to the on-axis fits in Table~\ref{tab:lfa-results}. The off-axis solutions also obtain comparable $\chi^2$/DoFs, but more predictive {\elpd} scores.

Overall, we find that an off-axis high-$\Gamma_0$ origin for {\lfa} cannot be ruled out. The models' underestimates of late-time radio emission may be due to {\afterglowpy}'s lack of jet spreading for the finite $\Gamma_0$ setting, and the observed X-ray excesses may be due an insufficient $\chi^2$ penalty on the single X-ray detection or from ongoing central engine activity \citep{Zhao_2020}.

\section{Conclusion}
\label{sec:conclusion}

In this work, we presented the identification and multiwavelength observations of {\lcr}, a red, cosmological fast optical transient detected without a GRB trigger. With {\lcr}, there are now 10 total afterglows discovered without associated detected GRBs, and six such events with a measured redshift. Using {\afterglowpy} and {\emcee}, we modeled the multi-wavelength emission of {\lcr} and three similarly discovered afterglows, {\blt}, {\any}, and {\lfa}. We found that a classical on-GRB origin cannot be ruled out for {\lcr}, {\blt}, and {\any}.
However, we also found that {\blt} and {\any} could also be described with non-classical solutions (off-axis and/or low-$\Gamma_0$). The multimodalities in the solution may be due to a lack of detailed early-time data, but could also arise from {\emcee}/{\afterglowpy} biasing our posteriors to particular locations in parameter space. 

Of all afterglows explored in this work, only {\lfa} has a convincing non-classical origin, largely motivated by the slow optical rise time. We found that {\lfa} is consistent with both on-axis low Lorentz factor ($\Gamma_0 = 5 - 13$) and off-axis high Lorentz factor ($\Gamma_0 \approx 100$) jets. 
The long-lasting fluctuations in {\lfa}'s radio light curve may favor the low-$\Gamma_0$ solution, implying a smaller radius and therefore slower expansion speed than other events.


We note that without the rise phase of the optical light curve, multiwavelength modeling of {\lfa} yields a result consistent with an on-axis classical GRB. 
Since early-time observations are more sensitive to initial physical conditions, such as the initial Lorentz factor, being able to capture early-time emission is extremely important to constraining an afterglow's origin. 
The upcoming Argus Array \citep{Argus} promises a high sensitivity, high cadence, and wide field of view, so should be well-suited to routinely detect the rising phase. 

Our analysis on {\lfa} makes it the second afterglow without an associated detected GRB that is consistent with both on-axis low-$\Gamma_0$ and off-axis high-$\Gamma_0$ solutions, the first being {\pim} \citep{Perley_2024}. 
To resolve the degeneracy, a detection of the prompt emission with wide-field X-ray surveys such as Einstein Probe \citep{Einstein_Probe} may be needed. Both dirty fireballs and off-axis GRBs would be expected to be accompanied by an X-ray flash \citep{Heise_2001, Zhang_2004, Sakamoto_2005, Soderberg_2007}, but off-axis afterglow emission should be smoother, while on-axis prompt emission should have shorter-timescale variability. 
In addition, off-axis GRBs are expected to be accompanied by cocoon emission that peaks in the UV \citep{Nakar_2016}. Such emission could be detected by the high cadence and sensitivity of the upcoming wide-field survey ULTRASAT \citep{Shvartzvald2024}.

\section{Acknowledgements}

M.L.L. was supported in part by a Roger and Mary Lou West Summer Fellowship. A.Y.Q.H. was supported in part by NASA Grant 80NSSC23K1155. We thank Murray Brightman for advice on Chandra data reduction. Research at Perimeter Institute is supported in part by the Government of Canada through the Department of Innovation, Science and Economic Development and by the Province of Ontario through the Ministry of Colleges and Universities. M.W.C acknowledges support from the National Science Foundation with
grant numbers PHY-2308862 and PHY-2117997. J.S.B. was partially supported by a grant from the Gordon and Betty Moore Foundation and the National Science Foundation (2206744). N.S. acknowledges support from the Knut and Alice Wallenberg Foundation through the “Gravity Meets Light” project. 

Based on observations obtained with the Samuel Oschin Telescope 48-inch and the 60-inch Telescope at the Palomar Observatory as part of the Zwicky Transient Facility project. ZTF is supported by the National Science Foundation under Grant No. AST-2034437 and a collaboration including Caltech, IPAC, the Weizmann Institute of Science, the Oskar Klein Center at Stockholm University, the University of Maryland, Deutsches Elektronen-Synchrotron and Humboldt University, the TANGO Consortium of Taiwan, the University of Wisconsin at Milwaukee, Trinity College Dublin, Lawrence Livermore National Laboratories, IN2P3, University of Warwick, Ruhr University Bochum, Cornell University, and Northwestern University. Operations are conducted by COO, IPAC, and UW.

The ZTF forced-photometry service was funded under the Heising-Simons Foundation grant \#12540303 (PI: Graham). The Gordon and Betty Moore Foundation, through both the Data-Driven Investigator Program and a dedicated grant, provided critical funding for SkyPortal.

The Liverpool Telescope is operated on the island of La Palma by Liverpool John Moores University in the Spanish Observatorio del Roque de los Muchachos of the Instituto de Astrofisica de Canarias with financial support from the UK Science and Technology Facilities Council.

This work made use of data from the GROWTH-India Telescope (GIT) set up by the Indian Institute of Astrophysics (IIA) and the Indian Institute of Technology Bombay (IITB). It is located at the Indian Astronomical Observatory (Hanle), operated by IIA. We acknowledge funding by the IITB alumni batch of 1994, which partially supports operations of the telescope. Telescope technical details are available online \citep{Kumar2022}. We thank the staff of IAO, Hanle and CREST, Hosakote, that made these obervations possible. The facilities at IAO and CREST are operated by the Indian Institute of Astrophysics, Bangalore.

Some of the data presented herein were obtained at the W. M. Keck Observatory, which is operated as a scientific partnership among the California Institute of Technology, the University of California, and NASA. The Observatory was made possible by the generous financial support of the W. M. Keck Foundation.
The authors wish to recognise and acknowledge the very significant cultural role and reverence that the summit of Maunakea has always had within the indigenous Hawaiian community. We are most fortunate to have the opportunity to conduct observations from this mountain.

The National Radio Astronomy Observatory is a facility of the National Science Foundation operated under cooperative agreement by Associated Universities, Inc.

GMRT is run by the National Centre for Radio Astrophysics of the Tata Institute of Fundamental Research

The Submillimeter Array is a joint project between the Smithsonian Astrophysical Observatory and the Academia Sinica Institute of Astronomy and Astrophysics and is funded by the Smithsonian Institution and the Academia Sinica.
This paper makes use of the following ALMA data: ADS/JAO.ALMA\#2022.A.00025.T. ALMA is a partnership of ESO (representing its member states), NSF (USA), and NINS (Japan), together with NRC (Canada), MOST and ASIAA (Taiwan), and KASI (Republic of Korea), in cooperation with the Republic of Chile. The Joint ALMA Observatory is operated by ESO, AUI/NRAO, and NAOJ. The National Radio Astronomy Observatory is a facility of the NSF operated under cooperative agreement by Associated Universities, Inc.

This work made use of data supplied by the UK Swift Science Data Centre at the University of Leicester.
The scientific results reported in this article are based in part on observations made by the {\it Chandra} X-ray Observatory. This research has made use of software provided by the Chandra X-ray Center (CXC) in the application packages CIAO and Sherpa.

\bibliography{main}{}
\bibliographystyle{aasjournal}

\appendix

\section{Optical data}
\label{sec:optical-data}

\startlongtable
\begin{deluxetable}{llll}
\tablecolumns{4}
\tablecaption{Optical photometry of {\lcr}. \label{tab:optical-photometry}}
\tablehead{\colhead{Start Date (MJD)} & \colhead{Instrument$^{a}$} & \colhead{Filter} & \colhead{Mag$^{b}$}}
\startdata
60112.26888 & P48 & $r$ & $>21.61$ \\
60112.30249 & P48 & $r$ & $>21.63$ \\
60112.31690 & P48 & $i$ & $>20.76$ \\
60112.36221 & P48 & $g$ & $>21.67$ \\
60112.40275 & P48 & $g$ & $>21.54$ \\
60113.27531 & P48 & $g$ & $19.63\pm0.05$ \\
60113.32655 & P48 & $g$ & $19.57\pm0.05$ \\
60113.33749 & P48 & $r$ & $19.29\pm0.04$ \\
60113.36493 & P48 & $r$ & $19.17\pm0.04$ \\
60114.26641 & P48 & $g$ & $20.53\pm0.15$ \\
60114.27940 & P48 & $g$ & $20.73\pm0.16$ \\
60114.32899 & P48 & $r$ & $20.35\pm0.12$ \\
60114.34106 & P48 & $r$ & $20.36\pm0.10$ \\
60114.35403 & P48 & $r$ & $20.24\pm0.10$ \\
60114.40855 & P48 & $g$ & $20.75\pm0.17$ \\
60114.73413 & GIT & $r$ & $20.67\pm0.06$ \\
60114.77990 & GIT & $g$ & $21.37\pm0.15$ \\
60114.80649 & GIT & $i$ & $20.70\pm0.16$ \\
60114.92765 & LT & $g$ & $21.32\pm0.08$ \\
60114.93103 & LT & $r$ & $20.88\pm0.06$ \\
60114.93440 & LT & $i$ & $20.68\pm0.06$ \\
60114.94962 & LT & $r$ & $21.04\pm0.06$ \\
60114.95300 & LT & $z$ & $20.47\pm0.10$ \\
60114.95733 & LT & $u$ & $22.31\pm0.68$ \\
60115.27025 & P48 & $g$ & $>21.65$ \\
60115.29196 & P48 & $i$ & $>21.09$ \\
60115.31154 & P48 & $r$ & $>21.54$ \\
60115.35355 & P48 & $r$ & $20.98\pm0.19$ \\
60115.43086 & P48 & $g$ & $>21.30$ \\
60115.70108 & GIT & $r$ & $21.26\pm0.08$ \\
60116.00182 & LT & $g$ & $21.77\pm0.22$ \\
60116.00592 & LT & $r$ & $21.53\pm0.14$ \\
60116.00928 & LT & $i$ & $21.40\pm0.16$ \\
60116.01265 & LT & $z$ & $21.14\pm0.20$ \\
60116.71280 & GIT & $r$ & $21.79\pm0.08$ \\
60117.02526 & LT & $g$ & $22.14\pm0.21$ \\
60117.03004 & LT & $r$ & $21.89\pm0.18$ \\
60117.03410 & LT & $i$ & $21.64\pm0.19$ \\
60117.68238 & GIT & $g$ & $>21.86$ \\
60117.74601 & GIT & $r$ & $21.92\pm0.17$ \\
60117.97296 & LT & $i$ & $22.21\pm0.25$ \\
60117.97904 & LT & $r$ & $22.04\pm0.20$ \\
60117.98509 & LT & $g$ & $22.34\pm0.24$ \\
60118.70342 & GIT & $r$ & $22.24\pm0.10$ \\
60118.99971 & LT & $i$ & $22.62\pm0.28$ \\
60119.00955 & LT & $r$ & $22.93\pm0.27$ \\
60119.01938 & LT & $g$ & $23.28\pm0.38$ \\
60119.82540 & HCT & $r$ & $22.74\pm0.18$ \\
60120.02784 & LT & $i$ & $23.15\pm0.40$ \\
60120.03768 & LT & $r$ & $22.68\pm0.24$ \\
60120.04750 & LT & $g$ & $23.59\pm0.40$ \\
\enddata
\tablenotetext{a}{P48: Palomar Observatory 48-inch Samuel Oschin Telescope; GIT: GROWTH-India Telescope; LT: Liverpool Telescope.}
\tablenotetext{b}{Not corrected for Milky Way extinction}
\end{deluxetable}

\section{Radio data}
\label{sec:radio-data}

\startlongtable
\begin{deluxetable}{cllll}
\tablecolumns{5}
\tablecaption{Radio observations of {\lcr}$^a$ \label{tab:radio}}
\tablehead{
\colhead{Epoch} & \colhead{Start MJD} & \colhead{Instrument$^{b}$} & \colhead{$\nu_\mathrm{obs}$ (GHz)} & \colhead{$f_\nu$ ($\mu$Jy)}}
\startdata
1 & 60116.26547 & VLA & 8.5 & $109\pm9$ \\
1 & 60116.26547 & VLA & 9.5 & $140\pm9$ \\
1 & 60116.26547 & VLA & 10.5 & $164\pm9$ \\
1 & 60116.26547 & VLA & 11.5 & $282\pm11$ \\
2 & 60117.26563 & VLA & 4.5 & $267\pm11$ \\
2 & 60117.26563 & VLA & 5.5 & $256\pm11$ \\
2 & 60117.26563 & VLA & 6.5 & $257\pm10$ \\
2 & 60117.26563 & VLA & 7.5 & $375\pm10$ \\
2 & 60117.28324 & VLA & 8.5 & $561\pm11$ \\
2 & 60117.28324 & VLA & 9.5 & $625\pm10$ \\
2 & 60117.28324 & VLA & 10.5 & $666\pm11$ \\
2 & 60117.28324 & VLA & 11.5 & $710\pm12$ \\
2 & 60117.30085 & VLA & 2.2 & $48\pm21$ \\
2 & 60117.30085 & VLA & 2.8 & $35\pm19$ \\
2 & 60117.30085 & VLA & 3.2 & $151\pm16$ \\
2 & 60117.30085 & VLA & 3.8 & $227\pm16$ \\
 & 60119.14653 & SMA & 230.0 & $<600$ \\
3 & 60119.29234 & VLA & 13.0 & $631\pm11$ \\
3 & 60119.29234 & VLA & 15.0 & $672\pm10$ \\
3 & 60119.29234 & VLA & 17.0 & $699\pm13$ \\
3 & 60119.30164 & VLA & 4.5 & $298\pm14$ \\
3 & 60119.30164 & VLA & 5.5 & $382\pm12$ \\
3 & 60119.30164 & VLA & 6.5 & $356\pm11$ \\
3 & 60119.30164 & VLA & 7.5 & $374\pm11$ \\
3 & 60119.31363 & VLA & 8.5 & $516\pm12$ \\
3 & 60119.31363 & VLA & 9.5 & $582\pm12$ \\
3 & 60119.31363 & VLA & 10.5 & $621\pm12$ \\
3 & 60119.31363 & VLA & 11.5 & $677\pm13$ \\
3 & 60119.32570 & VLA & 2.2 & $108\pm21$ \\
3 & 60119.32570 & VLA & 2.8 & $112\pm18$ \\
3 & 60119.32570 & VLA & 3.2 & $192\pm16$ \\
3 & 60119.32570 & VLA & 3.8 & $218\pm15$ \\
4 & 60122.25104 & VLA & 4.5 & $144\pm14$ \\
4 & 60122.25104 & VLA & 5.5 & $200\pm14$ \\
4 & 60122.25104 & VLA & 6.5 & $242\pm12$ \\
4 & 60122.25104 & VLA & 7.5 & $290\pm12$ \\
4 & 60122.25556 & VLA & 5.0 & $151\pm14$ \\
4 & 60122.25556 & VLA & 7.0 & $235\pm13$ \\
4 & 60122.25754 & VLA & 1.4 & $83\pm25$ \\
4 & 60122.25754 & VLA & 1.8 & $52\pm18$ \\
4 & 60122.28969 & VLA & 2.2 & $102\pm21$ \\
4 & 60122.28969 & VLA & 2.8 & $100\pm18$ \\
4 & 60122.28969 & VLA & 3.2 & $151\pm16$ \\
4 & 60122.28969 & VLA & 3.8 & $153\pm16$ \\
4 & 60122.30117 & VLA & 8.5 & $352\pm13$ \\
4 & 60122.30117 & VLA & 9.5 & $396\pm8$ \\
4 & 60122.30117 & VLA & 10.5 & $394\pm13$ \\
4 & 60122.30117 & VLA & 11.5 & $412\pm14$ \\
4 & 60122.31543 & VLA & 19.0 & $410\pm14$ \\
4 & 60122.31543 & VLA & 21.0 & $432\pm21$ \\
4 & 60122.31543 & VLA & 23.0 & $372\pm36$ \\
4 & 60122.31543 & VLA & 25.0 & $376\pm19$ \\
4 & 60122.34082 & VLA & 30.0 & $292\pm21$ \\
4 & 60122.34082 & VLA & 32.0 & $340\pm25$ \\
4 & 60122.34082 & VLA & 34.0 & $370\pm24$ \\
4 & 60122.34082 & VLA & 36.0 & $316\pm26$ \\
4 & 60122.36696 & VLA & 13.0 & $405\pm13$ \\
4 & 60122.36696 & VLA & 15.0 & $387\pm13$ \\
4 & 60122.36696 & VLA & 17.0 & $395\pm17$ \\
4 & 60122.37847 & VLA & 5.0 & $195\pm14$ \\
4 & 60122.37847 & VLA & 7.0 & $299\pm12$ \\
5 & 60123.23863 & VLA & 4.5 & $322\pm13$ \\
5 & 60123.23863 & VLA & 5.5 & $435\pm13$ \\
5 & 60123.23863 & VLA & 6.5 & $559\pm12$ \\
5 & 60123.23863 & VLA & 7.5 & $613\pm12$ \\
5 & 60123.24306 & VLA & 5.0 & $347\pm12$ \\
5 & 60123.24306 & VLA & 7.0 & $514\pm12$ \\
5 & 60123.24511 & VLA & 1.4 & $85\pm21$ \\
5 & 60123.24511 & VLA & 1.8 & $84\pm17$ \\
5 & 60123.27727 & VLA & 2.2 & $112\pm17$ \\
5 & 60123.27727 & VLA & 2.8 & $123\pm15$ \\
5 & 60123.27727 & VLA & 3.2 & $222\pm14$ \\
5 & 60123.27727 & VLA & 3.8 & $246\pm12$ \\
5 & 60123.29294 & VLA & 8.5 & $690\pm14$ \\
5 & 60123.29294 & VLA & 9.5 & $705\pm13$ \\
5 & 60123.29294 & VLA & 10.5 & $643\pm10$ \\
5 & 60123.29294 & VLA & 11.5 & $635\pm15$ \\
5 & 60123.31137 & VLA & 19.0 & $382\pm14$ \\
5 & 60123.31137 & VLA & 21.0 & $354\pm19$ \\
5 & 60123.31137 & VLA & 23.0 & $378\pm20$ \\
5 & 60123.31137 & VLA & 25.0 & $372\pm16$ \\
5 & 60123.33328 & VLA & 31.0 & $321\pm19$ \\
5 & 60123.35726 & VLA & 13.0 & $456\pm12$ \\
5 & 60123.35726 & VLA & 15.0 & $379\pm12$ \\
5 & 60123.35726 & VLA & 17.0 & $335\pm15$ \\
5 & 60123.36806 & VLA & 5.0 & $372\pm13$ \\
5 & 60123.36806 & VLA & 7.0 & $629\pm12$ \\
 & 60124.15000 & ALMA & 90.5 & $163\pm29$ \\
 & 60124.15000 & ALMA & 92.4 & $166\pm25$ \\
 & 60124.15000 & ALMA & 102.5 & $147\pm26$ \\
 & 60124.15000 & ALMA & 105.5 & $132\pm23$ \\
6 & 60125.05966 & VLA & 4.5 & $191\pm12$ \\
6 & 60125.05966 & VLA & 5.5 & $163\pm11$ \\
6 & 60125.05966 & VLA & 6.5 & $232\pm11$ \\
6 & 60125.05966 & VLA & 7.5 & $265\pm12$ \\
6 & 60125.06250 & VLA & 5.0 & $172\pm12$ \\
6 & 60125.06250 & VLA & 7.0 & $212\pm12$ \\
6 & 60125.06815 & VLA & 1.4 & $96\pm20$ \\
6 & 60125.06815 & VLA & 1.8 & $94\pm17$ \\
6 & 60125.08433 & VLA & 2.2 & $109\pm18$ \\
6 & 60125.08433 & VLA & 2.8 & $153\pm15$ \\
6 & 60125.08433 & VLA & 3.2 & $150\pm13$ \\
6 & 60125.08433 & VLA & 3.8 & $167\pm13$ \\
6 & 60125.10000 & VLA & 8.5 & $324\pm14$ \\
6 & 60125.10000 & VLA & 9.5 & $361\pm13$ \\
6 & 60125.10000 & VLA & 10.5 & $340\pm13$ \\
6 & 60125.10000 & VLA & 11.5 & $353\pm15$ \\
6 & 60125.11298 & VLA & 13.0 & $336\pm12$ \\
6 & 60125.11298 & VLA & 15.0 & $298\pm11$ \\
6 & 60125.11298 & VLA & 17.0 & $277\pm15$ \\
6 & 60125.12500 & VLA & 5.0 & $212\pm12$ \\
6 & 60125.12500 & VLA & 7.0 & $256\pm12$ \\
& 60129.04549 & ALMA & 90.5 & $139\pm20$ \\
& 60129.04549 & ALMA & 92.4 & $120\pm18$ \\
& 60129.04549 & ALMA & 102.5 & $<69$ \\
& 60129.04549 & ALMA & 105.5 & $<60$ \\
7 & 60130.99178 & VLA & 4.5 & $148\pm9$ \\
7 & 60130.99178 & VLA & 5.5 & $115\pm9$ \\
7 & 60130.99178 & VLA & 6.5 & $108\pm12$ \\
7 & 60130.99178 & VLA & 7.5 & $117\pm13$ \\
7 & 60130.99306 & VLA & 5.0 & $141\pm15$ \\
7 & 60130.99306 & VLA & 7.0 & $89\pm13$ \\
7 & 60130.99826 & VLA & 1.4 & $68\pm22$ \\
7 & 60130.99826 & VLA & 1.8 & $143\pm19$ \\
7 & 60131.01449 & VLA & 2.2 & $158\pm21$ \\
7 & 60131.01449 & VLA & 2.8 & $142\pm16$ \\
7 & 60131.01449 & VLA & 3.2 & $124\pm14$ \\
7 & 60131.01449 & VLA & 3.8 & $161\pm13$ \\
7 & 60131.03141 & VLA & 8.5 & $135\pm13$ \\
7 & 60131.03141 & VLA & 9.5 & $107\pm12$ \\
7 & 60131.03141 & VLA & 10.5 & $92\pm10$ \\
7 & 60131.03141 & VLA & 11.5 & $100\pm15$ \\
7 & 60131.04550 & VLA & 20.0 & $94\pm15$ \\
7 & 60131.04550 & VLA & 24.0 & $92\pm17$ \\
7 & 60131.06055 & VLA & 13.0 & $137\pm12$ \\
7 & 60131.06055 & VLA & 15.0 & $131\pm11$ \\
7 & 60131.06055 & VLA & 17.0 & $103\pm14$ \\
7 & 60131.07292 & VLA & 5.0 & $162\pm14$ \\
7 & 60131.07292 & VLA & 7.0 & $147\pm12$ \\
8 & 60143.98697 & VLA & 4.5 & $113\pm13$ \\
8 & 60143.98697 & VLA & 5.5 & $124\pm12$ \\
8 & 60143.98697 & VLA & 6.5 & $123\pm13$ \\
8 & 60143.98697 & VLA & 7.5 & $129\pm13$ \\
8 & 60143.98958 & VLA & 5.0 & $91\pm15$ \\
8 & 60143.98958 & VLA & 7.0 & $102\pm13$ \\
8 & 60143.99346 & VLA & 1.4 & $104\pm23$ \\
8 & 60143.99346 & VLA & 1.8 & $68\pm20$ \\
8 & 60144.00969 & VLA & 2.2 & $41\pm20$ \\
8 & 60144.00969 & VLA & 2.8 & $80\pm16$ \\
8 & 60144.00969 & VLA & 3.2 & $77\pm14$ \\
8 & 60144.00969 & VLA & 3.8 & $68\pm13$ \\
8 & 60144.02565 & VLA & 8.5 & $110\pm15$ \\
8 & 60144.02565 & VLA & 9.5 & $110\pm15$ \\
8 & 60144.02565 & VLA & 10.5 & $99\pm15$ \\
8 & 60144.02565 & VLA & 11.5 & $78\pm16$ \\
8 & 60144.03977 & VLA & 13.0 & $40\pm16$ \\
8 & 60144.03977 & VLA & 15.0 & $51\pm10$ \\
8 & 60144.03977 & VLA & 17.0 & $41\pm12$ \\
8 & 60144.05556 & VLA & 5.0 & $152\pm13$ \\
8 & 60144.05556 & VLA & 7.0 & $152\pm12$ \\
9 & 60175.63000 & GMRT & 1.4 & $135\pm26$ \\
9 & 60175.87542 & VLA & 5.0 & $32\pm10$ \\
9 & 60175.87542 & VLA & 7.0 & $0\pm30$ \\
9 & 60175.88190 & VLA & 3.5 & $25\pm9$ \\
9 & 60175.90322 & VLA & 10.0 & $0\pm29$ \\
 & 60177.63000 & GMRT & 0.6 & $<90$ \\
 & 60178.63000 & GMRT & 0.4 & $<420$ \\
10 & 60187.11613 & VLA & 5.0 & $46\pm6$ \\
10 & 60187.11613 & VLA & 7.0 & $32\pm6$ \\
10 & 60187.13373 & VLA & 9.0 & $23\pm6$ \\
10 & 60187.13373 & VLA & 11.0 & $22\pm6$ \\
10 & 60187.15885 & VLA & 13.0 & $35\pm9$ \\
10 & 60187.15885 & VLA & 15.0 & $35\pm9$ \\
10 & 60187.15885 & VLA & 17.0 & $0\pm50$ \\
11 & 60216.00837 & VLA & 5.0 & $23\pm3$ \\
11 & 60216.00837 & VLA & 7.0 & $19\pm2$ \\
\enddata
\tablenotetext{a}{Upper limits reported as 3$\times$ image RMS. Epochs of VLA observations are numbered. We report statistical errors, but include systematic errors in our reported reduced $\chi^2$/DoFs.
\tablenotetext{b}{VLA: Karl G. Jansky Very Large Array; SMA: Submillimeter Array; ALMA: Atacama Large Millimeter/sub-millimeter Array; GMRT: Giant Metrewave Radio Telescope.}
}
\end{deluxetable}

\section{Corner Plots}
\label{app:corner-plots}

\begin{figure*}[ht]
    \centering
    \includegraphics[width=\linewidth]{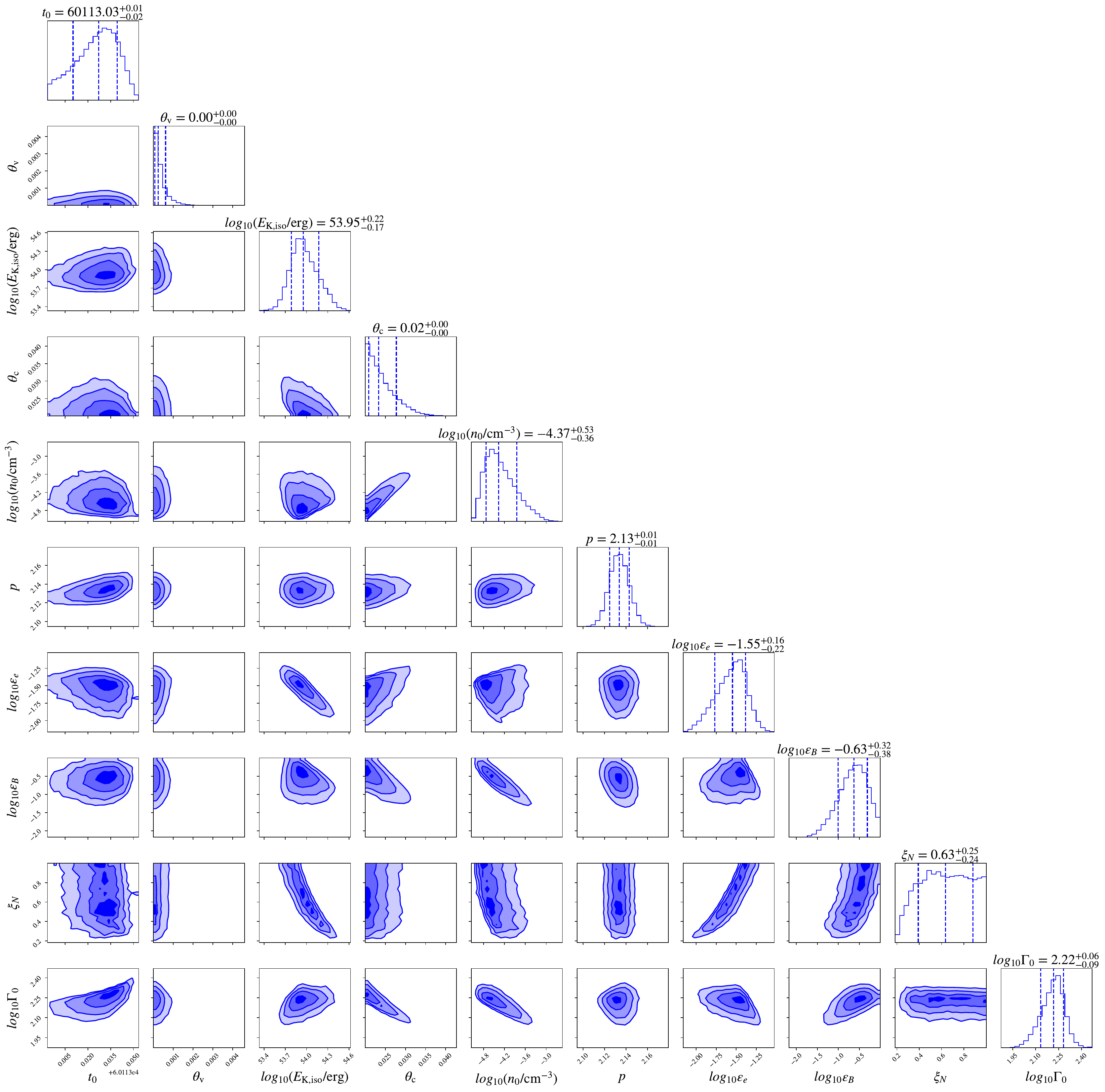}
    \caption{Corner plots ($68\%$ uncertainties) of the on-axis, $\Gamma_0 \approx 166$, top hat configuration for {\lcr}. Ran with 64 walkers and 75,000 iterations; discarded 25,000.}
    \label{fig:lcr-corner}
\end{figure*}

\begin{figure*}[ht]
    \centering
    \includegraphics[width=\linewidth]{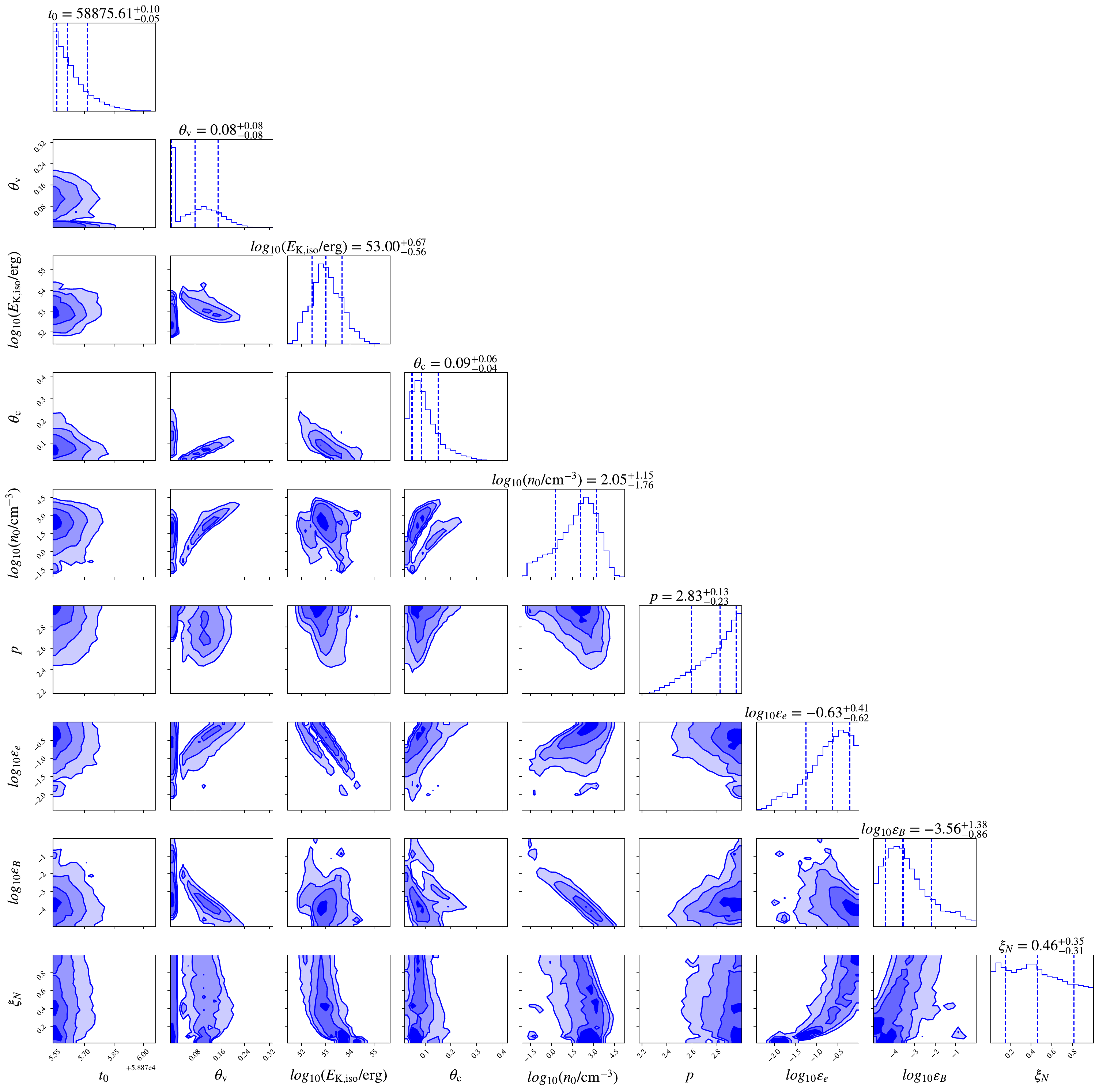}
    \caption{Corner plots ($68\%$ uncertainties) of the on-axis, $\Gamma_0 = \infty$, top hat configuration for {\blt}. Ran with 64 walkers and 225,000 iterations; discarded 125,000.}
    \label{fig:blt-corner}
\end{figure*}

\begin{figure*}[ht]
    \centering
    \includegraphics[width=\linewidth]{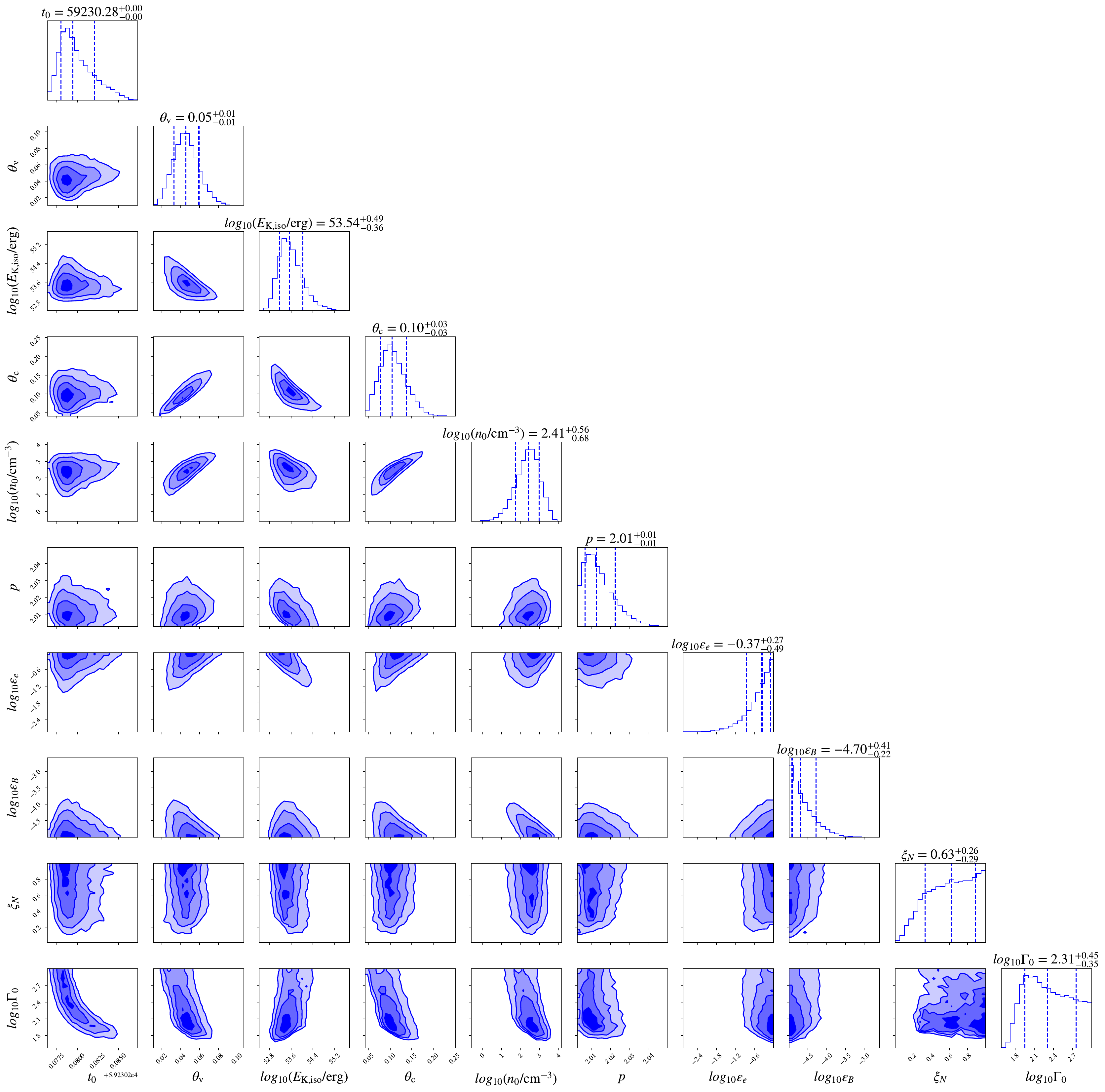}
    \caption{Corner plots ($68\%$ uncertainties) of the on-axis, $\Gamma_0 \neq \infty$, top hat configuration for {\any}. Ran with 64 walkers and 75,000 iterations; discarded 25,000.}
    \label{fig:any-corner}
\end{figure*}

\begin{figure*}[ht]
    \centering
    \includegraphics[width=\linewidth]{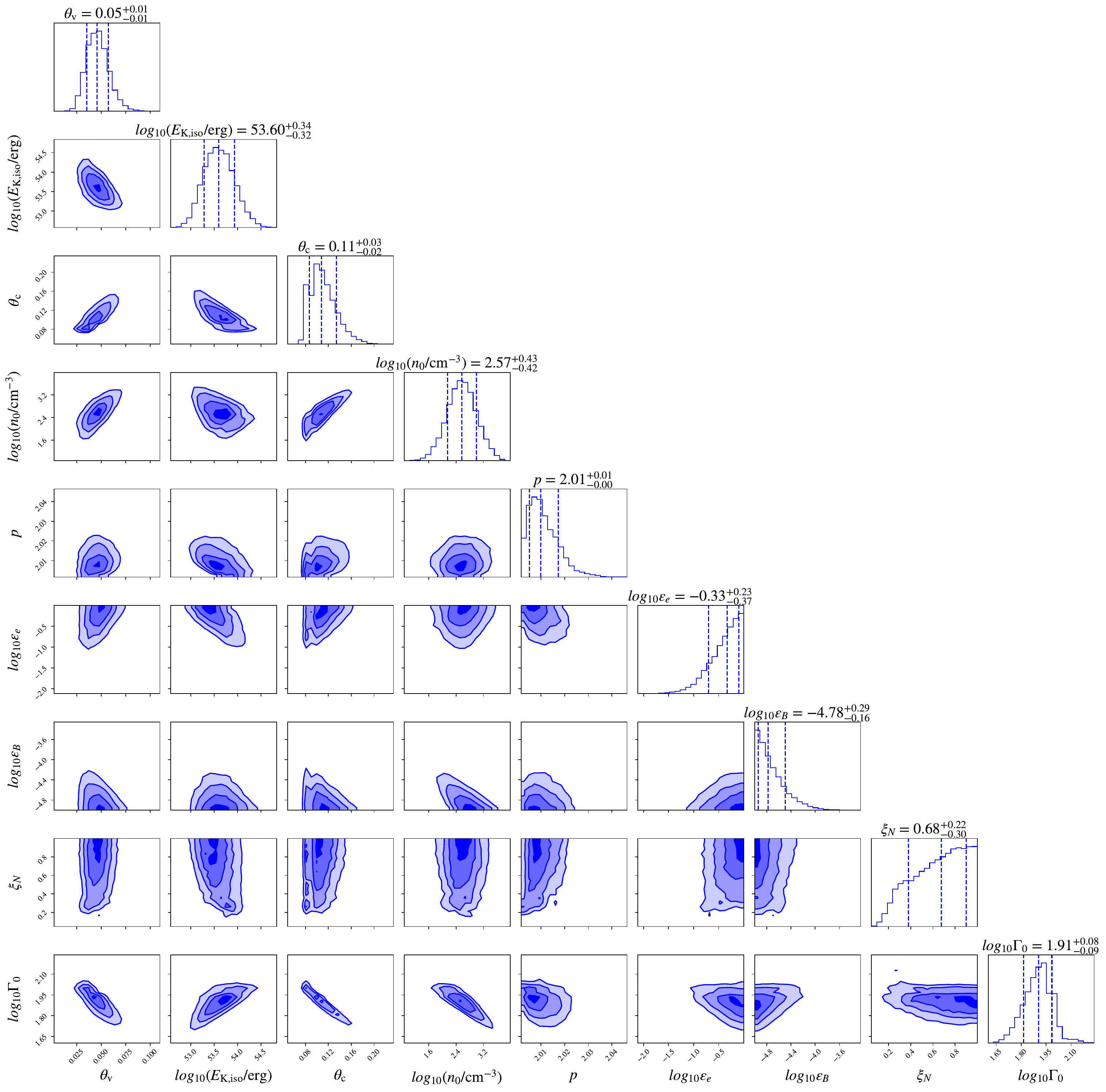}
    \caption{Corner plots ($68\%$ uncertainties) of the on-axis, fixed $t_0=59230.290$ MJD, $\Gamma_0 \neq \infty$, top hat configuration for {\any}. Ran with 64 walkers and 75,000 iterations; discarded 25,000.}
    \label{fig:any-t0-corner}
\end{figure*}

\begin{figure*}[ht]
    \centering
    \includegraphics[width=\linewidth]{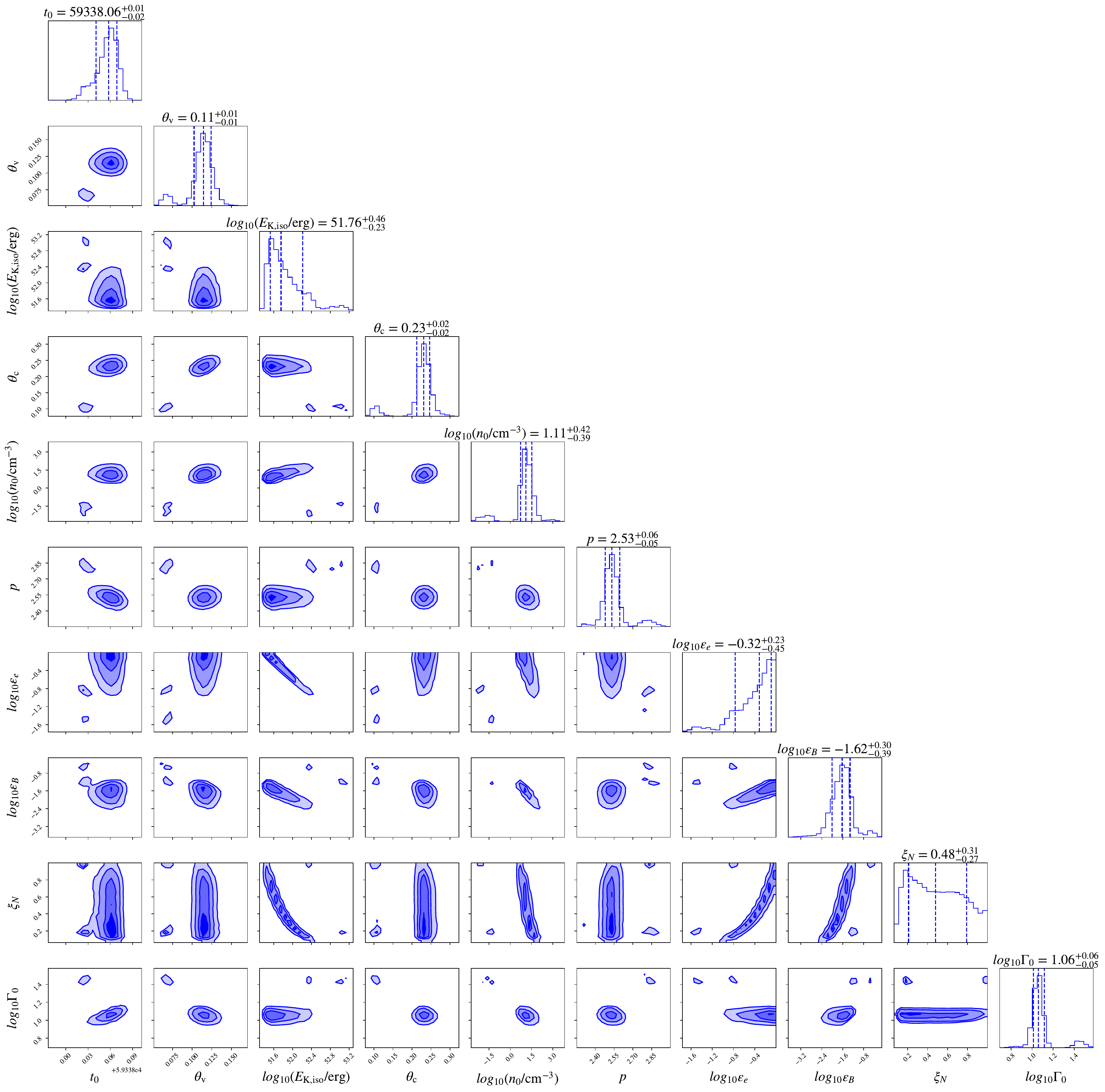}
    \caption{Corner plots ($68\%$ uncertainties) of the on-axis, $\Gamma_0 \neq \infty$, top hat configuration for {\lfa}. Ran with 64 walkers and 225,000 iterations; discarded 125,000.}
    \label{fig:lfa-corner}
\end{figure*}

\begin{figure*}[ht]
    \centering
    \includegraphics[width=\linewidth]{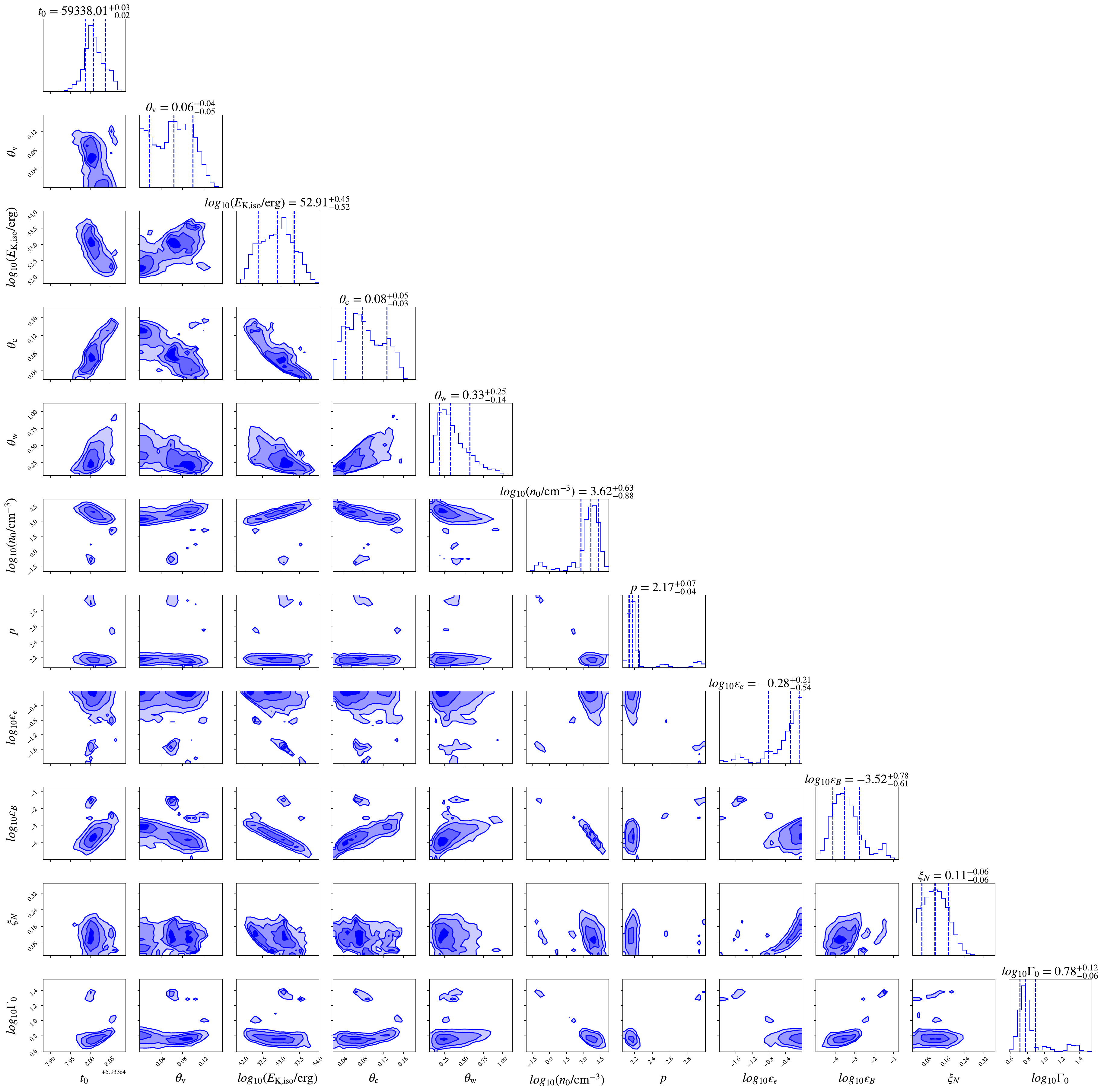}
    \caption{Corner plots ($68\%$ uncertainties) of the on-axis, $\Gamma_0 \neq \infty$, Gaussian configuration for {\lfa}. Ran with 64 walkers and 225,000 iterations; discarded 125,000.}
    \label{fig:lfa-corner-gauss}
\end{figure*}

\begin{figure*}[ht]
    \centering
    \includegraphics[width=\linewidth]{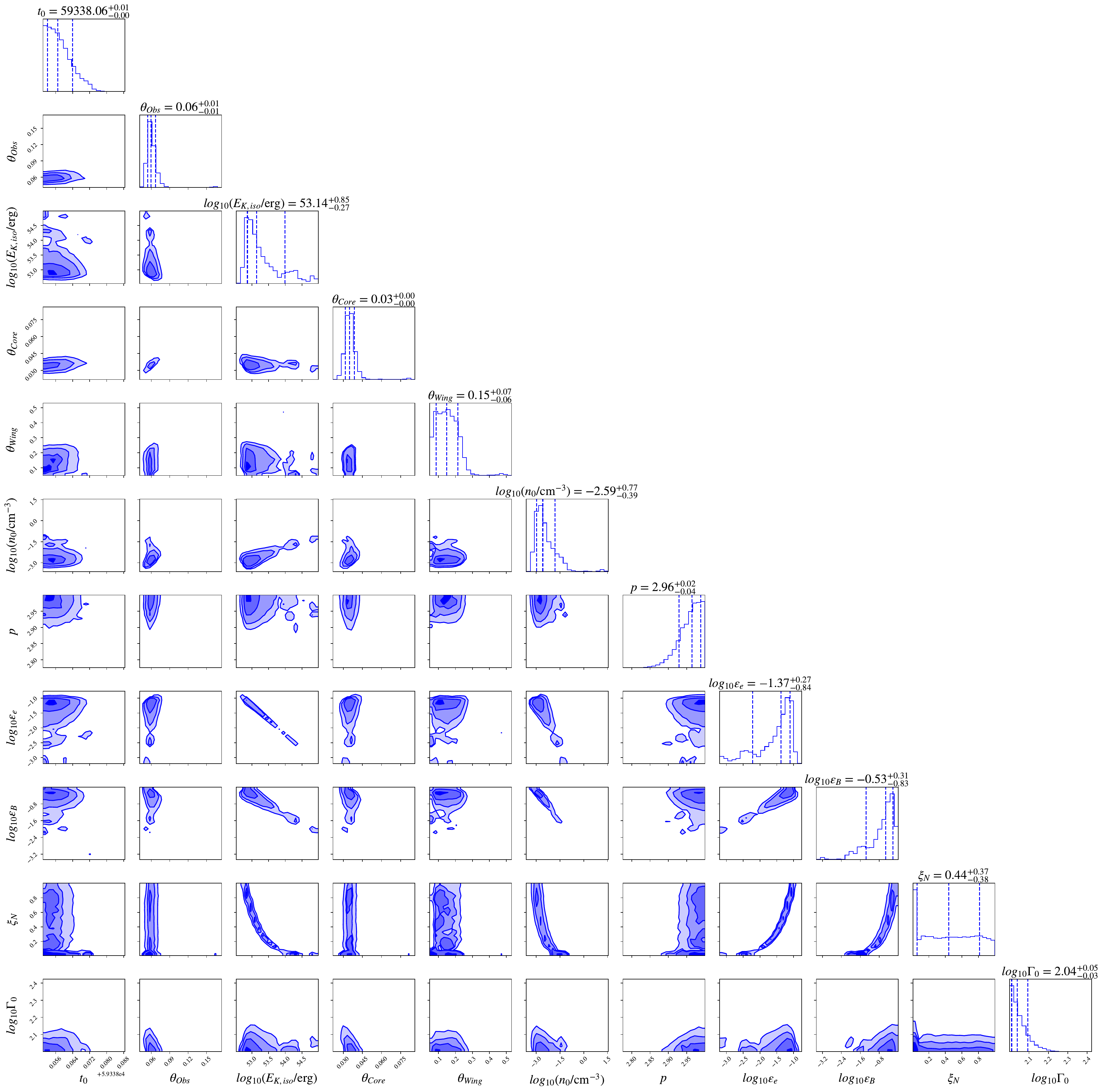}
    \caption{Corner plots ($68\%$ uncertainties) of the off-axis, $\Gamma_0 \neq \infty$, Gaussian configuration for {\lfa}. Ran with 64 walkers and 225,000 iterations; discarded 125,000.}
    \label{fig:lfa-off-axis-corner-gauss}
\end{figure*}

\begin{figure*}[ht]
    \centering
    \includegraphics[width=\linewidth]{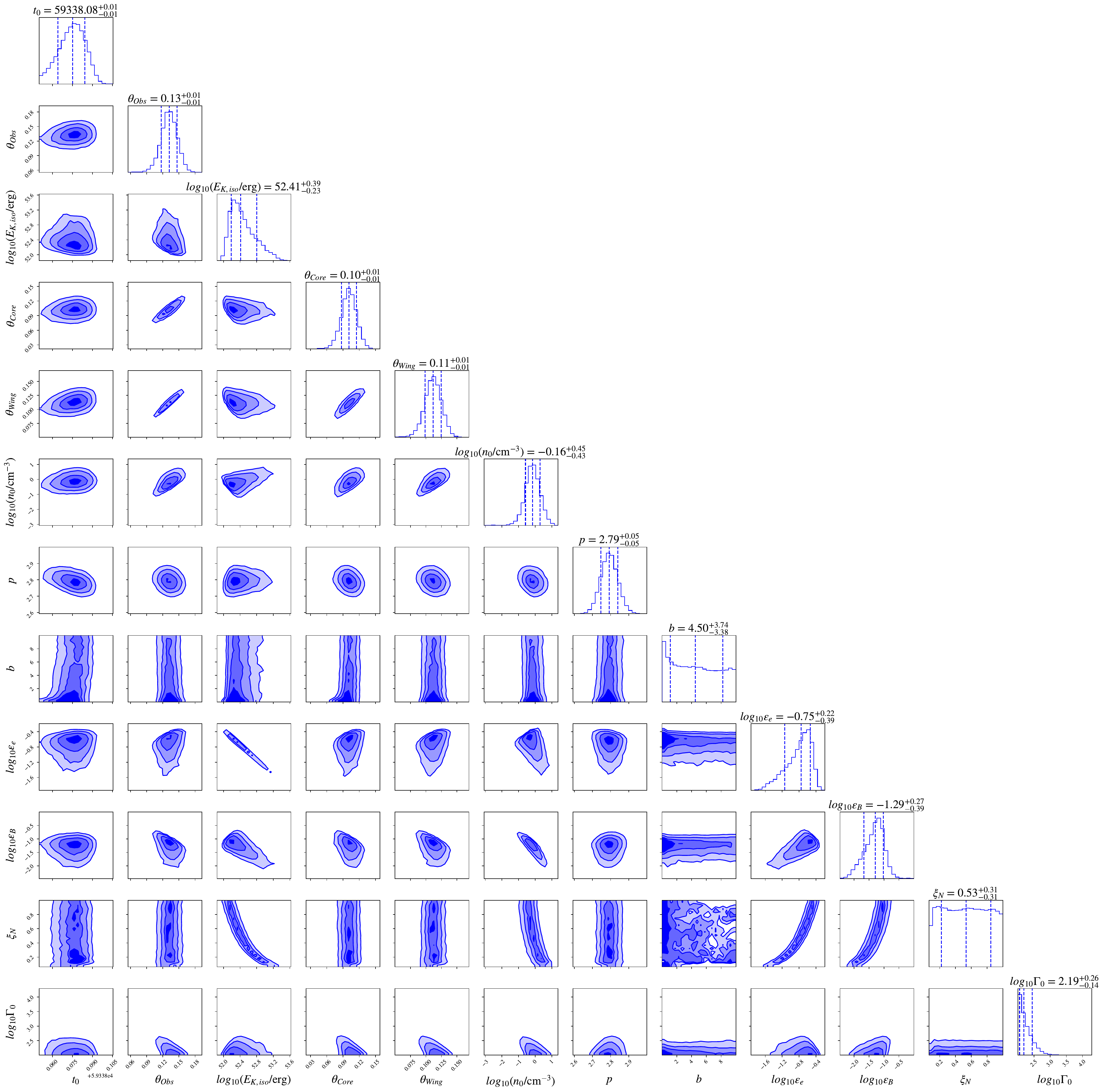}
    \caption{Corner plots ($68\%$ uncertainties) of the off-axis, $\Gamma_0 \neq \infty$, power law configuration for {\lfa}. Ran with 64 walkers and 225,000 iterations; discarded 125,000.}
    \label{fig:lfa-off-axis-corner-power}
\end{figure*}

\clearpage
\section{Selected configurations}
\label{app:selected}

\begin{table}[ht]
\normalsize
\centering
\begin{tabular}{llll}
\hline
Parameter & Gaussian & Top hat & Top hat\\
&  & (ICC) & ($\Gamma_0=\infty$)\\
\hline
$t_0$ [MJD] & $60113.03_{-0.02}^{+0.01}$ & $60113.02_{-0.02}^{+0.01}$ & $60112.99_{-0.00}^{+0.00}$ \\

$\theta_\text{v}$ [rad & $0.00_{-0.00}^{+0.00}$ & $0.00_{-0.00}^{+0.00}$ & $0.00_{-0.00}^{+0.00}$ \\

$\log_{10}(E_\text{K,iso}/\text{erg})$ & $53.93_{-0.17}^{+0.22}$ & $54.14_{-0.27}^{+0.67}$ & $54.51_{-0.20}^{+0.25}$ \\

$\theta_\text{c}$ [rad] & $0.02_{-0.00}^{+0.00}$ & $0.03_{-0.01}^{+0.00}$ & $0.02_{-0.00}^{+0.00}$ \\

$\theta_\text{w}$ & $0.02_{-0.00}^{+0.00}$ & - & - \\

$\log_{10}(n_0/\text{cm}^{-3})$ & $-4.20_{-0.41}^{+0.54}$ & $-3.81_{-0.74}^{+1.21}$ & $-5.22_{-0.19}^{+0.25}$\\

$p$ & $2.14_{-0.01}^{+0.01}$ & $2.14_{-0.01}^{+0.01}$ & $2.09_{-0.01}^{+0.01}$ \\

$\log_{10}\epsilon_e$ & $-1.52_{-0.22}^{+0.16}$ & $-1.42_{-0.29}^{+0.48}$ & $-2.23_{-0.28}^{+0.24}$\\

$\log_{10}\epsilon_B$ & $-0.73_{-0.40}^{+0.34}$ & $-1.16_{-2.89}^{+0.66}$ & $-0.21_{-0.26}^{+0.15}$ \\

$\xi_N$ & $0.63_{-0.25}^{+0.24}$ & $0.60_{-0.25}^{+0.27}$ & $0.20_{-0.09}^{+0.14}$ \\

$\log_{10}\Gamma_0$ & $2.19_{-0.08}^{+0.07}$ & $2.15_{-0.09}^{+0.10}$ & $\infty$ \\

$\eta_\gamma$ ($z=1.0272$) & $< 1.5-3.5\%$ & $< 0.3-2.8\%$ & $< 0.4-1.0\%$ \\

$\eta_\gamma$ ($z=1.6$) & $< 3.4-7.8\%$ & $< 0.8-6.2\%$ & $< 0.8-2.3\%$ \\

$\chi^2$/DoF & $10.8$ & $7.9$ & $14.3$ \\

{\elpd} & $(-5.7 \pm 7.9) \times 10^{2}$ & $31.8 \pm 62.3$ & $(-2.5 \pm 2.2) \times 10^{2}$\\
\hline
\end{tabular}
\caption{Final parameters ($68\%$ uncertainty) for selected configurations of {\lcr}. We calculate $\eta_\gamma$ using the $1\sigma$ distribution of {\Ekiso} and the {\Egiso} limits from Table~\ref{table:radiative-energies}. We present the {\elpd} and minimum $\chi^2$/DoF over 5,000 posterior samples. Ran with 64 walkers and 75,000 iterations; discarded 25,000.}
\label{tab:lcr-selected}
\end{table}
\vspace{-0.35cm}

\begin{table}[ht]
\normalsize
\centering
\begin{tabular}{llll}
\hline
Parameter & Gaussian$^\ddag$ & Top hat$^\ddag$ & Top hat$^\ddag$\\
& ($\Gamma_0\neq\infty$) & (ICC, $\Gamma_0\neq\infty$) & ($\Gamma_0\neq\infty$)\\
\hline
$t_0$ [MJD] & $58875.67_{-0.08}^{+0.11}$ & $58875.78_{-0.12}^{+0.13}$ & $58875.66_{-0.08}^{+0.11}$ \\

$\theta_\text{v}$ [rad] & $0.09_{-0.04}^{+0.04}$ & $0.01_{-0.00}^{+0.01}$ & $0.07_{-0.07}^{+0.05}$ \\

$\log_{10}(E_\text{K,iso}/\text{erg})$ & $52.77_{-0.53}^{+0.58}$ & $53.56_{-0.75}^{+0.84}$ & $52.97_{-0.88}^{+0.71}$\\

$\theta_\text{c}$ [rad] & $0.05_{-0.02}^{+0.03}$ & $0.08_{-0.01}^{+0.01}$ & $0.07_{-0.03}^{+0.09}$\\

$\theta_\text{w}$ & $0.16_{-0.08}^{+0.14}$ & - & -\\

$\log_{10}(n_0/\text{cm}^{-3})$ & $1.91_{-1.02}^{+1.00}$ & $0.47_{-0.64}^{+1.16}$ & $1.85_{-1.24}^{+1.10}$ \\

$p$ & $2.96_{-0.06}^{+0.03}$ & $2.95_{-0.08}^{+0.04}$ & $2.95_{-0.06}^{+0.03}$\\

$\log_{10}\epsilon_e$ & $-0.47_{-0.43}^{+0.28}$ & $-0.94_{-1.06}^{+0.55}$ & $-0.58_{-0.44}^{+0.33}$\\

$\log_{10}\epsilon_B$ & $-2.51_{-1.33}^{+1.13}$ & $-2.98_{-1.32}^{+1.39}$ & $-2.74_{-1.39}^{+1.31}$\\

$\xi_N$ & $0.49_{-0.29}^{+0.32}$ & $0.23_{-0.21}^{+0.48}$ & $0.51_{-0.32}^{+0.32}$\\

$\log_{10}\Gamma_0$ & $2.11_{-0.41}^{+0.50}$ & $2.49_{-0.42}^{+0.35}$ & $2.00_{-0.50}^{+0.54}$\\

$\eta_\gamma$ & $<4.3-36.5\%$ & $<0.4-13.4\%$ & $<2.0-44.8\%$ \\

$\eta_\gamma$ (\textit{Fermi}) & $<0.4-5.4\%$ & $<0.04-1.5\%$ & $<0.2-7.5\%$ \\

$\chi^{2 ^ \dag}$ & $3.6$ & $3.5$ & $3.4$ \\

{\elpd} & $38.9 \pm 38.2$ & $47.1 \pm 35.1$ & $5.3 \pm 60.5$\\
\hline
\end{tabular}
\caption{Final parameters ($68\%$ uncertainty) for selected configurations of {\blt}. We calculate $\eta_\gamma$ using the $1\sigma$ distribution of {\Ekiso} and the {\Egiso} limits from Table~\ref{table:radiative-energies}. We present the {\elpd} and minimum $\chi^2$/DoF over 5,000 posterior samples. Ran with 64 walkers and 75,000 iterations; discarded 25,000.\\
$^\dag$ $\chi^2$ does not account for non-detections.\\
$^\ddag$ Fails to account for the radio non-detection at $\sim25$\,d and struggles or fails with the radio non-detection at $\sim100$\,d}
\label{tab:blt-selected}
\end{table}

\begin{table}[ht]
\normalsize
\centering
\begin{tabular}{llll}
\hline
Parameter & Gaussian & Top hat & Top hat\\
&  & (ICC) & ($\Gamma_0=\infty$)\\
\hline
$t_0$ [MJD] & $59230.28_{-0.00}^{+0.00}$ & $59230.29_{-0.00}^{+0.00}$ & $59230.28_{-0.00}^{+0.00}$ \\

$\theta_\text{v}$ [rad] & $0.04_{-0.01}^{+0.01}$ & $0.03_{-0.00}^{+0.01}$ & $0.04_{-0.02}^{+0.03}$ \\

$\log_{10}(E_\text{K,iso}/\text{erg})$ & $53.31_{-0.32}^{+0.42}$ & $53.87_{-0.53}^{+0.62}$ & $53.94_{-0.73}^{+1.33}$ \\

$\theta_\text{c}$ [rad] & $0.09_{-0.02}^{+0.02}$ & $0.03_{-0.01}^{+0.01}$ & $0.08_{-0.05}^{+0.07}$ \\

$\theta_\text{w}$ & $0.34_{-0.18}^{+0.25}$ & - & -\\

$\log_{10}(n_0/\text{cm}^{-3})$ & $2.79_{-0.54}^{+0.50}$ & $-1.04_{-0.48}^{+0.52}$ & $1.66_{-1.80}^{+1.22}$ \\

$p$ & $2.01_{-0.01}^{+0.01}$ & $2.04_{-0.02}^{+0.03}$ & $2.01_{-0.01}^{+0.01}$\\

$\log_{10}\epsilon_e$ & $-0.25_{-0.39}^{+0.18}$ & $-0.30_{-0.30}^{+0.21}$ & $-0.79_{-1.35}^{+0.64}$\\

$\log_{10}\epsilon_B$ & $-4.78_{-0.16}^{+0.32}$ & $-2.28_{-1.39}^{+1.14}$ & $-4.49_{-0.40}^{+1.02}$\\

$\xi_N$ & $0.64_{-0.28}^{+0.25}$ & $0.26_{-0.12}^{+0.20}$ & $0.59_{-0.31}^{+0.28}$\\

$\log_{10}\Gamma_0$ & $2.28_{-0.36}^{+0.45}$ & $2.90_{-0.09}^{+0.07}$ & $\infty$ \\

$\eta_\gamma$ & $< 21.0-59.4\%$ & $< 4.4-39.5\%$ & $<0.7-46.9\%$\\

$\chi^2$/DoF & $16.9$ & $10.4^\dag$ & $12.4$ \\

{\elpd} & $6.4 \pm 34.1$ & $34.2 \pm 22.4$ & $21.2 \pm 26.5$\\
\hline
\end{tabular}
\caption{Final parameters ($68\%$ uncertainty) for selected configurations of {\any}. We calculate $\eta_\gamma$ using the $1\sigma$ distribution of {\Ekiso} and the {\Egiso} limits from Table~\ref{table:radiative-energies}. We present the {\elpd} and minimum $\chi^2$/DoF over 5,000 posterior samples. Ran with 64 walkers and 75,000 iterations; discarded 25,000.\\
$^\dag$ Although this is the smallest $\chi^2$ of all {\any} configurations shown in this work, the top hat with ICC model underestimates the X-ray by $\sim 1/2$ an order of magnitude; the smaller $\chi^2$ is likely because this model has a slightly better agreement with X-band (9.0 - 9.7\,GHz) observations.}
\label{tab:any-selected}
\end{table}

\begin{table}[ht]
\normalsize
\centering
\begin{tabular}{llll}
\hline
Parameter & Gaussian & Top hat\\
& ($\Gamma_0=\infty$) & (ICC)\\
\hline
$t_0$ [MJD] & $59337.19_{-0.17}^{+0.19}$ & $59338.06_{-0.01}^{+0.01}$ \\

$\theta_\text{v}$ [rad & $0.20_{-0.01}^{+0.01}$ & $0.05_{-0.00}^{+0.00}$ \\

$\log_{10}(E_\text{K,iso}/\text{erg})$ & $54.95_{-0.31}^{+0.20}$ & $52.66_{-0.28}^{+0.42}$ \\

$\theta_\text{c}$ [rad] & $0.02_{-0.00}^{+0.01}$ & $0.12_{-0.01}^{+0.01}$ \\

$\theta_\text{w}$ & $0.10_{-0.05}^{+0.05}$ & - \\

$\log_{10}(n_0/\text{cm}^{-3})$ & $5.14_{-0.33}^{+0.25}$ & $-0.52_{-0.27}^{+0.38}$ \\

$p$ & $2.13_{-0.04}^{+0.05}$ & $2.64_{-0.05}^{+0.05}$ \\

$\log_{10}\epsilon_e$ & $-0.14_{-0.17}^{+0.10}$ & $-0.56_{-0.37}^{+0.21}$ \\

$\log_{10}\epsilon_B$ & $-4.71_{-0.20}^{+0.26}$ & $-1.88_{-0.69}^{+0.57}$ \\

$\xi_N$ & $0.02_{-0.01}^{+0.01}$ & $0.55_{-0.31}^{+0.30}$ \\

$\log_{10}\Gamma_0$ & $\infty$ & $1.34_{-0.02}^{+0.03}$ \\

$\eta_\gamma$ & $<0.01-0.03\%$ & $<1.0-4.8\%$ \\

$\chi^2$/DoF & $5.4$ & $4.5$ \\

{\elpd} & $(-1.0 \pm 0.3) \times 10^{2}$ & $(1.3 \pm 0.2) \times 10^{2}$ \\
\hline
\end{tabular}
\caption{Final parameters ($68\%$ uncertainty) for selected configurations of {\lfa}. We calculate $\eta_\gamma$ using the $1\sigma$ distribution of {\Ekiso} and the {\Egiso} limits from Table~\ref{table:radiative-energies}. We present the {\elpd} and minimum $\chi^2$/DoF over 5,000 posterior samples. Ran with 64 walkers and 75,000 iterations; discarded 25,000.}
\label{tab:lfa-selected}
\end{table}

\end{document}